\documentclass[prl,twocolumn,floatfix,superscriptaddress,citeautoscript]{revtex4-2}

\usepackage{graphicx}
\usepackage{dcolumn}   % needed for some tables
\usepackage{bm}        % for math
\usepackage{amssymb}   % for math
\usepackage{verbatim}
\usepackage{multirow}
\usepackage{xcolor} % for \textcolor
\usepackage{amsmath}
\usepackage{physics}
\usepackage{amsfonts}
\usepackage{hyperref}
\usepackage[normalem]{ulem}
\hypersetup{colorlinks=true, pdfstartview=FitV, linkcolor=blue, citecolor=black, plainpages=false, pdfpagelabels=true, urlcolor=blue}
\usepackage[all]{hypcap}

\begin{document}

%\title{Implementing quantum stabilizers with superconducting hardware}
%\title{Protecting quantum information in complex superconducting circuits}
\title{Hardware implementation of quantum stabilizers in superconducting circuits}

\author{K. Dodge}
\thanks{These authors contributed equally}
\affiliation{Department of Physics, Syracuse University, Syracuse, NY 13244-1130}
\author{Y. Liu}
\thanks{These authors contributed equally}
\affiliation{Department of Physics, Syracuse University, Syracuse, NY 13244-1130}
\author{A. R. Klots}
\affiliation{Google Quantum AI, Santa Barbara, California 93111, USA}
\author{B. Cole}
\affiliation{Department of Physics, Syracuse University, Syracuse, NY 13244-1130}
\author{A. Shearrow}
\affiliation{Department of Physics, University of Wisconsin-Madison, Madison, Wisconsin 53706}
\author{M. Senatore}
\affiliation{Department of Physics, Syracuse University, Syracuse, NY 13244-1130}
\author{S. Zhu}
\affiliation{Department of Physics, University of Wisconsin-Madison, Madison, Wisconsin 53706}
\author{L. B. Ioffe}
\affiliation{Google Quantum AI, Santa Barbara, California 93111, USA}
\author{R. McDermott}
\affiliation{Department of Physics, University of Wisconsin-Madison, Madison, Wisconsin 53706}
\author{B. L. T. Plourde}
\email[]{bplourde@syr.edu}
\affiliation{Department of Physics, Syracuse University, Syracuse, NY 13244-1130}

\date{\today}

\begin{abstract}
%PRL --> < 600 characters
Stabilizer operations are at the heart of quantum error correction and are typically implemented in software-controlled entangling gates and measurements of groups of qubits. Alternatively, qubits can be designed so that the Hamiltonian corresponds directly to a stabilizer for protecting quantum information. We demonstrate such a hardware implementation of stabilizers in a superconducting circuit composed of chains of $\pi$-periodic Josephson elements. With local on-chip flux- and charge-biasing, we observe a progressive softening of the energy band dispersion with respect to flux as the 
%that is exponential in the 
number of frustrated plaquette elements  is increased, in close agreement with our numerical modeling.
\end{abstract}

\maketitle

%Introduction
Protecting fragile information in quantum processors requires some form of quantum error correction (QEC). With typical ``software" QEC techniques such as the surface code \cite{Fowler2012}, stabilizing a single logical qubit requires many physical qubits, each of which is typically implemented as a weakly nonlinear oscillator. Error correction and computation is achieved by a string of operations and measurements that allow identification of bit-flip and phase-flip errors. An alternative is to implement quantum stabilizers directly in hardware. Here, error correction arises from the natural quantum dynamics, reducing the need for repeated entangling gates, measurements, and a multitude of control lines and complex classical control hardware. In this approach, the highly non-trivial Hamiltonian results in a tiny protected subspace within a huge Hilbert space.

Both approaches can be characterized by the error suppression factor $\Lambda$, the rate at which the logical error decreases with system size. The long time required by each round of software error correction for current transmon qubit arrays implies that $\Lambda$ is only marginally greater than one \cite{Google2021QEC}. In this work, we experimentally demonstrate the potential to achieve much larger $\Lambda\gtrsim$100 with the Hamiltonian approach. The price that one pays is the appearance of relatively low energy modes with gaps $\lesssim 1 \rm{\;GHz}$ that make initialization challenging; these gaps can be made higher through parameter optimization. Before building a scalable logical qubit with hardware QEC, it is crucial to demonstrate the effectiveness of protection based on Hamiltonian engineering as system size increases. In this manuscript, we observe and quantify the stabilizing interaction Hamiltonian between unprotected elements. We perform spectroscopic measurements with local flux control and observe signatures of stabilizer terms in the Hamiltonian. Specifically, we find 
%exponential 
a progressive flattening of the energy bands with respect to flux as system size increases, consistent with linear flux dispersion for a system size of one, quadratic for two, and cubic for three. In addition, we observe a characteristic periodic modulation with offset charge as we tune between regimes with different levels of protection.

\begin{figure*}
\centering
\includegraphics[width=6.8in]{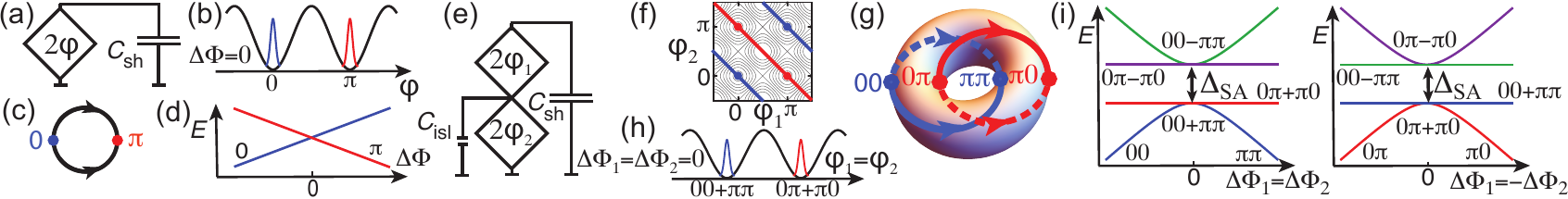}
  \caption{{\bf Concatenation of $\pi$-periodic plaquettes.} (a) Schematic of single plaquette shunted by $C_{\rm sh}$. (b) $\cos2\varphi$ potential at frustration ($\Delta\Phi=\Phi-\Phi_0/2=0$) with localized wavefunctions in 0 and $\pi$ wells, drawn in the $0,\pi$ basis for vanishingly small tunneling. (c) Sketch of CW and CCW tunneling paths for $\varphi$ going between 0,$\pi$ wells indicated by blue/red dots. (d) Linear flux dispersion of 0 and $\pi$ levels for vanishing tunnel splitting. (e) Schematic of two plaquettes shunted by $C_{\rm sh}$ with small capacitance $C_{\rm isl}$ from intermediate island to ground. Potential with respect to phase across each plaquette displayed on (f) contour plot, and (g) surface of torus; blue (red) lines correspond to hybridized even (odd) parity states; arrows indicate CW/CCW tunneling paths between wells of same parity. (h) 1D cut of effective potential at double frustration. (i) Quadratic dispersion of even (odd)-parity levels and flat dispersion of odd (even)-parity levels near double frustration for simultaneous scan of plaquette fluxes along $\Delta\Phi_1=\Delta\Phi_2$ ($\Delta\Phi_1=-\Delta\Phi_2$) on left (right) (sketches do not include higher levels within a well).
\label{fig:plq-schem}}
\end{figure*}

A variety of qubit designs with intrinsic protection against decoherence have been studied previously \cite{Doucot2012,gyenis2021review}, including the 0$-\pi$ qubit \cite{Brooks13,groszkowski2018,gyenis2021}, the two-Cooper-pair tunneling qubit \cite{Smith2020}, the bifluxon qubit \cite{kalashnikov2020}, and rhombi arrays \cite{Ioffe2002,Gladchenko2009,Bell2014}. In this last work, previous devices had limited symmetry due to the inability to tune each element to the optimum flux independently; in addition, the devices were sensitive to offset charge fluctuations on internal nodes in each element, and the suppression of tunneling between the logical states was limited. Similar to previous protected qubit designs, our device is based on $\pi$-periodic Josephson elements \cite{Smith2022}, for which the Josephson energy is proportional to $\cos2\varphi$, where $\varphi$ is the superconducting phase difference across the element. Here, charge transport consists of coherent tunneling of $4e$, as opposed to $2e$ for a conventional junction. We implement each element as a plaquette formed from a dc Superconducting QUantum Interference Device (SQUID), consisting of two conventional Josephson junctions and a non-negligible loop inductance. When flux-biased at frustration, $\Phi_0/2$ ($\Phi_0 \equiv h/2e$), the first harmonic of the Josephson energy (proportional to $\cos\varphi$) vanishes. This leaves a second order term $E_2\cos2\varphi$, with sequential minima separated by $\pi$; $E_2$ depends on the Josephson energy of the individual junctions $E_J$ and the energy of the SQUID inductance $E_L$ (Supplement \cite{supplement}, Sec. XI); $\varphi$ is thus a compact variable residing on a circle. Biasing below (above) $\Phi_0/2$ raises (lowers) the $\pi$ wells relative to the $0$ wells; for flux bias at 0$\,{\rm mod}\,\Phi_0$, the potential becomes proportional to $\cos\varphi$. A small asymmetry between the two junctions has a similar but less severe effect on the $\cos2\varphi$ potential compared to a small flux deviation from frustration (Supplement \cite{supplement}, Sec. I). 

For a single frustrated plaquette with a large capacitive shunt $C_{\rm sh}$ [Fig.~\ref{fig:plq-schem}(a)], tunneling between the ground states in the $0,\pi$ wells is suppressed. In the phase basis, wavefunctions localized in the $0,\pi$ wells are thus disjoint and well protected against bit-flip errors. At the same time, the wavefunctions are spread out in the charge basis, corresponding for the 0($\pi$) states to superpositions of even (odd) multiples of Cooper pairs on the logical island where the plaquette connects to $C_{\rm sh}$. For bias away from frustration, the energy levels disperse linearly [Fig.~\ref{fig:plq-schem}(d)], with no protection against phase flips due to flux noise. 

We next consider concatenation of multiple plaquettes while maintaining the large shunt $C_{\rm sh}$ across the array. At double frustration, when two plaquettes are simultaneously biased to $\Phi_0/2$, there are four minima in the two-dimensional surface defined by the phase drops across each plaquette: $00,\pi\pi,0\pi,\pi 0$. For the two-plaquette circuit this has the topology of a torus, since $\varphi$ for each plaquette is a compact variable with $2\pi$ periodicity [Fig.~\ref{fig:plq-schem}(f,g)]. If the capacitance of the intermediate island between plaquettes to ground $C_{\rm isl}$ is sufficiently small, with charging energy $E_C^{\rm isl}=(2e)^2/2C_{\rm isl}> E_J$, quantum fluctuations of the island phase cause hybridization along the direction between wells of the same parity; that is, $00$ will hybridize with $\pi\pi$ and $0\pi$ with $\pi 0$. Levels with the same parity develop a splitting near double frustration, with ground states corresponding to the symmetric superpositions $00+\pi\pi$ ($0\pi+\pi 0$) for even (odd) parity. Excited states are given by the antisymmetric superpositions 00-$\pi\pi$ ($0\pi$-$\pi 0$) for even (odd) parity; these states are separated by an energy $\Delta_{\rm SA}$ from the symmetric ground state of the same parity. The hybridized ground state wavefunctions of opposite parity are the logical states for the device [Fig.~\ref{fig:plq-schem}(h)] and form interlocking rings on the torus [Fig.~\ref{fig:plq-schem}(g)]. Due to delocalization and intertwining of the hybridized ground state wavefunctions, local perturbations affect the logical states symmetrically. Larger $E_C^{\rm isl}$ increases $\Delta_{\rm SA}$ and further flattens the bands [Fig.~\ref{fig:plq-schem}(i)], thus protecting against dephasing from flux noise. 

Treating each plaquette as a spin-1/2 particle, the $\Delta_{\rm SA}$ splitting corresponds to an $XX$ stabilizer term in the Hamiltonian of frustrated plaquettes $i,j$: $H_{XX}=-(\Delta_{\rm SA}^{(ij)}/2)X_i X_j$, where $X_i$ is the Pauli $\sigma_x$ matrix for plaquette $i$. The error suppression factor $\Lambda$ can be approximated as the ratio of $\Delta_{\rm SA}^{(ij)}$ to twice the scale $h_Z$ of dephasing fluctuations for single plaquette $i$, $\delta H(t) = h_Z(t) Z_i$, which, for this device, will be dominated by flux noise (Supplement \cite{supplement}, Sec.~XII). $C_{\rm sh}$ still suppresses tunneling between logical states of opposite parity, protecting against bit-flip errors.

%Experimental scheme
In our experiments, we target a three-plaquette circuit with $E_J\sim E_L\sim 1.5\,{\rm K}$ ($k_B$=1), where $E_L$ is the energy $\left(\Phi_0/2\pi\right)^2/L$ of the inductance $L$ on each plaquette arm. We aim for a charging energy of each plaquette junction $E_C=(2e)^2/2 C_j\sim 3.5\,{\rm K}$, where $C_j$ is the junction capacitance. These values can be achieved with conventional Al-AlO$_x$-Al junctions. We implement the inductors with chains of large-area junctions, similar to fluxonium \cite{Manucharyan2009}, thus eliminating charge fluctuations on the internal nodes between each small junction and inductor within a plaquette. The shunt capacitor $C_{\rm sh}$=1.2~pF is capacitively coupled to a resonator. There are four flux-bias lines, each of which couples strongly to one or two plaquettes. There are three  charge-bias lines: one to the logical island that forms $C_{\rm sh}$, and one to each intermediate island between plaquettes (Supplement \cite{supplement}, Sec. II-V).

\begin{figure}
\centering
\includegraphics[width=3.35in]{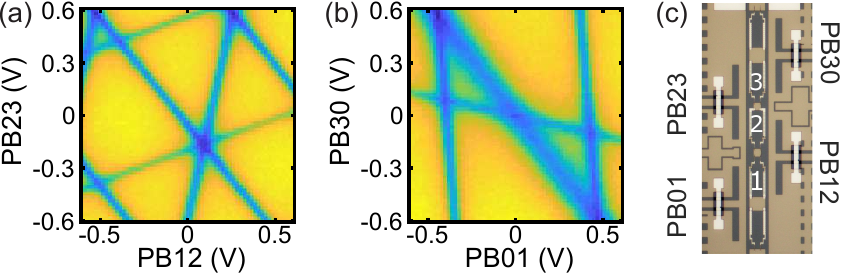}
  \caption{{\bf Multi-plaquette flux biasing.} 2D flux-modulation scans of readout cavity dispersive shift for (a) PB23 vs. PB12, (b) PB30 vs. PB01. (c) Optical micrograph of device.
\label{fig:cavity-flux-scans}}
\end{figure}

%Flux scans of cavity
For device tune-up, we scan various pairs of flux-bias lines while monitoring the dispersive shift of the readout resonator. Each blue line in Fig.~\ref{fig:cavity-flux-scans}(a,b) corresponds to one plaquette passing through frustration. A crossing of two (three) lines indicates double (triple) frustration. The spacing between parallel sets of lines defines the period $\Phi_0$. We fit the slopes and spacing of the lines to extract the inductance matrix mapping bias levels on each flux line to net flux coupled to each plaquette (Supplement \cite{supplement}, Sec. VI). By inverting this matrix, we determine  bias parameters for moving along arbitrary flux vectors. 

\begin{figure}
\centering
\includegraphics[width=3.4in]{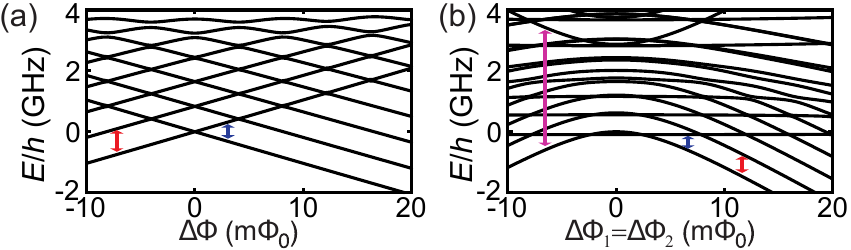}
  \caption{{\bf Level transitions.} Simulated level diagrams near (a) single and (b) double frustration; lines indicate example plasmons (red), heavy fluxons (blue), light fluxons (magenta).
  \label{fig:plq-E-levels}}
\end{figure}

%energy levels and transitions
We next map out the flux dispersion of the level transitions for different frustration conditions. With our ability to adjust the various plaquette fluxes independently using local flux-biasing, we maintain some plaquettes at unfrustration (0$\,{\rm mod}\,\Phi_0$), where the plaquette behaves like a conventional Josephson element, while we scan the flux of other plaquettes near frustration. In Fig.~\ref{fig:plq-E-levels}, we consider the expected level structure and define the types of possible transitions. We refer to transitions between levels in the same well as {\it plasmons}; transitions between different wells are referred to as {\it heavy fluxons} because of the vanishingly small gap associated with the corresponding anticrossing, a consequence of the large effective mass from $C_{\rm sh}$. Transitions between hybridized levels of the same parity but opposite symmetry, for example, $00$+$\pi\pi$ to $00$-$\pi\pi$, disperse sharply with flux; these are known as {\it light fluxons} due to the low effective mass in the $\varphi_2=-\varphi_1$ direction from the smallness of $C_{\rm isl}$.

\begin{figure*}[!ht]
\centering
\includegraphics[width=6.8in]{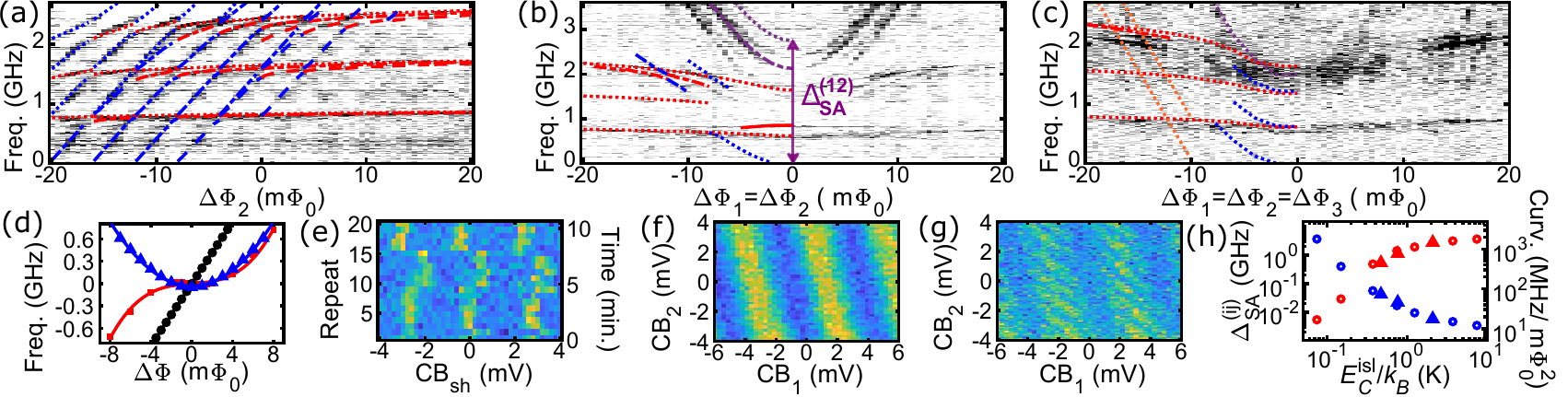}
  \caption{{\bf Spectroscopy at different frustration points.} Spectroscopy at (a) plaquette 2 single frustration, (b) plaquette (12) double frustration, and (c) triple frustration.  
  Lines indicate modeled transitions with: red = plasmons, blue = heavy fluxons, purple = light fluxons, dotted = transitions out of 0 level, dash-dotted = transitions out of 1 level, dashed = transitions out of 2 level, solid red line = plasmon transition between antisymmetric levels in even-parity well, orange = light fluxon plus cavity photon (Supplement \cite{supplement}, Sec.~X). (d) Comparison of dispersion of lowest heavy fluxon from modeled levels with linear, quadratic, and cubic fits for single (black circle), double (blue triangle), and triple (red square) frustration; frequency axis inverted for single and triple frustration for $\Delta\Phi<0$. (e) Repeated scans of cavity response vs. offset charge bias to $C_{\rm sh}$ island at plaquette 2 single frustration. 2D scan of spectroscopy at 0-1 transition frequency while scanning bias voltages to gate electrodes coupled to both intermediate islands for (f) plaquette (12) double frustration and (g) triple frustration. (h) Plot of $\Delta_{\rm SA}^{(ij)}$ and curvature of fluxon transition between even/odd-parity ground states vs. $E_C^{\rm isl}$ showing measured values for plaquette (12), (23), and (13) double frustration (solid triangles) plus modeled values for a range of $C_{\rm isl}$ (open circles). 
\label{fig:multi-frustration}}
\end{figure*}

%Spectroscopy at single frustration
To perform spectroscopy, we drive a microwave probe tone into the charge bias line coupled to $C_{\rm sh}$ while monitoring the cavity dispersive shift. Near single frustration, we initialize in the $\pi$~well prior to each spectroscopy pulse by setting the bias to $0.1\,\Phi_0$ from frustration, thus moving out of the protected space; we then quickly ramp the bias to the measurement point and apply spectroscopy and readout pulses (Supplement \cite{supplement}, Sec. VII). In Fig.~\ref{fig:multi-frustration}(a), we show single-frustration measurements for plaquette 2. Features that disperse gradually correspond to plasmons within the $\pi$~well where the qubit is initialized. We continue to observe transitions out of the $\pi$ well even when the device is biased past frustration, where the $\pi$ well is higher in energy than the 0 well, due to suppressed tunneling between states of opposite parity. In addition to the 0-1, 0-2, and 0-3 transitions, we observe transitions out of excited states in the well, such as 1-2, 1-3, and 1-4, and even 2-3 and 2-4, due to insufficient cooling into the ground state of the $\pi$ well. Because of the spurious excitations to multiple levels, we are unable to apply initialization techniques that are commonly used for other low-gap qubits, such as heavy fluxonium \cite{Gusenkova2021,Vool2018}. Nevertheless, we observe only weak transitions out of the 0 well, indicating that we are predominantly preparing the circuit in the $\pi$~well. In addition to the plasmons, we also observe heavy fluxons that disperse linearly with flux, which arise from transitions between various levels in the $\pi$ and 0~wells, where the barrier to tunneling is small because the initial state is an excited level or the wells are tilted by the flux bias; note that we do not observe the heavy fluxon between the protected ground states in the 0 and $\pi$ wells, which are the logical levels. We observe similar behavior for plaquettes 1 and 3 (Supplement \cite{supplement}, Sec.~X).

% Fig4-multi-frustrations-BP.pdf

The curves included in Fig.~\ref{fig:multi-frustration}(a) are generated from detailed numerical modeling of the device energy levels (Supplement \cite{supplement}, Sec.~IX). With the ability to calculate the level spectrum, we adjust the circuit parameters to fit the measured transitions from the spectroscopic data (Supplement \cite{supplement}, Sec.~X). We observe excellent agreement, even capturing splittings that result when a fluxon crosses a plasmon due to resonant tunnel coupling between aligned levels in the 0 and $\pi$~wells. In addition, these splittings depend on the offset charge on the $C_{\rm sh}$ island [Fig.~\ref{fig:multi-frustration}(e)] due to Aharonov-Casher (A-C) interference \cite{Aharonov1984,Bell2016} between tunneling paths clockwise (CW) or counterclockwise (CCW) in the $\cos2\varphi$ potential [Fig.~\ref{fig:plq-schem}(c)] (Supplement \cite{supplement}, Sec. VIII). At single frustration, as expected, the heavy fluxon dispersion is linear down to zero energy, thus offering no protection against flux noise. 
%Because only one plaquette is frustrated, there is no hybridization, and, thus, no light fluxons. 

%Spectroscopy at double frustration
Upon tuning to double frustration, we observe a qualitatively different behavior. We initialize in the $\pi \pi$ well of the two-plaquette potential, then quickly ramp near double frustration. We scan both plaquette fluxes in tandem along the direction between the regimes with a global potential minimum at $\pi\pi$ and 00 and passing through double frustration. Spectroscopy at plaquette (12) double frustration shows plasmons similar to the single frustration measurements [Fig.~\ref{fig:multi-frustration}(b)]. However, unlike single frustration, where suppressed tunneling between the $0,\pi$ wells allows the device to remain in the $\pi$~well even after the flux is ramped well past frustration, at double frustration, the large symmetric-antisymmetric gap $\Delta_{\rm SA}^{(12)}$ causes an adiabatic transition from $\pi\pi$ to 00 upon passing through double frustration. At higher frequencies, we observe steeply dispersing light fluxons, with the minimum at double frustration corresponding to $\Delta_{\rm SA}^{(12)}$ from hybridization of the 00 and $\pi\pi$ wells. For scans along the odd-parity flux direction, or if the circuit is initialized in an odd-parity well and scanned in the even-parity flux direction, the spectral features become swapped [Fig.~1(i), Supplement \cite{supplement}, Sec. X.C].

As with spectroscopy at single frustration, we include curves for the various transitions from numerical modeling and fitting for double frustration [Fig.~\ref{fig:multi-frustration}(b)]. Here, the larger Hilbert space requires a significant increase in computational resources. Our modeled transition curves agree well with the measured spectroscopy, capturing both the plasmons and heavy fluxons. We are unable to directly drive a microwave transition between the logical states in the $00$+$\pi\pi$ and $0\pi$+$\pi 0$ wells due to the vanishing matrix element, the basis of protection. However, the increasing flatness of the higher fluxon transitions as one moves lower in the spectrum indicates that the logical levels will be the flattest. This can also be seen in the blue modeled curves near the bottom of the figure highlighting the dispersion of the logical level transition, which exhibits quadratic curvature. Additionally, our modeling captures the light fluxons to the antisymmetric levels. 

The effectiveness of concatenation depends on $C_{\rm isl}$ of the intermediate island between the two frustrated plaquettes. For plaquette (12) double frustration, $\Delta_{\rm SA}^{(12)}$ is 2.7~GHz. At plaquette (23) double frustration, which involves a significantly larger $C_{\rm isl}$ because of the orientation of the plaquette 2 inductors, we observe a smaller $\Delta_{\rm SA}^{(23)}$ and a correspondingly larger curvature of the heavy fluxon transition. $\Delta_{\rm SA}^{(13)}$ is even smaller because of the excess capacitance to ground of the unfrustrated plaquette 2 (Supplement \cite{supplement}, Sec.~X). Figure~\ref{fig:multi-frustration}(h) shows the variation of $\Delta_{\rm SA}$ with $E_C^{\rm isl}$, including measured values of $\Delta_{\rm SA}^{(ij)}$ for each combination of double frustration, as well as numerically modeled values. For a typical flux noise level, $h_Z$ for these plaquettes will be $\sim$2~MHz, which, when combined with the measured $\Delta_{\rm SA}^{(12)}$, is consistent with $\Lambda\sim 700$. Note that this is an extracted parameter characterizing protection in one channel: dephasing. The complete $\Lambda$-parameter for a logical qubit must be derived from the scaling of $T_1$ and $T_2$ with system size, which is beyond the scope of this manuscript. Nonetheless, $\Lambda$ can also be expressed as the ratio of $T_2$ for a higher degree of frustration relative to $T_2$ at single frustration (Supplement \cite{supplement}, Sec. XII).

In addition to the symmetric/antisymmetric gap, another characteristic of the stabilizer term is the periodic modulation of $\Delta_{\rm SA}^{(ij)}$ with offset charge on the intermediate island between plaquettes $i$ and $j$. Destructive A-C interference of tunneling paths in the CW and CCW directions on the constant-parity circles for double frustration [Fig.~\ref{fig:plq-schem}(g)] causes $\Delta_{\rm SA}^{(ij)}$ to vanish for island offset charge near $e\,{\rm mod}\,2e$. We observe periodic modulation with charge bias to the islands with a spectroscopy pulse on the 0-1 transition [Fig.~\ref{fig:multi-frustration}(f)]. While the island offset charge is stable on timescales up to one hour, it is critical there are no jumps to near $e\,{\rm mod}\,2e$. Thus, it is important to actively stabilize these offset charges through periodic calibrations (Supplement \cite{supplement}, Sec.~VIII, IX).

By simultaneously frustrating all plaquettes, we measure spectroscopy near triple frustration [Fig.~\ref{fig:multi-frustration}(c)]. In this case, we are unable to numerically fit the level spectrum since the Hilbert space size becomes prohibitively large. Nonetheless, we are able to compute the spectrum using parameter values from previous fits to double and single frustration, although the calculation takes several weeks to complete. We obtain reasonable agreement with the measurements, although the spectral features are more challenging to resolve compared to other degrees of frustration; the higher transitions are off by $\sim$5-10\%, which is not unreasonable considering the circuit complexity and intertwined wavefunctions, given limitations on the number of quantum states needed for the computation to converge. Around 1.5~GHz, we observe a prominent central flat feature of width $\sim$7~m$\Phi_0$ around the 0-3 transition, which is uncharacteristic for parabolic, let alone linear, dispersion; below this, the 0-1 transition around 0.6~GHz is similarly flat. The transition between the logical states, which cannot be directly driven due to protection of these states from the environment, will be comparably flat (Supplement \cite{supplement}, Sec.~X.D).  Also, the light fluxon transitions are qualitatively different compared to double frustration. 
We 
%also 
additionally observe charge modulation with two different periods and slopes corresponding to separate tuning of offset charge on each intermediate island [Fig.~\ref{fig:multi-frustration}(g)], characteristic of a Hamiltonian with two stabilizer terms: $H_{XX}=-(\Delta_{\rm SA}^{(12)}/2)X_1 X_2-(\Delta_{\rm SA}^{(23)}/2)X_2 X_3$. For our present device $\Delta_{\rm SA}^{(23)}$ is smaller than $\Delta_{\rm SA}^{(12)}$ due to excess ground capacitance from plaquette 2, resulting in the logical level dispersion at triple frustration being only marginally flatter than at double frustration [Fig.~\ref{fig:multi-frustration}(d)] (Supplement \cite{supplement}, Sec.~XI).

%Outlook for implementing qubit... 
While our present device successfully demonstrates the implementation of stabilizer terms in hardware, development of protected qubits based on hybridized ground states of opposite parity requires larger gaps to the excited states. This, in conjunction with weaker radiative coupling to parasitic high-frequency modes from a more compact $C_{\rm sh}$, perhaps achieved using a parallel-plate rather than planar design, will avoid spurious excitations to multiple excited levels that complicate the initialization process for our present device. A device with higher excited-state energies that can be operated in the qubit regime requires larger $E_J$, ideally at least 3~K. We must also maintain even larger $E_C$ to have large $\Delta_{\rm SA}$ at double frustration with the resulting flat dispersion. 
%Achieving this with conventional Al electrodes is not possible due to the small superconducting gap and the electronic capacitance arising when the junction plasma frequency approaches the gap \cite{Eckern1984}. Thus, protected qubits incorporating this stabilizer mechanism will need junctions fabricated from a larger gap superconductor. 
For a qubit with these improved parameters subject to typical flux- and charge-noise levels, optimistic but feasible junction asymmetries, and dielectric loss from a parallel-plate $C_{\rm sh}$, we project $\Lambda\gtrsim$100, corresponding to $T_1\gg 1\,{\rm s}$ and $T_2\sim60\,{\rm ms}$ (Supplement \cite{supplement}, Sec.~XI), well beyond current state-of-the-art superconducting qubits. 

This work is supported by the U.S. Government under ARO grant W911NF-18-1-0106. Fabrication was performed in part at the Cornell NanoScale Facility, a member of the National Nanotechnology Coordinated Infrastructure (NNCI), which is supported by the National Science Foundation (Grant NNCI-2025233). Portions of this work were supported by the National Science Foundation, Quantum Leap Challenge Institute for Hybrid Quantum Architectures and Networks, Grant No. 2016136.

% \bibliography{plq-stabilizer}

\widetext
\clearpage
\begin{center}
\textbf{\large Supplementary Information: Hardware implementation of quantum stabilizers in superconducting circuits}
\end{center}

\def\thesection{\Roman{section}}

\setcounter{secnumdepth}{3}
\setcounter{equation}{0}
\setcounter{figure}{0}
\setcounter{table}{0}
\setcounter{page}{1}
\renewcommand{\theHtable}{Supplement.\thetable}
\renewcommand{\theHfigure}{Supplement.\thefigure}
\makeatletter
\renewcommand{\theequation}{S\arabic{equation}}
\renewcommand{\thefigure}{S\arabic{figure}}
\renewcommand{\thetable}{S\arabic{table}}

\newif\if@seccntdot

\def\@seccntformat#1{%
  \csname the#1\endcsname
  \if@seccntdot .\fi
  \quad
}

\makeatother
\renewcommand{\thetable}{\arabic{table}}
\renewcommand*{\thesection}{\Roman{section}.}
\renewcommand*{\thesubsection}{\Alph{subsection}.}
\renewcommand{\sectionautorefname}{14}
\renewcommand{\bibnumfmt}[1]{[S#1]}
\renewcommand{\citenumfont}[1]{S#1}

\section{$\pi$-periodic Josephson elements from dc SQUIDs}

\begin{figure*}[b]
\centering
\includegraphics[width=6.8in]{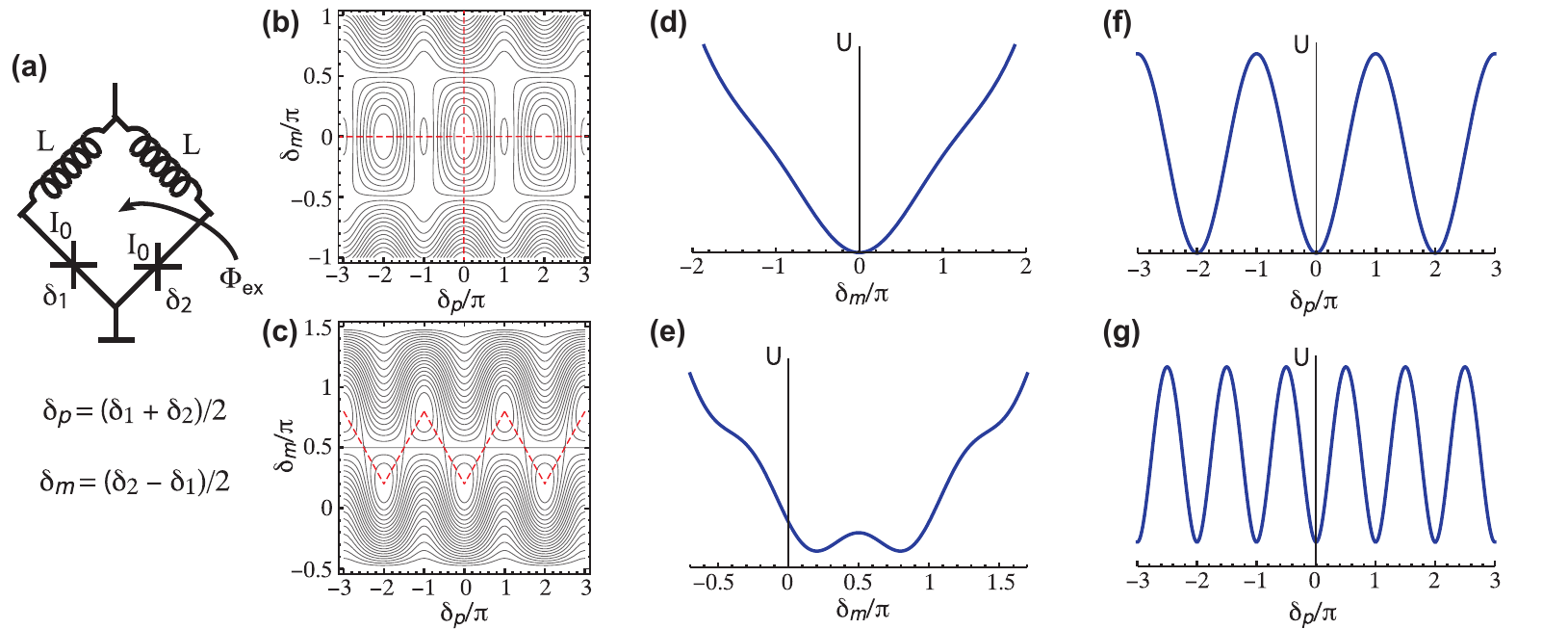}
  \caption{(a) Circuit schematic for dc SQUID plaquette. 2D potential as a function of common-mode ($\delta_p$) and differential ($\delta_m$) phase variables at external flux bias $\Phi_{\rm ex}$ of (b) 0, (c) $\Phi_0/2$. (d) Linecut along $\delta_m$ for $\delta_p=0$ for $\Phi_{\rm ex}=0$. (e) Linecut between adjacent minima vs. $\delta_m$ for $\Phi_{\rm ex}=\Phi_0/2$. (f) Linecut along $\delta_p$ at $\delta_m=0$ for $\Phi_{\rm ex}=0$. Linecut between adjacent minima vs. $\delta_p$ for $\Phi_{\rm ex}=\Phi_0/2$.
  \label{fig:SQUID-plq}}
\end{figure*}

%Similar to previous protected qubit designs, our circuit is based on $\pi$-periodic Josephson elements, where the Josephson energy varies like $\cos2\varphi$, where $\varphi$ is the difference in the superconducting phase across the Josephson element. In this case, charge transport consists of the coherent tunneling of $4e$, or pairs of Cooper pairs, rather than $2e$ in the case of a conventional Josephson junction. 
In our device, we implement each $\pi$-periodic Josephson element with a plaquette formed from a dc Superconducting QUantum Interference Device (SQUID), consisting of two conventional Josephson junctions and a non-negligible loop inductance [Fig.~\ref{fig:SQUID-plq}(a)]. Each junction has a critical current $I_0$ and $E_J=\Phi_0 I_0/2\pi$; the inductance in each arm of the SQUID $L$ is related to the inductive energy $E_L=(\Phi_0/2\pi)^2/L$. In order to understand the origin of the $\cos2\varphi$ potential, we consider the two-dimensional potential energy landscape as a function of the two junction phases, $\delta_1$ and $\delta_2$, which is determined by $E_J$, $E_L$, and the external flux bias $\Phi_{\rm ex}$ \cite{SLefevre1992}. For now, we consider symmetric plaquettes where both junction critical currents are identical; later in this section we will consider the effects of junction asymmetry. Following convention for dc SQUIDs we plot the potential energy in terms of the common-mode and differential phase variables: $\delta_p=(\delta_1+\delta_2)/2$ and $\delta_m=(\delta_2-\delta_1)/2$. The phase dependence of the Josephson energy for each junction results in a 2D washboard pattern of potential minima. At the same time, the inductive energy associated with circulating currents flowing through the inductors corresponds to a parabolic sheet with its minimum along a line running parallel to $\delta_p$. Changing $\Phi_{\rm ex}$ shifts where the minimum of this inductive parabolic sheet falls with respect to the minima of the Josephson washboard, and thus determines the pattern of the global minima in the potential.

For a flux bias at unfrustration $\Phi_{\rm ex}=0\,{\rm mod}\Phi_0$, the minima are centered on $\delta_m=0$ and are spaced by $2\pi$ in $\delta_p$ [Fig.~\ref{fig:SQUID-plq}(b)]. Along $\delta_m$, there is only the one minimum at $\delta_m=0$ [Fig.~\ref{fig:SQUID-plq}(d)], corresponding to no circulating current around the SQUID loop. Along $\delta_p$ for $\delta_m=0$, the potential follows a $\cos\delta_p$ dependence. Thus, at unfrustration, the plaquette behaves like a single Josephson junction with critical current $2I_0$. 
When flux biased at $\Phi_0/2$, the plaquette exhibits a staggered pattern of energy minima about a line along $\delta_p$ for $\delta_m=\pi/2$ [Fig.~\ref{fig:SQUID-plq}(c)]. Figure~\ref{fig:SQUID-plq}(e) shows a linecut along a line between two adjacent minima as a function of $\delta_m$; the two minima correspond to %in the direction along the difference of the two Josephson junction phases in the plaquette, while other metastable states lie at higher energies because of the inductive screening potential along this direction; these minima correspond to 
opposite senses of circulating current around the plaquette loop, similar to a flux qubit \cite{SMooij1999} or fluxonium \cite{SManucharyan2009}. However, unlike these other qubits, these plaquettes also have another independent phase degree of freedom from $\delta_p$, which corresponds to the phase drop across the plaquette. 
%two minima also correspond to a phase 
Along $\delta_p$, the potential is simply $E_2\cos2\varphi$, with sequential minima separated by $\pi$ [Fig.~\ref{fig:SQUID-plq}(g)], where the energy scale $E_2$ depends on the Josephson energies of the individual Josephson junctions $E_J$ and the inductive energy of the SQUID loop inductance $E_L$. 
%The common-mode phase is therefore a compact variable residing on a circle. Thus, if the circuit is initialized in a  well with a particular circulating current, it could tunnel to a well with opposite circulating current either by moving forward or backward by $\pi$ in the common-mode phase direction [Fig.~\ref{fig:SQUID-plq}(b)]. 

%Why doesn't this happen for any device with a dc SQUID, e.g., a flux-tunable transmon?
 While the behavior described here is generic for any dc SQUID, achieving a $\cos2\varphi$ potential at frustration with a significant barrier height $E_2$ requires a sufficiently large ratio $E_J/E_L$. In the conventional language of dc SQUIDs, screening effects are characterized by the parameter $\beta_L=2L I_0/\Phi_0=E_J/\pi E_L$. For SQUIDs in the limit $\beta_L \rightarrow 0$ and perfect symmetry, the critical current of the SQUID will modulate to zero at frustration. For such a device, not only is the first-order Josephson energy suppressed, but $E_2$ will be vanishingly small as well, and thus not support bound states in a $\cos2\varphi$ potential. In order to have a significant $E_2$, $E_J/E_L$ must be of order unity. The dc SQUID in Fig.~\ref{fig:SQUID-plq} has $E_J/E_L=\pi$ to highlight the development of the $\pi$-periodicity at frustration.

\begin{figure*}
\centering
\includegraphics[width=6.8in]{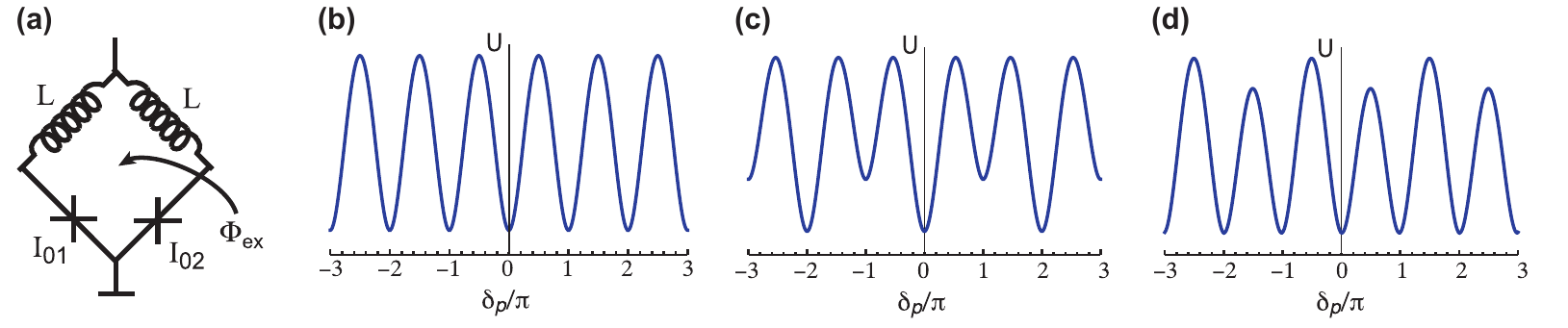}
  \caption{(a) Circuit schematic for dc SQUID plaquette with asymmetric Josephson junctions. Linecut between adjacent minima vs. $\delta_p$ for (b) $\Phi_{\rm ex}=\Phi_0/2$, $\alpha=0$, (c) $\Phi_{\rm ex}=0.45\Phi_0$, $\alpha=0$, (d) $\Phi_{\rm ex}=\Phi_0/2$, $\alpha=0.05$.
  \label{fig:SQUID-asymm}}
\end{figure*}

We next consider deviations from this ideal $\pi$-periodic plaquette behavior. With the flux bias moved below (above) frustration ($\Phi_0/2$), the $\pi$ wells are raised above (below) the $0$ wells [Fig.~\ref{fig:SQUID-asymm}(c)]. 
%at a flux bias of 0$\,{\rm mod}\,\Phi_0$, the potential along the common-mode phase direction becomes $\propto \cos\varphi$, as in a conventional Josephson junction.
To account for asymmetries between the two junctions in a plaquette we define $\alpha=\left(E_{J2}-E_{J1}\right)/\left(E_{J2}+E_{J1}\right)$, where $E_{J1}\,(E_{J2})$ is the Josephson energy of the left (right) junction. With a non-zero $\alpha$, the common-mode potential along $\delta_p$ for $\Phi_{\rm ex}=\Phi_0/2$ has equal minima for the 0 and $\pi$ wells, but now the barrier heights between wells become asymmetric.

\section{Device Fabrication}

This device was fabricated on a high resistivity ($\geq$10~k$\Omega$-cm) silicon wafer that was given a standard RCA clean followed by an etch step in a buffered-2\% per volume HF bath to remove native oxides immediately before loading into a vacuum chamber for the base-layer metal deposition. The base layer of 60-nm thick niobium is sputter-deposited and is then coated with DSK101-4 anti-reflective-coating (ARC) and DUV210-0.6 photoresist before performing deep-UV photolithography on a photostepper to define the ground plane, feedline, resonator, flux/charge bias lines, and the logical islands. The exposed wafer is then baked at 135$^\circ$C for 90~seconds, developed with AZ~726 MiF, briefly cleaned with an ARC etch to remove any remaining unwanted ARC and then dry etched using a BCl$_3$, Cl$_2$, and Ar in an inductively coupled plasma etcher. The wafer is then subject to another buffered HF dip to remove any further oxides that may have formed on the surface of the remaining niobium. 

The next set of lithography steps creates ground straps that connect ground planes on either side of the flux, charge, and feedlines. The first step uses lift-off resist LOR3A and then DUV210-0.6 photoresist to expose a region underneath the intended ground straps where we deposit SiO$_2$ to function as an insulating dielectric support for the aluminum ground straps to follow. The SiO$_2$ is evaporated in an electron beam evaporator at a rate ~3.5~${\rm \AA}$/s until 100~nm is deposited. The wafer is then placed in 1165 Remover (N-Methly-2-pyrrolidone (NMP)) at 65$^\circ$C to lift off the excess SiO$_2$ and resist and then another clean bath of NMP at 65$^\circ$C for further liftoff. The wafer is then sonicated for 10 seconds to remove any final remaining resist and SiO$_2$. The second layer of the ground strap process is exposed in the same way, using LOR3A and DUV photoresist, but this time the pattern lies over the existing SiO$_2$ and extends further so that once developed, there is an exposed region of the niobium ground plane for the aluminum to contact. The wafer is baked again and developed, and the ground straps are then deposited by electron beam evaporation of aluminum (100~nm thick). The wafer is once again subject to NMP to remove the remaining resist and excess aluminum. 

Once clean, the wafer is put through a light oxygen plasma resist strip before a bilayer resist stack of MMA/PMMA is spun for electron beam lithography to define the Josephson junctions. The Al-AlO$_x$-Al junctions are written at 100~keV to form a standard double-angle evaporation airbridge pattern. Following development, there is a brief ion mill step before the first electrode is deposited by electron beam evaporation. The bottom (top) electrode is 40~(80)~nm thick. Once the junctions are deposited, the wafer is covered in S1813 photoresist and then diced to (6.25~mm)$^2$ chips. After the dicing, the aluminum metallization is lifted off and the chips are then cleaned with a UV/ozone process before measurement.

\section{Device Layout}

\begin{figure}[hbt!]
\centering
\includegraphics[width=6.8in]{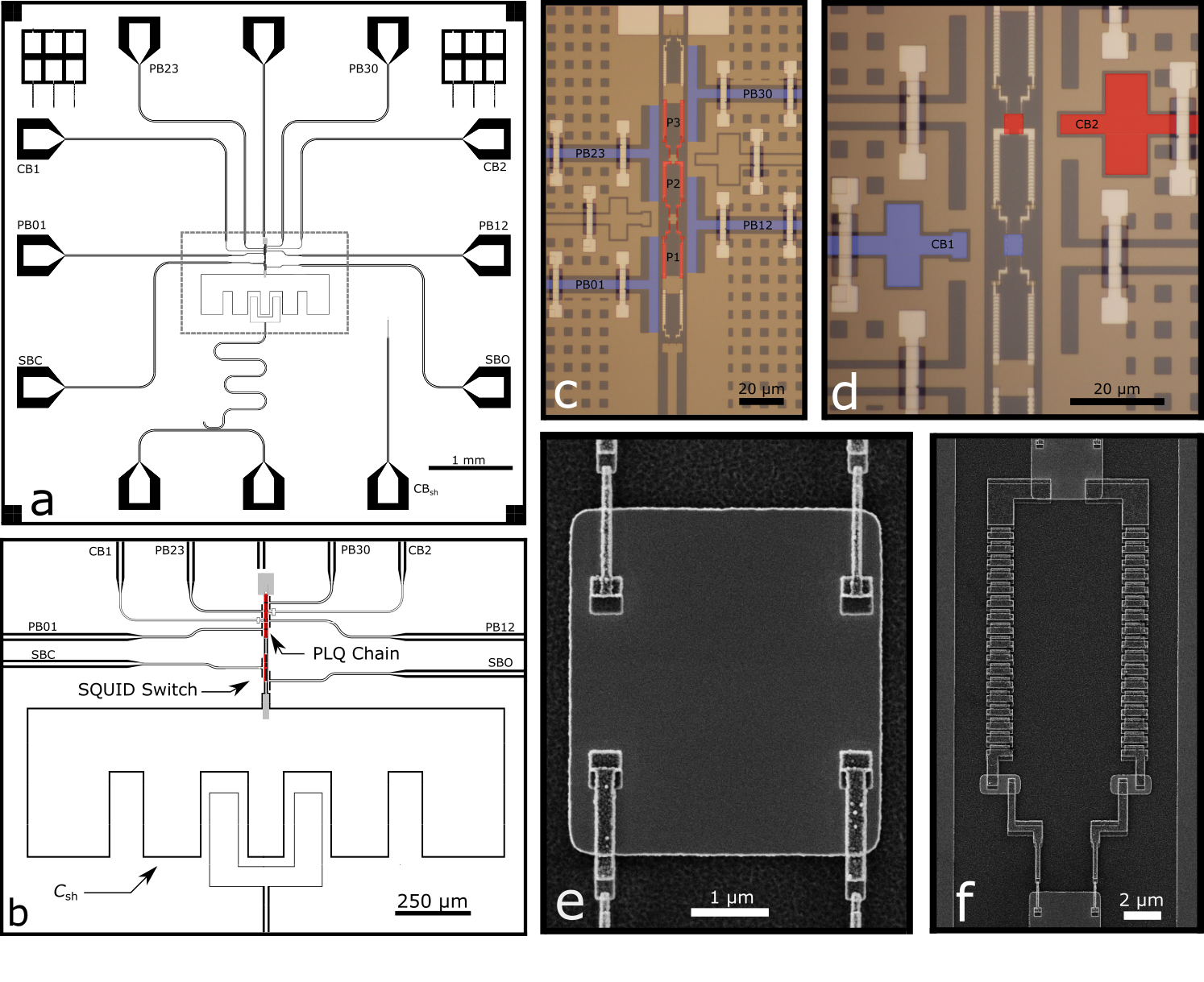}
  \caption{(a) Chip layout including on-chip diagnostic structures and launcher pads for each bias line and feedline. (b) Zoomed-in layout of the dashed box section highlighting the location of the plaquette chain, SQUID switch, and shunt capacitor island. Optical micrographs of plaquette chain: (c) red colorization indicates the 3 plaquettes in the chain; blue colorization highlights the flux bias lines, (d) offset charge bias lines for each of the two intermediate islands between plaquette pairs; blue (red) colorization indicates the charge bias line and island 1 (2). Scanning Electron Microscopy images of (e) intermediate island 1 between plaquettes 1 and 2, (f) an image of plaquette 2, including small junctions and junction chain inductors.
  \label{fig:colorized flux lines}}
\end{figure}

In order to allow for local flux-biasing of the different plaquettes and charge-biasing of the various superconducting islands, our device incorporates a series of on-chip bias lines, indicated in Fig.~\ref{fig:colorized flux lines}. The heart of the device contains a chain of three plaquettes, each with two small Josephson junctions (130~nm $\times$110~nm) and two junction-chain inductors (seventeen 140 nm $\times$1070~nm junctions in series). As discussed in the main paper, minimizing $C_{\rm isl}$ for each intermediate island between two adjacent plaquettes is critical for successful concatenation. Thus, ideally the four Josephson junctions in two adjacent plaquettes will all be located near the island between the plaquettes so that the junction electrode that is closest to the island will be as short as possible and contribute a minimal amount of excess capacitance to ground. However, in a chain of three plaquettes, this is only possible for one of the two intermediate islands. The other island will necessarily have to be connected to the two inductors for one of the plaquettes, and the capacitance to ground for these inductors will enhance the effective island capacitance. 
In addition, the 3-plaquette chain has dummy plaquettes at either end, which have the same geometry as the other plaquettes, but the small junctions and inductor-chain junctions are intentionally shorted out. The dummy plaquettes are included to symmetrize the geometry and minimize the inductive coupling of the on-chip flux-bias lines to the $LC$ mode of oscillation of the plaquette chain, sometimes referred to as the $M^{\prime}$ coupling, as defined in Ref.~\cite{SKoch2007}.

There are four on-chip flux-bias lines for controlling the flux bias to each of the three plaquettes, with the labeling as described in the main paper. Each flux-bias line has a coplanar geometry and splits into a T-shaped path adjacent to the plaquette chain, with the two ends of the T connected directly to the ground plane. In order to have a well-defined path for the return currents, and to suppress slot-line modes between different portions of the ground plane, we fabricated superconducting ground straps across each flux-bias line in multiple locations. In addition to the flux-bias lines, we also have three charge-bias lines for tuning the offset charge to the shunt capacitor electrode and each of the two intermediate islands between pairs of plaquettes. These charge-bias lines are isolated from ground, but also include similar ground straps to the flux-bias lines.

Our design also includes a pair of series dc SQUIDs between the plaquette chain and $C_{\rm sh}$ that could be used for gate operations in a future implementation of a protected qubit based on concatenated $\pi$-periodic plaquettes. For the experiments presented here, this SQUID switch, which has separate flux-bias lines from the plaquettes, was not used and the two loops of the SQUID switch were maintained at a flux bias of 0$\,{\rm mod}\,\Phi_0$ throughout the experiment. At this bias point, the SQUIDs behave primarily as superconducting shorts, although we must still account for the nonlinearity of the SQUID junctions in modeling the energy levels for our device.

The target shunt capacitance, $C_{\rm sh} \sim 1200\,{\rm fF}$ for our present device is rather large compared to more conventional superconducting qubits. Nonetheless, in the present experiment, we implemented $C_{\rm sh}$ with a planar superconducting Nb electrode with a small gap to the ground plane around the perimeter. 
For measuring our device, we have a coplanar waveguide (CPW) readout resonator with a fundamental resonance at 4.7~GHz. This is a 1/4-wave resonator with one end inductively coupled to a CPW feedline that is connected to our measurement circuitry; the other end of the resonator has a coupling capacitance $C_{\rm c} = 44\,{\rm fF}$ to our device.

The majority of our device is patterned in Nb, including the ground plane, bias lines, readout resonator, and shunt capacacitor. All Josephson junctions are fabricated from a standard Al-AlO$_x$-Al double-angle shadow-evaporation process. As an initial attempt at superconductor gap engineering for reducing quasiparticle poisoning of the plaquette chain, we include two patches of Al for suppressing the Nb gap underneath -- one patch is at the joint between the plaquette chain and the ground plane; the other patch is between the plaquette chain and shunt capacitor.

%\begin{figure}[hbt!]
%\centering
%\includegraphics[width=6.8in]{LargeLayoutTextv2.png}
%  \caption{(a) Chip layout including on-chip diagnostic structures and launcher pads for each bias line and feedline. (b) Zoomed-in layout of the dashed box section highlighting the location of the plaquette chain, SQUID switch, and shunt capacitor island. 
%  \label{fig:model circuit diagram}}
%\end{figure}

%\begin{figure}
%\centering
%\includegraphics[width=3.35in]{fig1-supp}
%  \caption{Device layout -- labels of various bias lines, plaquettes, %$C_{\rm sh}$, etc.
%  \label{fig:device-layout}}
%\end{figure}

%\begin{figure}
%\centering
%\includegraphics[width=6.8in]{SEMImagesv3.png}
%  \caption{Scanning Electron Microscopy images of (a) logical center island 1 between plaquettes 1 and 2 and (b) a full singular plaquette including small junctions and junction chain inductors.
%  \label{fig:fab-pics}}
%\end{figure}

\section{Device and Measurement Setup}

\begin{figure}
\centering
\includegraphics[width=6.8in]{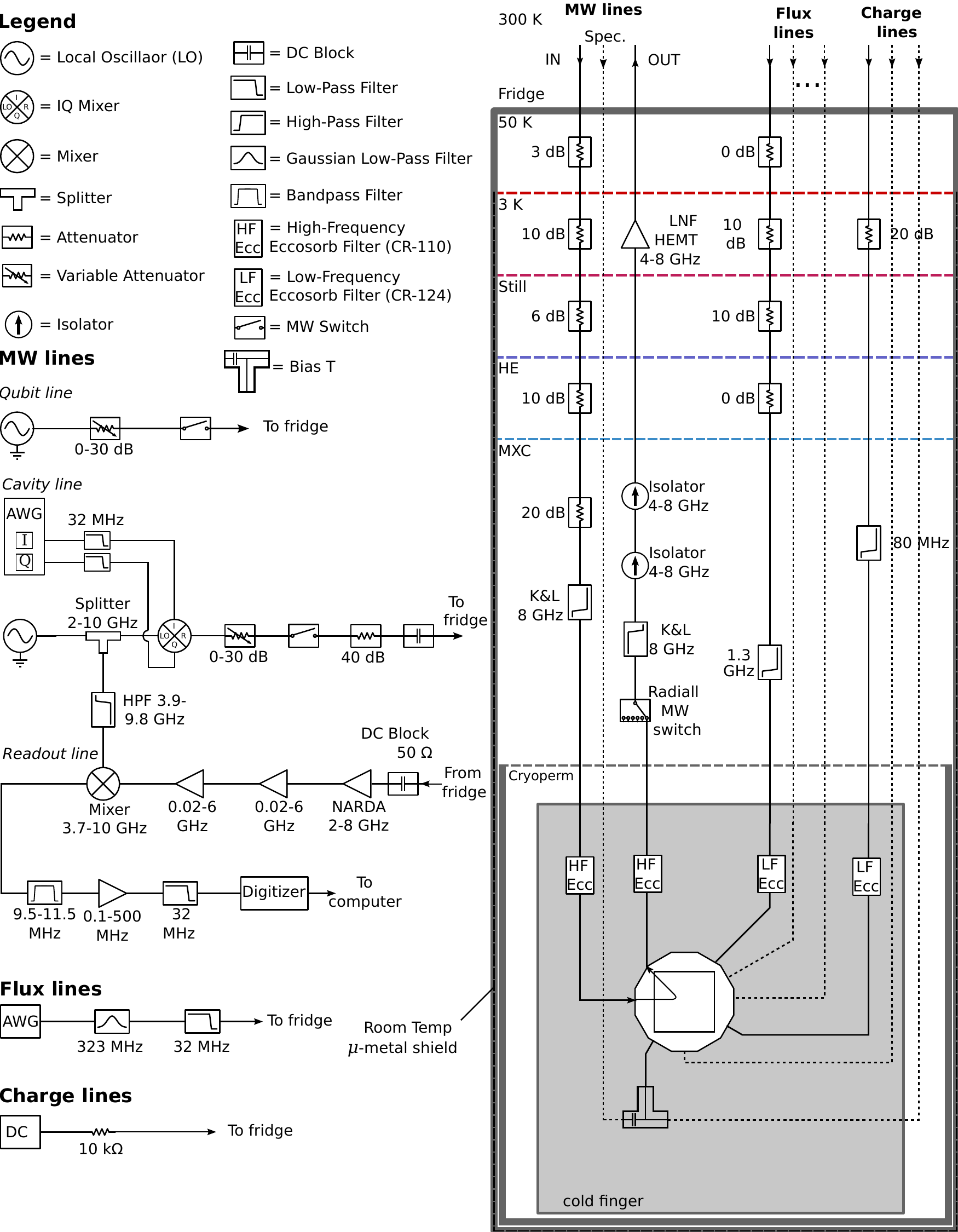}
  \caption{Schematic of dilution refrigerator wiring, filtering, and shielding, as well as configuration of room-temperature electronics.
  \label{fig:wiring-schem}}
\end{figure}

Measurements are performed on a cryogen-free dilution refrigerator running at a temperature below 15~mK. The device chip is wire-bonded into a machined Al sample box that is mounted on a cold-finger attached to the mixing chamber stage and surrounded by a Cryoperm magnetic shield. The detailed configuration of our cabling, attenuation, filtering, and shielding inside the refrigerator, and the room-temperature electronics hardware for control and readout, is shown in Fig.~\ref{fig:wiring-schem}.

%voltage division on charge-bias lines, thermalization...

\section{Device Parameters}

Establishing clear stabilizer behavior at double frustration requires plaquettes with a dominant $\pi$-periodic potential and large quantum fluctuations in the direction of constant $\varphi_1$+$\varphi_2$ in the space of common-mode phases across each plaquette. The $\pi$-periodicity comes from a dc SQUID consisting of two conventional Josephson junctions and a non-negligible loop inductance. We implement inductors in each plaquette with chains of large-area Josephson junctions, similar to typical fluxonium designs \cite{SManucharyan2009}. The inductive energy of the junction chain can be extracted with $E_L=(\Phi_0\Delta/4e)/R_n^L$, where $R_n^L$ is the junction chain resistance at room temperature. To have large quantum fluctuations in the direction of constant $\varphi_1+\varphi_2$ for effective hybridization between the two plaquettes, we need large $E_C$ and $E_C^{\rm isl}$ compared to the barrier height, which determines the coupling between the 00 and $\pi\pi$~wells and the 0$\pi$ and $\pi$0~wells. For our device, 
%we are considering energies in K and 
we target $E_J \sim 1.5 \,\rm{K}$, $E_L \sim 1.5\,\rm{K}$ and $E_C \sim 3.5 \,\rm{K}$ ($k_B=1$). For a junction with large $E_J$ and $E_C$, if the junction plasma frequency $\omega_p=\sqrt{2E_J E_C}/\hbar$ approaches $2\Delta$ of the junction electrodes, the junction acquires an extra capacitance from quasiparticles on either side of the junction. This specific electronic capacitance can be expressed as $C_{\rm elec}^{\rm sp}= 3 \hbar e J_C/16\Delta^2$ \cite{SEckern1984}, where $J_C$ is the critical current density of the junction and $\Delta$ is the superconducting gap. Our target $E_J$ is $\sim$1.5~K and junction area is ~$110\,{\rm nm}\times 130\,{\rm nm}$, and the corresponding $J_C \sim$4~$\mu \rm{A}/{\mu \rm{m}}^2$ and $C_{\rm elec} \sim 0.3$~fF. Our estimated specific geometric capacitance is $\sim$50~fF/$\mu{\rm m}^2$, so $C_{\rm geo} \sim 0.7$~fF. The total capacitance of the junction is $C_J = C_{\rm elec}+C_{\rm geo} = 1$~fF, and $E_C \sim$4~K. Junctions of this size are close to the lower limit where we can maintain reasonably small junction asymmetry with our fabrication. Thus, making smaller junctions to reduce $C_{\rm geo}$ is not practical. The simulated geometric charging energy of each intermediate island to ground $E_C^{\rm isl} = (2e)^2/2C_{\rm isl}$ is $\sim 4.6 \,\rm{K}$ for the island between plaquettes 1 and 2, and $\sim 0.74 \,\rm{K}$ for the island between plaquettes 2 and 3. The $E_C^{\rm isl}$ is significantly smaller for the island between plaquettes 2 and 3 because the inductor junction chains in plaquette 2 contribute to the capacitance to ground of the intermediate island. The intermediate island also has capacitance to ground through the junction capacitors~[Fig.~\ref{fig:Model-circuit-schematic}], thus $E_C$ of each of the four junctions in the two plaquettes reduces the total charging energy of the intermediate island below $E_C^{\rm isl}$. Thus, minimizing the capacitance of these junctions, hence targeting large $E_C$, is crucial for strong hybridization between plaquettes.

%  Fitted Ec^isl: The charging energy of each intermediate island to the ground ($E_C^{isl}$) is $\sim 2.5 \,\rm{K}$ between Plaquette 1 and Plaquette 2, and is $\sim 0.65 \,\rm{K}$ between Plaquette 2 and Plaquette 3.
To estimate $E_2$ for this device, we model a circuit that embeds a plaquette in an rf SQUID, vary the flux across the rf SQUID loop, calculate the energy levels, and obtain the Fourier components for the lowest energy level. The $E_2$ value then corresponds to the Fourier component for the $\cos2\varphi$ term. The extracted $E_2$ for this circuit is $\sim 0.05$~K. For effective concatenation, both $E_C$ and $E_C^{\rm isl}$ need to be large compared to $E_2$. %The $E_C/E_2 \sim 8.3$, $E_C^{isl}/E_2 \sim 10$ for Plaquette 1 and Plaquette 2, $E_C^{isl}/E_2 \sim 1.6$ for Plaquette 2 and 3. 
% (**Might need to do some calculate to say what is the required $E_C/E_2$)
For this target value of $E_2$, we require a rather large shunt capacitor with $C_{\rm sh}$=1200 fF in order to suppress single Cooper pair tunneling on/off the logical island. The charging energy for this shunt capacitor is $E_C^{\rm sh} = 0.003$~K, and $E_C^{\rm sh}/E_2 = 0.06$, so the coupling between the even- and odd-parity states will be suppressed. 

When there are small asymmetries in the circuit, particularly between the $E_J$ values of the two junctions in a plaquette, the even- and odd-parity states experience slightly different potentials and the computational states do not have their minimum gap exactly at frustration. Based on our room-temperature measurements of the resistance of nominally identical junctions, as well as low-temperature measurements of the critical current modulation depth for dc SQUIDs fabricated with junctions that are identical to those in the plaquettes, we expect $\alpha\sim0.02$.

\section{flux scans: calibrating inductance matrix}

%Scans of cavity for different combinations of flux-bias lines beyond the examples in the main paper; extracting inductance matrix -- comparison with InductEx simulations -- table of values? or 2 tables, one for measured mutuals and one for simulated values?

%Orthogonalization, scanning along pure flux axes for each circuit element, show example(s)?... Maybe this can just be the two versions of the cavity fine structure plots described next?

%Cavity fine structure from higher circuit levels crossing cavity... scanned along raw experimental axes, and then scanned along pure flux vectors...

With the double SQUID switch and three plaquettes, the device has a total of five flux-tunable loops. The six on-chip bias lines allow us to tune the flux in each of the loops independently, provided we account for the different mutual inductances between the various bias lines and flux-tunable loops. 
%Figure~\ref{fig:Qubit_Cav_Flux} shows the readout cavity modulating as the voltage on the $PB12$ flux line is tuned. The voltage is applied by an AWG channel at room-T, then the configuration of attenuation in the fridge determines the amount of current that flows through the flux line for a given voltage. 
%When the flux through one of the SQUID loops approaches frustration, the resonant frequency of the cavity will decrease due to the increased inductance shifting the qubit frequency lower. 
%Since $PB12$ has large mutuals to both Plaquettes 1 and 2, as well as a smaller mutual to Plaquette 3, it is difficult to tell from this scan which dip in frequency corresponds to which of the SQUID loops being frustrated.
%
As described in the main manuscript, we map out the flux-bias parameter space by performing two-dimensional scans of the dispersive shift of the readout cavity for different pairs of flux-bias lines. 
When the flux through one of the loops approaches frustration, the resonance frequency of the cavity will decrease in response to the transitions of the plaquette circuit shifting to lower frequencies. 
We measure transmission through the feedline at a fixed cavity frequency near the resonance when one of the loops is frustrated. This results in high transmission when the plaquettes are away from frustration, while near frustration, we are driving on resonance and get low transmission.

%In order to get a better handle on the multi-dimensional flux space of the device, we can look at two-dimensional flux scans. These scans are done by driving at a fixed cavity frequency that is near the resonant frequency of the cavity when one of the loops is frustrated. This drive results in high transmission through the feedline when the plaquettes are away from frustration, while near frustration, we are driving on resonance and get low transmission through the feedline. The two SQUID switch loops are easier to distinguish because they are spatially separated from the plaquette SQUIDs, and thus have smaller cross mutual inductances to the plaquette loops. Figure~\ref{fig:Qubit_SQUID_2D} shows the modulation of the cavity as we vary the voltage on both of the SQUID switch flux lines. The blue lines of two different slope correspond to frustration for each of the SQUID loops in the SQUID switch; crossing blue lines occur when both SQUID loops are simultaneously frustrated. For the upcoming spectroscopy measurements in the following chapter, both SQUIDs in the SQUID switch will be continually biased at unfrustration so that the SQUID switch has a minimal impact on the level spectrum.

Following a series of two-dimensional scans of various combinations of pairs of flux-bias lines, as in Fig.~2(a,b) in the main manuscript, we fit the slopes and periods of the frustration lines, then calculate the mutual inductance matrix from the following relation:
%shows two examples of tuning different pairs of plaquette bias lines. We see that there are three sets of blue lines that correspond to frustration for the three plaquette loops. Two of the lines crossing correspond to double frustration for a pair of plaquettes and the intersection of all three corresponds to triple frustration. The period between a given set of parallel lines is 1~$\Phi_0$ for that particular loop. The identification of the different frustration lines with each plaquette can be deduced by the relative mutual inductances to the bias lines. For example, the nearly vertical lines in Fig.~\ref{fig:Qubit_SQUID_2D}a belong to Plaquette 1 because they tune rapidly with $PB01$, which is close to Plaquette 1, and minimally with $PB30$, which is far away. From the various periods and slopes of these lines, we can calculate the mutual inductance matrix,
%
\begin{equation}
\vec{\Phi} =  \mathbf{\rm{L}}\vec{I} + \vec{x} ,
  \label{eq:mutual}
\end{equation}
\noindent where $\vec{\Phi}$ is a length-3 vector of the plaquette fluxes, $\vec{x}$ is a length-3 vector of the flux offsets to each plaquette at zero bias, $\vec{I}$ is a length-4 vector of the bias currents in the four plaquette flux-bias lines, and $\mathbf{\rm{L}}$ is a $3\times 4$ matrix of the mutual inductances. The flux offsets at zero bias are due to small background magnetic flux that gets trapped in place when the ground plane goes superconducting during the initial cooldown of the device. These flux offsets can be stable for weeks at a time, although small changes that necessitate recalibration can occur occasionally. 
%often corresponding to when something is changed with the room-temperature electronics. 

\begin{figure}
\centering
\includegraphics[width=6.8in]{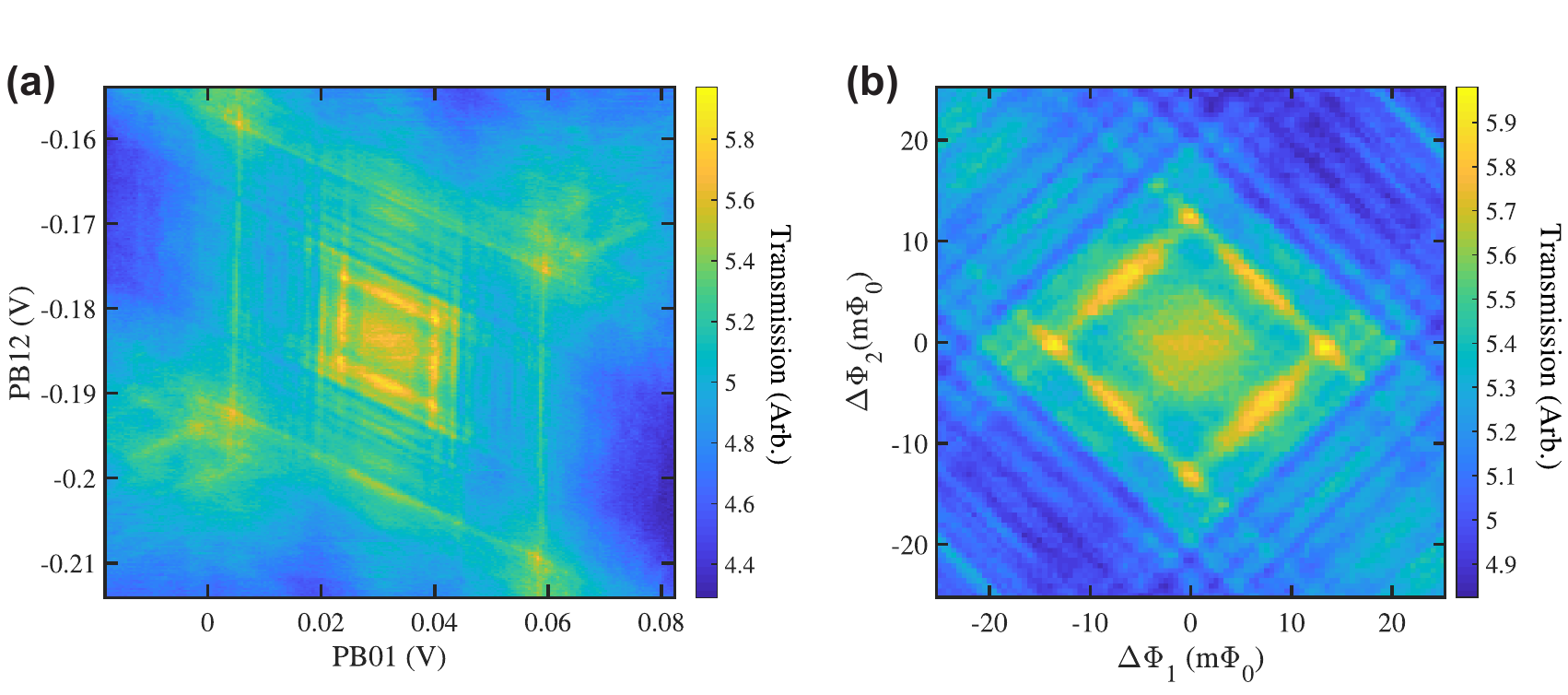}
  \caption{Measurements of readout cavity modulation through feedline transmission near plaquette (12) double frustration. Colorscale corresponds to the transmission through the feedline at a fixed frequency near the readout cavity resonance: (a) measurements with fine flux-bias steps on the PB01 and PB12 lines, (b) measurements with fine steps in the pure plaquette 1 and plaquette 2 directions based on the calibrated inductance matrix, centered on plaquette (12) double frustration. 
  %Measuring feedline transmission. Colorscale corresponds to |$S_{21}$| at a fixed frequency near the readout cavity resonance.
\label{fig:Qubit_FineStrucure}}
\end{figure}

For our spectroscopy measurements at different degrees of frustration for the plaquettes, we must be able to control the fluxes to an accuracy better than 1~m$\Phi_0$. The resolution of these two-dimensional scans over multiple $\Phi_0$ is not sufficient to determine the flux offsets and mutuals to this level. To achieve this, we zoom in near one of the double frustration points with finer voltage steps on the two flux-bias lines [Fig.~\ref{fig:Qubit_FineStrucure}(a)]. Here, we see fine structure in the feedline transmission that is symmetric around frustration that 
%we can use to very accurately find the flux offsets for two of the plaquettes. This structure 
comes from higher energy levels of the device crossing the cavity when a plaquette is tuned near frustration. We calculate the currents through the flux lines that the applied voltages create by analyzing the resistor network that is formed by the attenuators on the line.
%(Fig.~\ref{fig:Qubit_sim_Plaq2_Cav}). 
From the slopes and offsets relative to the symmetry points in these high-resolution flux scans, we extract the locations of double frustration with high accuracy; in addition, we refine the calculation of the flux periods and slopes in order to compute the mutual inductances with the required precision. Table~\ref{tab:inductance-matrix} shows the extracted mutual inductance matrix from our measurements, along with a comparison to the inductance matrix obtained from simulating the layout with the numerical software package InductEx~\cite{SInductex}. 

%{\bf We should say something about how we convert from the experimental bias %voltages on the APS2 (which are on the axis labels) to the currents in the %flux-bias lines for computing the mutual inductances.}

%\begin{figure}
%\centering
%\includegraphics[width=3.35in]{fig1-supp}
%  \caption{2D flux scans of cavity...
%  \label{fig:flux-scans}}
%\end{figure}

Using the experimental mutual inductance matrix and vector of offset fluxes, we can apply combinations of currents in the flux-bias lines to cancel out the various crosstalk fluxes and take steps in the pure flux direction for any plaquette or combination of plaquettes. Thus, we are able to scan along arbitrary vectors in the three-dimensional flux space for the three plaquettes. Figure ~\ref{fig:Qubit_FineStrucure}(b) is another high-resolution scan near plaquette (12) double frustration, but now the fluxes have been orthogonalized and the axes step through the pure fluxes through plaquettes 1 and 2 while the flux in plaquette 3 is maintained at unfrustration.

%\begin{figure}
%\centering
%\includegraphics[width=6.8in]{Qubit_OrthoFlux_3rd1289}
%  \caption{Taking fine steps in the pure Plaquette 1 and Plaquette 2 directions, centered on Plaquette 1+2 double frustration. Measuring feedline transmission. Colorscale corresponds to |$S_{21}$| at a fixed frequency near the readout cavity resonance.
%\label{fig:Qubit_OrthoFlux}}
%\end{figure}

\begin{table}[]
\begin{tabular}{|lllll|}
\hline
\multicolumn{5}{|l|}{Simulated Inductance Matrix (pH)}                                                                   \\ \hline
\multicolumn{1}{|l|}{}      & \multicolumn{1}{l|}{PB01} & \multicolumn{1}{l|}{PB12}  & \multicolumn{1}{l|}{PB23}  & PB30 \\ \hline
\multicolumn{1}{|l|}{Plaq1} & \multicolumn{1}{l|}{0.59} & \multicolumn{1}{l|}{0.76}  & \multicolumn{1}{l|}{-0.17} & 0.11 \\ \hline
\multicolumn{1}{|l|}{Plaq2} & \multicolumn{1}{l|}{0.15} & \multicolumn{1}{l|}{-0.69} & \multicolumn{1}{l|}{-0.55} & 0.24 \\ \hline
\multicolumn{1}{|l|}{Plaq3} & \multicolumn{1}{l|}{0.07} & \multicolumn{1}{l|}{-0.02} & \multicolumn{1}{l|}{0.60}  & 0.76 \\ \hline

\multicolumn{5}{|l|}{Extracted Inductance Matrix (pH)}                                                                   \\ \hline
\multicolumn{1}{|l|}{}      & \multicolumn{1}{l|}{PB01} & \multicolumn{1}{l|}{PB12}  & \multicolumn{1}{l|}{PB23}  & PB30 \\ \hline
\multicolumn{1}{|l|}{Plaq1} & \multicolumn{1}{l|}{0.64} & \multicolumn{1}{l|}{0.66}  & \multicolumn{1}{l|}{-0.15} & 0.05 \\ \hline
\multicolumn{1}{|l|}{Plaq2} & \multicolumn{1}{l|}{0.20} & \multicolumn{1}{l|}{-0.66} & \multicolumn{1}{l|}{-0.54} & 0.15 \\ \hline
\multicolumn{1}{|l|}{Plaq3} & \multicolumn{1}{l|}{0.13} & \multicolumn{1}{l|}{-0.24} & \multicolumn{1}{l|}{0.67}  & 0.59 \\ \hline
\end{tabular}
\caption{Inductance matrix (top) from InductEx simulations of device layout and (bottom) extracted from measurements of two-dimensional flux scans of readout cavity.}
\label{tab:inductance-matrix}
\end{table}

\section{Spectroscopy measurements}

%Discussion of need to overlap spectroscopy and readout pulses when plaquette 3 is involved? 
%Measurements of $T_1$? 
%Thermalization time -- do we have any measurements we can include to show this? Interwell relaxation time -- at single frustration, double frustration. Maybe these various lifetime measurements should be in a later section? discussion of readout at same/different flux?

\begin{figure}[!ht]
\centering
\includegraphics[width=\textwidth]{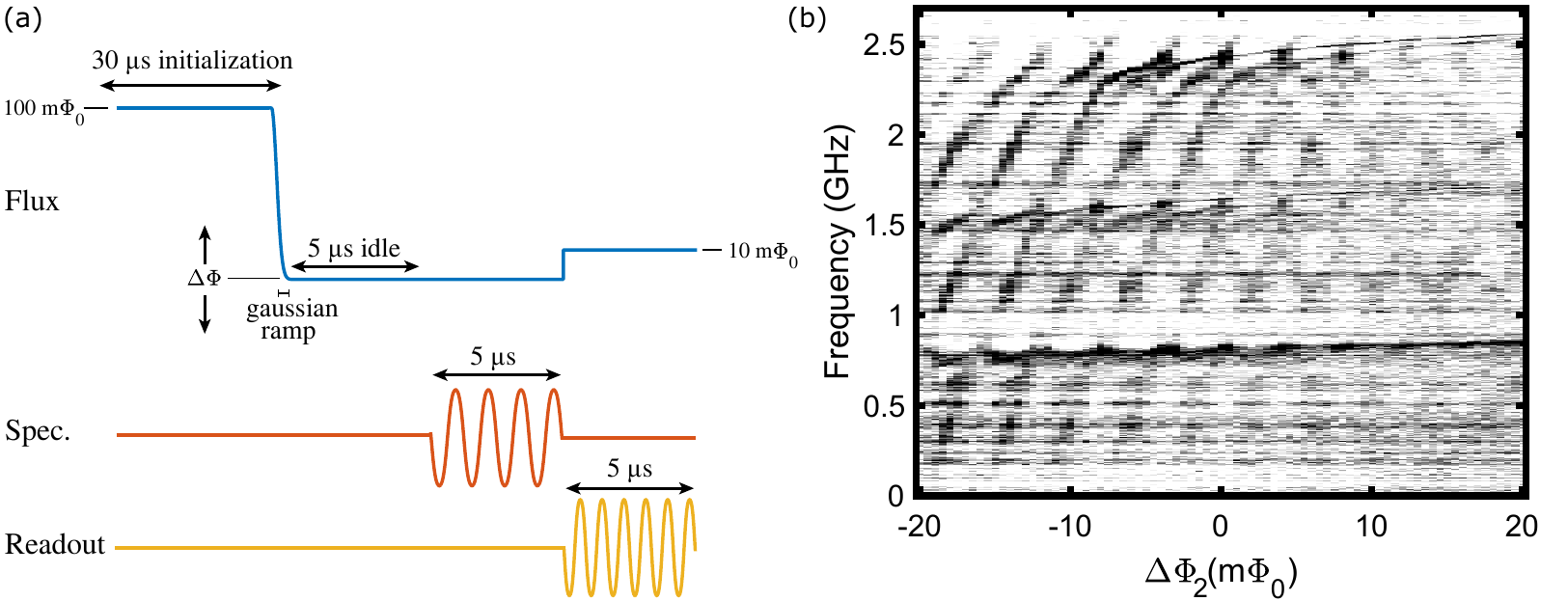}
  \caption{(a) Schematic of pulse sequence for spectroscopy measurements. The blue line represents the flux bias of the plaquette, the red line indicates the timing of the spectroscopy pulse applied to the charge-bias line coupled to the $C_{\rm sh}$ island, the yellow line shows the timing of the readout pulse. (b) Spectroscopy as a function of flux at plaquette 2 single frustration. Colorscale corresponds to the quadrature distance between the measurement of transmission through the feedline with and without a spectroscopy tone.
\label{fig:Initialization}}
\end{figure}

For spectroscopy measurements at single frustration, before each spectroscopy pulse, we initialize the circuit in the $\pi$~well by setting the plaquette flux bias to 0.1~$\Phi_0$ away from frustration, while maintaining the other two plaquettes at unfrustration. We then ramp the flux bias to each $\Delta\Phi$ coordinate on the flux axis using a gaussian edge with a 167~ns standard deviation, idle for 5~$\mu$s, then apply a 5~$\mu$s spectroscopy pulse to the $C_{\rm sh}$ charge-bias line followed by a 5~$\mu$s cavity readout pulse [Fig.~\ref{fig:Initialization}(a)]. For measurements at double or triple frustration, we perform a similar initialization sequence, but in the $\pi\pi$ ($\pi\pi\pi$) well for double (triple) frustration.

We choose the 0.1~$\Phi_0$ initialization point so that there is a single well for the system to relax into. Initialization points further from frustration would also produce a deep single well, but the larger flux amplitude would enhance flux distortions on the trajectory back near frustration for the spectroscopy measurements. We determine the 30-$\mu$s initialization time following measurements where we vary this wait time. For wait times much less than 30~$\mu$s, we observe significant excitations out of the 0 well in subsequent spectroscopy, indicating that the system hasn't fully reset into the $\pi$ well. Waiting longer than 30~$\mu$s doesn't provide any further benefit for initializing the system.

We use a 167-ns gaussian edge pulse shape so that we are moving sufficiently fast to be non-adiabatic, at least for measurements near single frustration, but not so fast that there are significant fourier components near the qubit transition frequency that could cause spurious excitations. With this particular edge time, following initialization in the $\pi$ well, if we ramp to a flux past frustration where the $\pi$ well becomes metastable, we still observe transitions out of the $\pi$ well. The 5-$\mu$s idle time before the spectroscopy pulse is applied provides time for the flux to settle. Flux distortions are commonly observed in low-temperature measurements with fast flux pulses, with various possible causes, including impedance mismatches on the line and eddy currents in the normal copper traces in the sample box~\cite{SFoxen2018}. In principle, it is possible to measure these distortions and compensate for them by applying a pre-distortion to the pulse waveform~\cite{SRol2020}. For the measurements presented here, the short idle time is sufficient for the flux to settle, as determined by varying this time; for short idle times, the frequencies of the spectroscopy features drift with respect to flux, but by 5~$\mu$s these settle to an asymptotic level.

%For performing spectroscopy near single frustration of any of the three plaquettes with the remaining two plaquettes maintained at unfrustration, we drive a microwave probe tone into the charge bias line coupled to $C_{\rm sh}$ then read out the dispersive shift of the cavity. Before each spectroscopy/readout pulse, we initialize in the $\pi$~well as described above; we then quickly ramp the flux bias to $\Delta \Phi = \Phi - \Phi_0/2$, which is the flux coordinate specified in our various spectroscopy plots and apply the spectroscopy tone. 
Following the spectroscopy pulse, for scans near plaquette 1 or 2 single frustration, the flux bias is then brought with a square pulse to a common readout point 10~m$\Phi_0$ to the right of frustration. For scans near plaquette 3 single frustration, we read out at the same flux point as the spectroscopy because we need to overlap the spectroscopy pulse with the readout pulse in this case. This is likely due to a shorter $T_1$ lifetime for the plasmon states for plaquette 3 compared to that for plaquettes 1 and 2. 
%, and this is short compared to the readout pulse. 
For all the spectroscopy scans, we plot the quadrature distance between the heterodyne measurement of transmission through the feedline at the cavity resonance for the readout flux bias with and without a 5-$\mu$s spectroscopy pulse.

Figure~\ref{fig:Initialization}(b) shows an example of a spectroscopy measurement at single frustration for plaquette 2. The features that disperse gradually with flux correspond to the plasmon excitations within the $\pi$~well where the qubit is initialized, as described in Fig.~3 in the main paper; in addition to the 0-1, 0-2, and 0-3 transitions, we also observe transitions out of excited plasmon levels, indicating that the device is not fully initialized into the ground state of the $\pi$ well.  
%, and even higher, 
%excited state in the well, such as 1-2, 1-3, 1-4, due to insufficient initialization into the ground state and excessive spurious excitation. 
%These excitations out of the ground state are most likely related to the low energy scale of the plasmon transitions, which at $\sim 800$ MHz are comparable to the frequency scale of the thermal background of the qubit environment; additionally, the large area of our shunt capacitor will support quite low-frequency spurious antenna modes, which can absorb photons from the qubit environment and drive unwanted qubit transitions~\cite{Rafferty2021}.  Although we are unable to prepare fully in the ground state of the $\pi$~well, we only observe weak transitions out of the 0~well, indicating that our initialization is predominantly preparing the circuit in the $\pi$~well. Besides the plasmon features, 
In addition, we observe heavy fluxon transitions that disperse linearly with flux, and with a much steeper slope than the plasmons, that arise from transitions between levels in the $\pi$ and 0~wells. We observe qualitatively similar behavior for single frustration of plaquettes 1 and 3, as can be seen in the spectroscopy plots in Fig.~\ref{fig:FigS-Single-Frus}.

For spectroscopy measurements at double and triple frustration, we add an extra step to stabilize the offset charge on the intermediate island(s) between the frustrated plaquettes. Details on this procedure are described in the next section.

\section{Offset charge scans}

As described in the main paper, Aharonov-Casher interference of the CW and CCW tunneling paths at various degrees of frustration results in a periodic modulation of the energy-level structure with respect to the offset charge on the $C_{\rm sh}$ island and the two intermediate islands between pairs of plaquettes. The modulation with offset charge on the $C_{\rm sh}$ island is difficult to observe directly in spectroscopy because the tunnel splittings for the low-lying levels are small. However, levels near the top of the barrier, which are also close to the readout cavity, exhibit a large modulation that leads to a significant periodic charge tuning of the readout cavity dispersive shift. The large physical footprint of the $C_{\rm sh}$ island results in a large effective charge sensing area, so that offset charge jumps occur on a timescale of a few minutes, as presented in Fig.~4(e) of the main paper.

\begin{figure}[htb!]
\centering
\includegraphics[width=\textwidth]{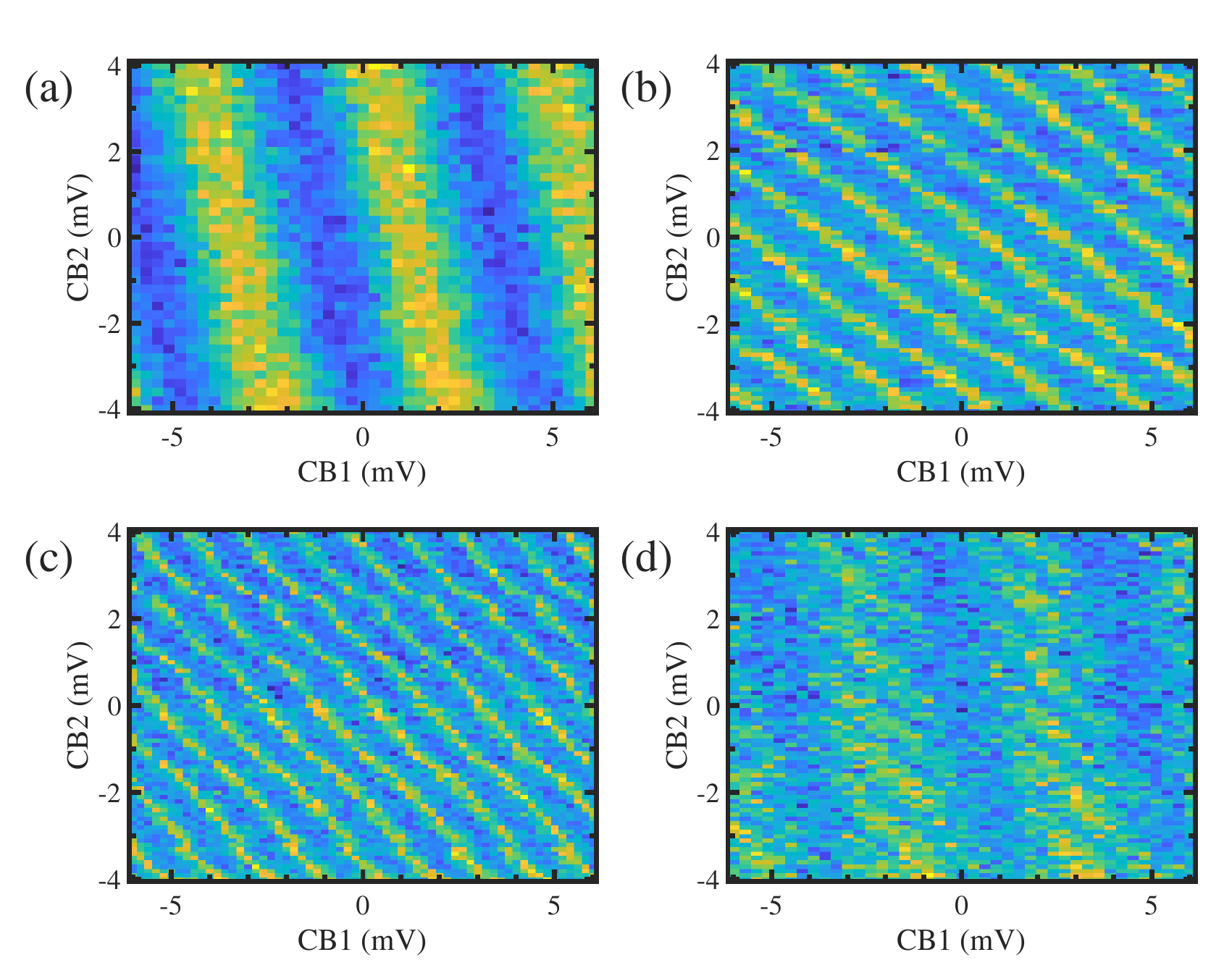}
  \caption{Spectroscopy measurements at the 0-1 transition frequency while scanning the two intermediate island charge-bias lines for: (a) plaquette (12), (b) plaquette (23), (c) plaquette (13) double frustration, and (d) triple frustration.
\label{fig:Qubit_chargeI1_chargeI2}}
\end{figure}

For measuring the modulation with respect to offset charge on the intermediate islands between plaquettes, we perform the spectroscopy sequence at the various combinations of double and triple frustration while applying a spectroscopy pulse at the 0-1 transition frequency for each particular frustration point and scanning the two island charge biases [Fig.~\ref{fig:Qubit_chargeI1_chargeI2}]. Each of these two charge lines, CB1 and CB2, couples to the intermediate island adjacent to it, but there is also non-negligible crosstalk to the other intermediate island. Thus, in general, we observe a periodic modulation due to the charge sensitivity of the relevant $\Delta_{\rm SA}^{(ij)}$ for the particular frustration point, and these modulation features have a slope in the two-dimensional charge-bias space that depends on the capacitance between each bias line and the intermediate island(s) between the particular pair(s) of frustrated plaquettes. At plaquette (12) double frustration, the modulation is significantly faster with respect to CB1 because the capacitive crosstalk between CB2 and the intermediate island between plaquettes 1 and 2 is relatively weak [Fig.~\ref{fig:Qubit_chargeI1_chargeI2}(a)]. By contrast, at plaquette (23) double frustration, the modulation is faster with respect to both CB1 and CB2 since the junction-chain inductors of plaquette 2 contribute to the intermediate island capacitance between plaquettes 2 and 3 and enhance its capacitance to both charge-bias lines [Fig.~\ref{fig:Qubit_chargeI1_chargeI2}(b)]. The modulation is even faster for plaquette (13) double frustration, since now the effective intermediate island includes all of plaquette 2, which is unfrustrated, thus enhancing the capacitance to both charge-bias lines [Fig.~\ref{fig:Qubit_chargeI1_chargeI2}(c)]. At triple frustration, we observe a double charge modulation, with one set of nearly vertical features corresponding to the modulation with respect to the offset charge on the intermediate island between plaquettes 1 and 2, and faster, more diagonal features from the modulation with respect to the offset charge on the intermediate island between plaquettes 2 and 3 [Fig.~\ref{fig:Qubit_chargeI1_chargeI2}(d)]. From the slope and period of these various modulation features, we can extract the capacitance matrix between the charge bias lines and the plaquette islands (Table~\ref{tab:Extr. Cap. Matrix}). These capacitances agree reasonably well with the Q3D numerical simulations of our device geometry~\cite{SQ3D}. 

\begin{table}[]
\begin{tabular}{|cccc|}
\hline
\multicolumn{4}{|c|}{Simulated capacitance matrix (aF)}                                                                               \\ \hline
\multicolumn{1}{|c|}{}           & \multicolumn{1}{c|}{$\rm CB_{\rm sh}$} & \multicolumn{1}{c|}{CB1} & CB2  \\ \hline
\multicolumn{1}{|c|}{$\rm Isl_{\rm sh}$} & \multicolumn{1}{c|}{57}                  & \multicolumn{1}{c|}{419}               & 353                \\ \hline
\multicolumn{1}{|c|}{Isl1}    & \multicolumn{1}{c|}{0}                   & \multicolumn{1}{c|}{45}               & 12                \\ \hline
\multicolumn{1}{|c|}{Isl2}    & \multicolumn{1}{c|}{0}                   & \multicolumn{1}{c|}{27}               & 88                \\ \hline
\multicolumn{4}{|c|}{Extracted capacitance matrix (aF)}                                                                               \\ \hline
\multicolumn{1}{|c|}{}           & \multicolumn{1}{c|}{$\rm CB_{\rm sh}$}   & \multicolumn{1}{c|}{CB1} & CB2 \\ \hline
 \multicolumn{1}{|c|}{$\rm Isl_{\rm sh}$} & \multicolumn{1}{c|}{57}                  & \multicolumn{1}{c|}{501}               & 327                \\ \hline
\multicolumn{1}{|c|}{Isl1}    & \multicolumn{1}{c|}{0}                   & \multicolumn{1}{c|}{35}               & 8                \\ \hline
\multicolumn{1}{|c|}{Isl2}    & \multicolumn{1}{c|}{0}                   & \multicolumn{1}{c|}{73}               & 120               \\ \hline
\end{tabular}
\caption{Simulated and extracted capacitance matrix.}
\label{tab:Extr. Cap. Matrix}
\end{table}

\begin{figure}[ht]
\centering
\includegraphics[width=\textwidth]{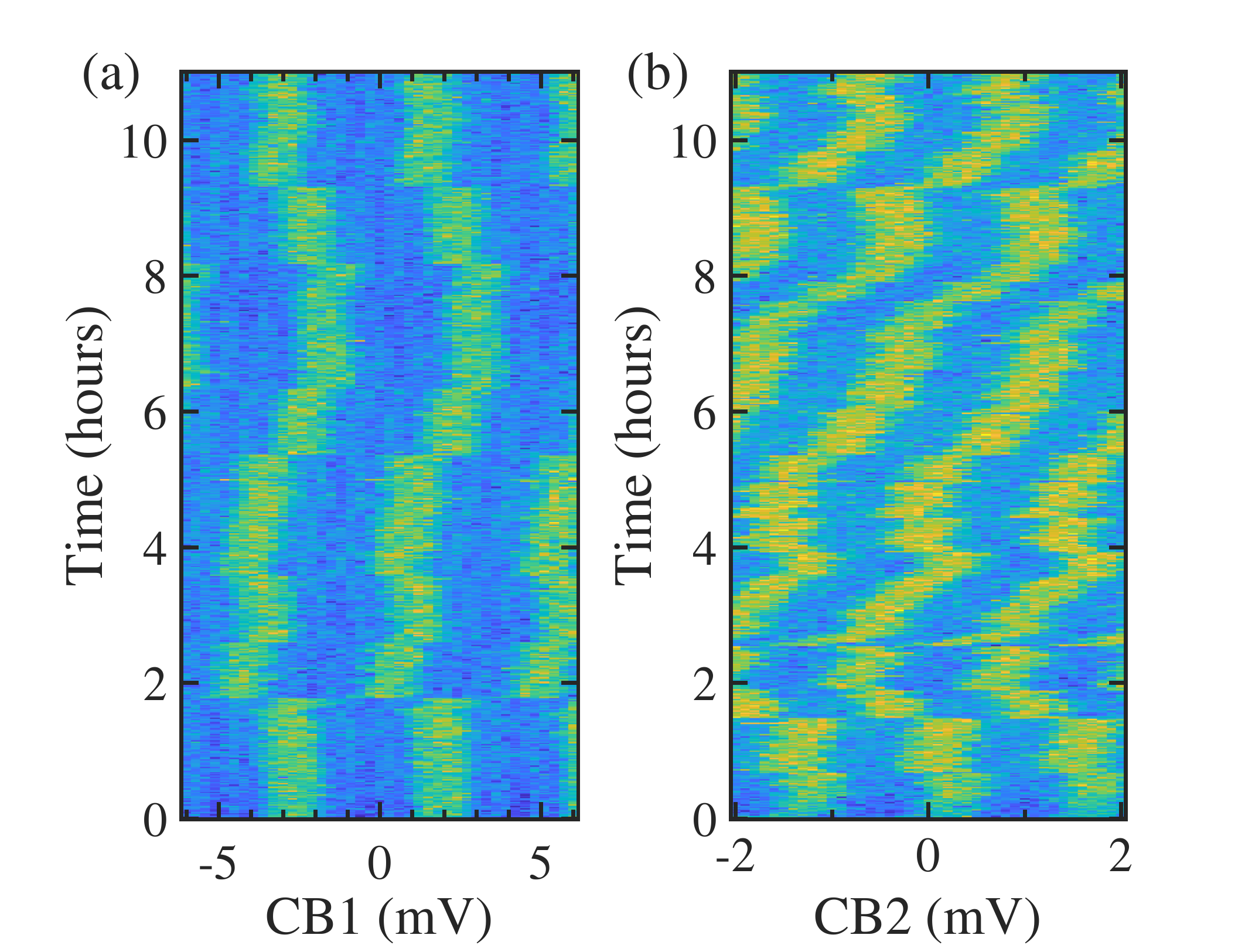}
  \caption{Simultaneous measurements of the offset charge on the intermediate islands between (a) plaquettes 1 and 2, (b) plaquettes 2 and 3, near double frustration for (a) plaquette (12), (b) plaquette (23) over an 11-hour span. Details on the particular measurement sequence here can be found in the text.
\label{fig:Qubit_chargeI1_chargeI2_stabalize}}
\end{figure}

While the offset charge jumps on the $C_{\rm sh}$ island occur every few minutes, we expect the offset charge jumps on the intermediate islands between plaquettes to be less frequent because of the much smaller charge sensing areas~\cite{SChristensen2019,SWilen2021}. In order to monitor offset charge jumps on both intermediate islands nearly simultaneously, we first scan the offset charge bias to island 1 (between plaquettes 1 and 2) at plaquettes (12) double frustratation while plaquette 3 is biased 50~m$\Phi_0$ away from frustration; we then shift the fluxes slightly and scan the offset charge bias to island 2 (between plaquettes 2 and 3) at plaquette (23) double frustration while plaquette 1 is biased 50~m$\Phi_0$ from frustration. We alternate back and forth between these two scans repeatedly over 11 hours (Fig. ~\ref{fig:Qubit_chargeI1_chargeI2_stabalize}). As expected, given the smaller charge sensing areas, the offset charge on the intermediate islands jumps less frequently than on the $C_{\rm sh}$ island. The island between plaquettes 1 and 2, which has the smallest charge sensing area, is the most stable, with roughly 1 hour between large offset charge jumps.

Despite the relative stability of the offset charge on the intermediate islands, we still need to actively stabilize the charge for long spectroscopy scans vs. flux at double and triple frustration, such as Fig.~4(b,c) in the main paper, to correct for occasional offset charge jumps. Thus, approximately every twenty minutes we interrupt the spectroscopy sequence to run a one-dimensional scan of the relevant intermediate island offset charge bias(es) while applying a spectroscopy pulse at the corresponding 0-1 transition frequency; each scan takes about 30 seconds. We then fit the resulting modulation signal with a cosine function to determine the appropriate adjustment to the charge bias to apply to maintain a constant total intermediate island offset charge.

\begin{figure*}
\centering
\includegraphics[width=\textwidth]{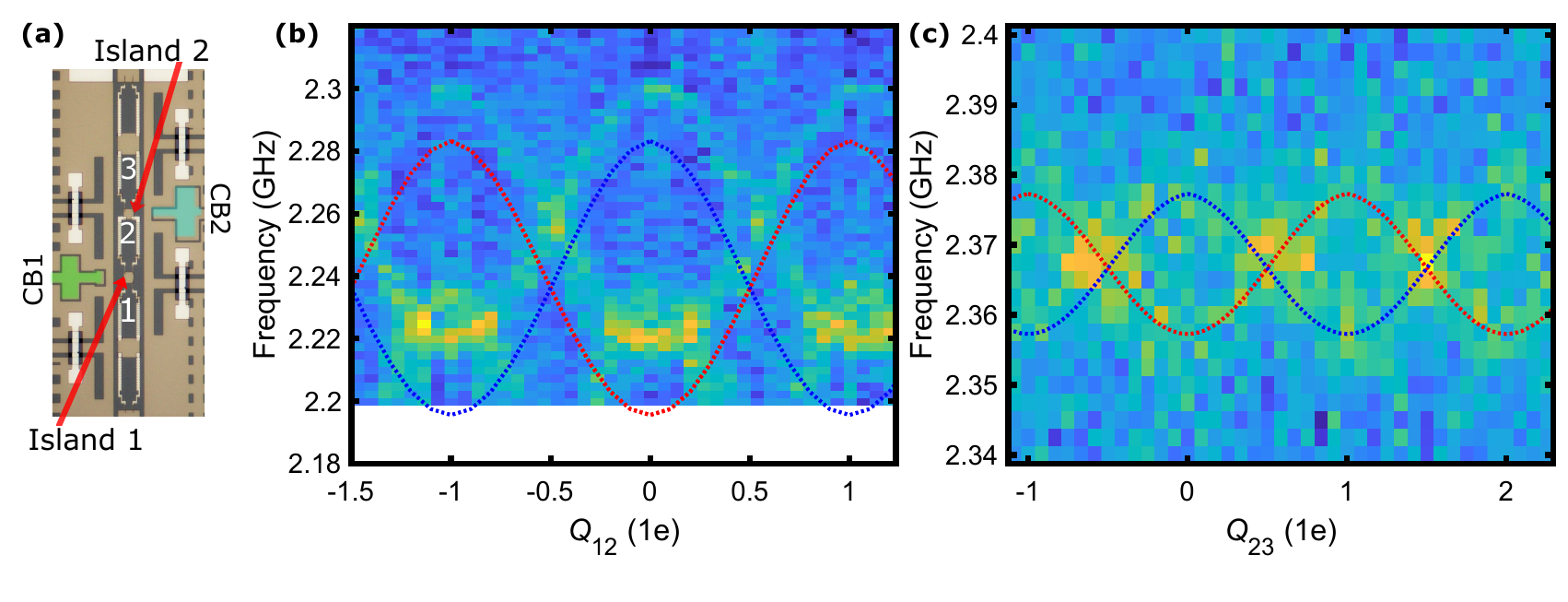}
  \caption{(a) Circuit image, with intermediate islands and charge bias lines indicated. (b) Plaquette (12) double frustration charge modulation data of $\ket{0_{ES}} \rightarrow \ket{3_{ES}}$ transition at 17~m$\Phi_0$. The red and blue dotted lines are the fitted transitions that correspond to different quasiparticle parity on intermediate island 1. (c) Plaquette (23) double frustration charge modulation data of $\ket{0_{ES}} \rightarrow \ket{3_{ES}}$ transition at 11 m$\Phi_0$. The red and blue dotted lines are the fitted transitions that correspond to different quasiparticle parity on intermediate island 2.
\label{fig:Fitting-DF-Qi}}
\end{figure*}

In addition to the random offset charge jumps due to charge dynamics in the qubit environment, for example, from the impact of high energy particles \cite{SWilen2021}, the various islands of our device are also subject to quasiparticle (QP) poisoning when a QP tunnels on or off the island. Because of the large physical footprint for the $C_{\rm sh}$ island, there will be spurious antenna resonances, as described in Refs.~\cite{SRafferty2021,SLiu2022}, at frequencies extending from above the Al superconducting gap to below 100~GHz that couple resonantly to stray photons in the device environment and generate QPs at the plaquette junctions. In Fig.~\ref{fig:Fitting-DF-Qi}, we show measurements of spectroscopy near the 0-3 transition as a function of frequency and intermediate island offset charge near plaquette (12) and (23) double frustration. In both cases, we observe a periodic charge modulation of the transition, but with two bands that are offset by $e$, indicating QP poisoning on the intermediate island between the plaquettes on a timescale faster than the spectroscopy measurement. 
%{\bf Is this the correct periodicity? Somehow it looks like $2e$ in the figure...} 
Such QP poisoning will need to be significantly suppressed in future devices for the successful implementation of a protected qubit.

\section{Modeling of energy levels}
\label{Sec:model}

\begin{figure}
\includegraphics[width=\textwidth]{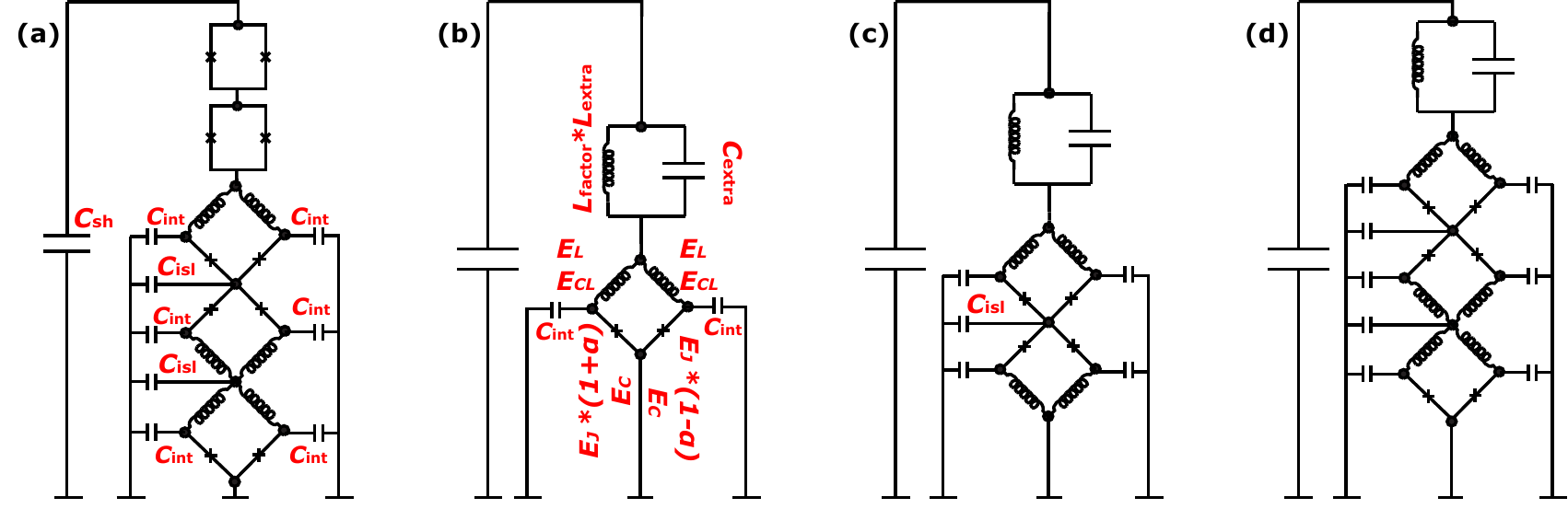}
  \caption{Circuit schematics for (a) 3-plaquette chip circuit plus SQUID switch, (b) single-frustration modeling, (c) double-frustration modeling, (d) triple-frustration modeling.
\label{fig:Model-circuit-schematic}}
\end{figure}

\begin{figure}
\includegraphics[width=\textwidth]{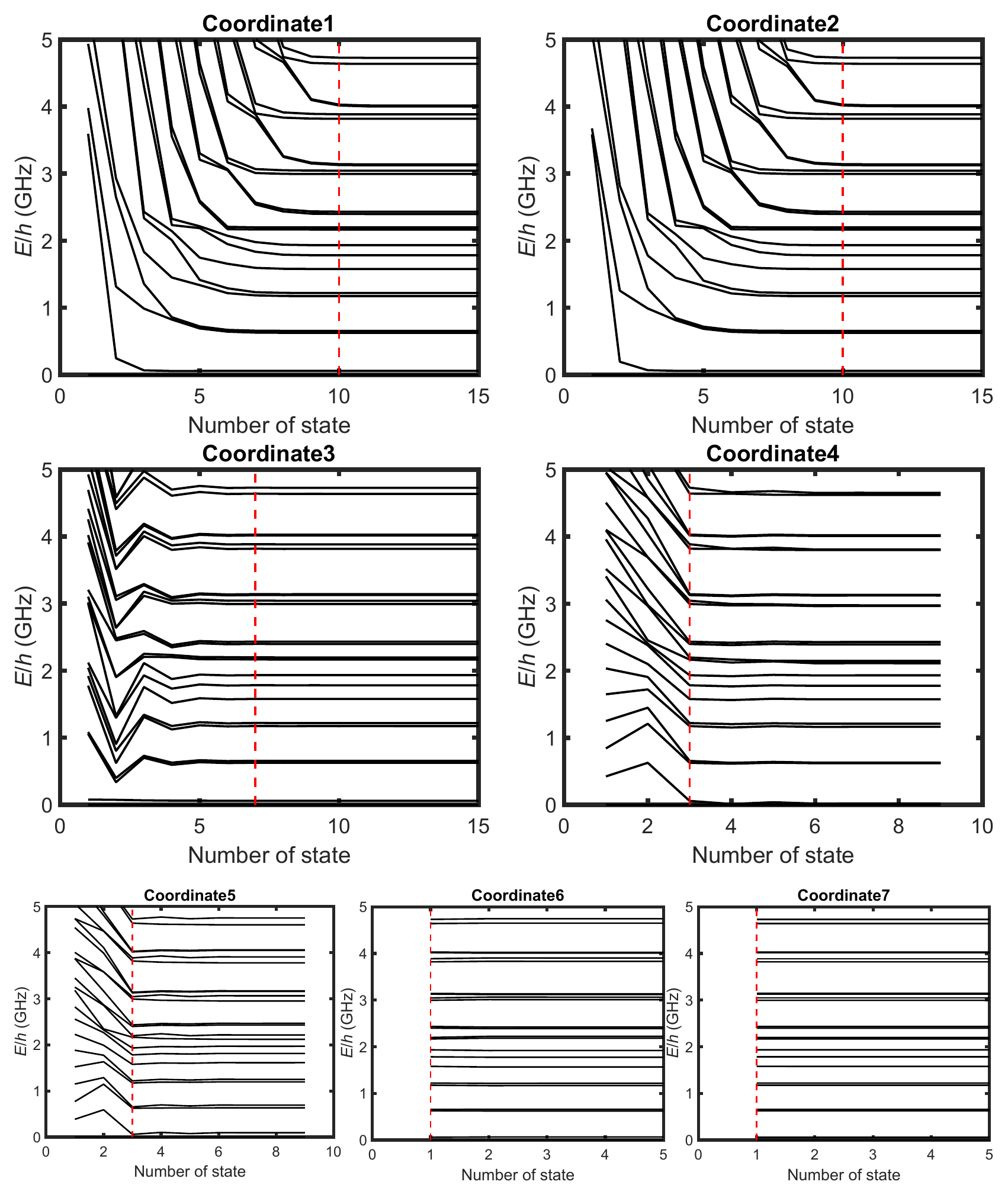}
  \caption{Convergence of transition frequencies at double frustration with respect to the number of states used for each coordinate in the simulation. The red dashed lines correspond to the number of states chosen for each coordinate for subsequent device simulations. Coordinates 1 and 2 are cyclic coordinates and coordinates 3-7 are oscillator coordinates.
\label{fig:Number-of-states}}
\end{figure}

\begin{figure}
\includegraphics[width=\textwidth]{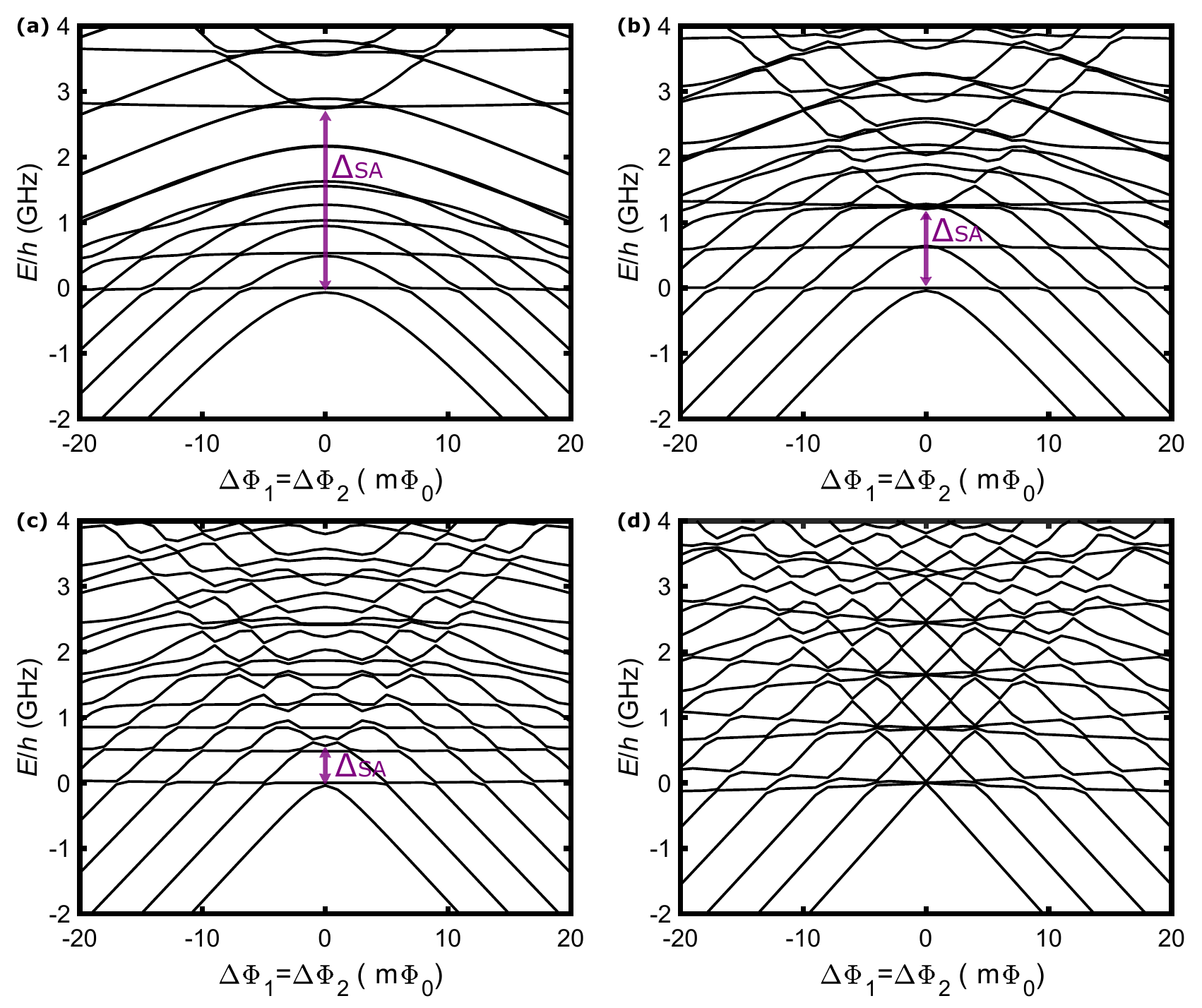}
  \caption{Double plaquette flux dispersion with different intermediate island capacitance (a) $C_{\rm isl}=$~1~fF, which results in strong hybridization and $\Delta_{\rm SA} \approx 2.9 ~\text{GHz}$. (b) $C_{\rm isl}=$~5~fF. The hybridization is reduced and $\Delta_{\rm SA} \approx 1.3 ~\text{GHz}$. (c) $C_{\rm isl}=$~10~fF. The hybridization is significantly suppressed. $\Delta_{\rm SA} \approx 0.5 ~\text{GHz}$, and the flux dispersion is close to linear. (d) $C_{\rm isl}=$~50~fF. The hybridization is almost suppressed, the $\Delta_{\rm SA} \approx 0 ~\text{GHz}$, and the flux dispersion is essentially linear. 
  %{\bf Let's update the flux axes to match the notation in Fig. 4 of the main paper, so $\Delta\Phi_1=\Delta\Phi_2$; also, it looks like the positions of the (a), (b),... labels need to be adjusted. } {\color{orange}\bf {Done}}
\label{fig:FigS-Double-Plq-Vs-Cisl}}
\end{figure}

\begin{figure}
\centering
\includegraphics[width=\textwidth]{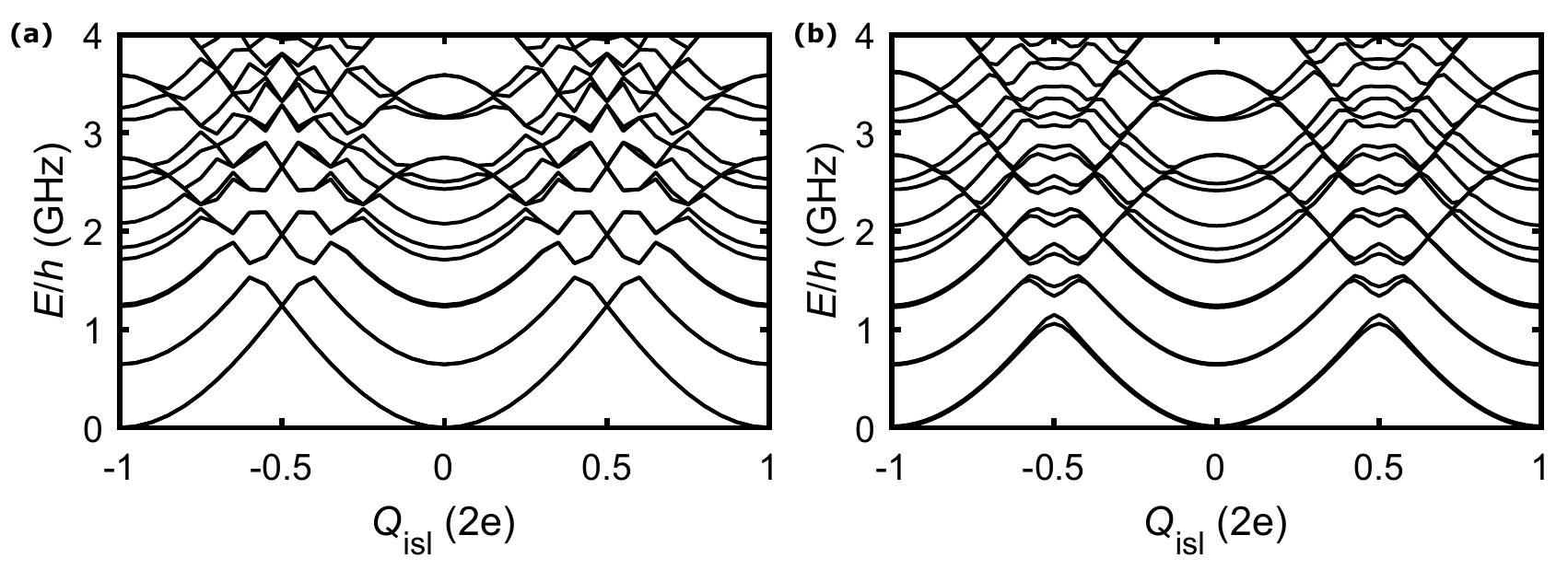}
  \caption{
  Simulation of intermediate island charge modulation at double frustration 
  %of the symmetric/antisymmetric energy levels for even- and odd-parity 
  for (a) $\alpha$ = 0 (even- and odd-parity levels are on top of each other) with levels crossing at $1e\,$mod$2e$ and the symmetric-antisymmetric gap closing because of complete destructive interference; (b) $\alpha$ of plaquette 1 is 0.01, $\alpha$ of plaquette 2 is 0.03; here, 
  %the charge modulation for the symmetric/antisymmetric energy levels for even- and odd-parity have $4e$ dependence. They cross at $1e$ and the 
  both of the gaps for even- and odd-parity states are not fully closed at $1e\,$mod$2e$ due to incomplete destructive interference.}
  \label{fig:FigS-DF-vs-Qi-at-frustration}
\end{figure}

% \begin{figure}
% \includegraphics[width=\textwidth]{FigS-Model-Double-2D-potential.pdf}
%   \caption{(a) Circuit schematic. (b) 2D potential when $\alpha = 0$ with unwrapping the compact phase variables so that they are no longer on a torus, but instead extend to $\pm\infty$. Red lines correspond to hybridization for even-parity wavefunctions, while blue lines correspond to hybridization for odd-parity wavefunctions. (c) 2D potential when $\alpha$ is non-negligible. (d) $\alpha = 0$: Left (right) plot is the 1D cut potential along the direction of hybridization for even- (odd-) parity wavefunctions, and they both have a $\cos 2\varphi$ potential. (e) $\alpha \neq 0$: Left/right plot is the 1D cut of the potential along the direction of hybridization for even- (odd-) parity wavefunctions. The even-parity wavefunctions experience a $\cos 2\varphi$ potential, while the odd-parity wavefunctions experiences a potential with a $\cos\varphi$ component mixed in with the $\cos 2\varphi$ potential.
% \label{fig:Model-Double-2D-potential}}
% \end{figure}

%Details of Andrey's code, running on various machines, number of states for each coordinate, oscillator vs. cyclic coordinates...
Modeling multi-plaquette %charge-parity qubits 
devices is challenging -- our 3-plaquette chip with a SQUID switch has eleven phase degrees of freedom, and the size of the truncated Hilbert space is $\sim 2 \times 10^5$. Instead of choosing generalized coordinates manually, we use the SuperQuantPackage~\cite{Sandreyklots2022Jul} to model the energy level spectra of the devices.

The SuperQuantPackage software framework was developed by Andrey Klots with the supervision of Lev Ioffe. This package is capable of modeling the energy spectrum of superconducting circuits with arbitrary configurations of Josephson junctions, capacitors, and inductors. 
%Andrey
The original flux coordinates of the nodes undergo a linear transformation that splits them into two classes: oscillator-like coordinates for which we choose a harmonic oscillator basis, and charge coordinates that correspond to clusters of nodes with quantized net charge and for which a natural charge basis is used. This automatically diagonalizes the inductive and capacitive parts of the Hamiltonian. At the same time, Josephson terms of the Hamiltonian assume a relatively simple and sparse form. Automatic assignment of physically meaningful coordinates does not require labor-intensive manual symmetry analysis of each configuration of all studied complex circuits. Meanwhile, it allows for efficient diagonalization of the Hamiltonians and relatively quick numerical convergence.  
%\sout{The circuit is divided into a series of independent harmonic oscillator coordinates, cyclic coordinates corresponding to islands with quantized charge, and a set of Josephson junctions that facilitate the interaction between the two sets of coordinates. With this choice of basis, the coordinates are maximally separated, and many portions of the Hamiltonian are automatically diagonalized, thus allowing us to extract the eigenvalues with minimal computational expense.}}

Despite this optimization of the numerics, modeling the full circuit of our most complex devices [Fig.~\ref{fig:Model-circuit-schematic}(a)] would require at least several months on the most powerful processors available to our research group. Thus, we must devise strategies for simplifying the modeled circuit to make the calculation practical. A plaquette or SQUID biased at unfrustration behaves like a superconducting inductive short with an effective shunt capacitance. For example, when modeling single frustration, we simplify the circuit to a single plaquette connected in series with a $LC$ resonator that represents the other unfrustrated plaquettes and SQUID-switch elements [Fig.~\ref{fig:Model-circuit-schematic}(b)]. The $LC$ resonator inductance and capacitance are shown in Fig.~\ref{fig:Model-circuit-schematic}(b) as $L_{\rm extra}$ and $C_{\rm extra}$. As shown in Fig.~\ref{fig:Model-circuit-schematic}, each plaquette contains two arms, each having one Josephson junction in series with a linear inductor. The Josephson junction is characterized by $E_J$ and $E_C$, where $E_J$ is the average energy of the Josephson junctions and $E_C$ is the charging energy set by the junction capacitance. The linear inductor is characterized by $E_L$ and $E_{CL}$, where $E_L$ is the average inductive energy of the junction-chain inductor and $E_{CL}$ is the charging energy across the junction-chain inductor. From our fabrication uniformity tests, our nominally identical junctions exhibit a spread in $E_J$ of a few percent. We account for this asymmetry between the two junctions in a plaquette with the parameter $\alpha = (E_{JL}-E_{JR})/(E_{JL}+E_{JR})$, with $E_{JL}$ and $E_{JR}$ the Josephson energy of the left and right junction, respectively. Each arm has capacitance to ground, and this is characterized by $C_{\rm int}$. $C_{\rm sh}$ is the capacitance of the shunt. We introduce the parameter $L_{\rm factor}$ to account for variations in $L_{\rm extra}$ due to small flux offsets in the bias of the nominally unfrustrated plaquettes or SQUIDs. After this, we can input the circuit elements in the SuperQuantPackage. 

The matrix is typically quite sparse, thus we can use the scipy.sparse.linalg.eigsh() function to find the eigenvalues and eigenvectors efficiently. This function only requires calculating the first few lowest eigenvalues, so the calculation speeds up. Typically, we only require the first $\sim$16-32 eigenvalues, but when we need to calculate the transitions involving the readout cavity, we need to calculate the first 40 eigenvalues. 

The next step is finding the minimum number of states for each coordinate. We start by using three states for each cyclic coordinate and one state for each oscillator coordinate. We then vary the number of states from 1-20 for each coordinate, while tracking how the transition frequencies change. When the transition frequencies change by less than 5\%, we choose the corresponding number of states for that particular coordinate for the next iteration. Using the new number of states, we repeat the same procedure until the process converges. In Fig.~\ref{fig:Number-of-states}, we show the convergence for each of the coordinates as a function of the number of states for double frustration.

We model single frustration by considering a single plaquette connected in series with an $LC$ circuit [Fig.~\ref{fig:Model-circuit-schematic}(b)]. For double frustration, we model the unfrustrated elements as a single $LC$ circuit in series with the two plaquettes that are modeled near double frustration [Fig.~\ref{fig:Model-circuit-schematic}(c)].

\subsection{Double plaquette flux dispersion vs. intermediate island capacitance}

We modeled double frustration flux dispersion with different intermediate island capacitance, and we present some of these results in Fig.~\ref{fig:FigS-Double-Plq-Vs-Cisl}. When $C_{\rm isl} = 1$~fF, the wavefunction is hybridized strongly between the $00$ and $\pi\pi$ wells, resulting in a large $\Delta_{\rm SA}\approx 2.9\,{\rm~GHz}$ and rather flat ground-state energy band [Fig.~\ref{fig:FigS-Double-Plq-Vs-Cisl}(a)]. The ground state energy band is relatively flat near frustration. When $C_{\rm isl}=$ 5~fF, the effective mass along this direction is larger, but there is still somewhat effective hybridization between the $00$ and $\pi\pi$ well [Fig.~\ref{fig:FigS-Double-Plq-Vs-Cisl}(b)]. In this case, $\Delta_{\rm SA} \approx 1.3~\text{GHz}$, and the energy band curvature is larger. When $C_{\rm isl}=$ 10~fF, the hybridization is significantly weaker [Fig.~\ref{fig:FigS-Double-Plq-Vs-Cisl}(c)]. $\Delta_{\rm SA} \approx 0.5 ~\text{GHz}$ and the antisymmetric level is now lower than the first excited plasmon state. Also, the energy bands have a nearly linear dispersion near double frustration. When $C_{\rm isl}=$ 50~fF, the effective mass is so large that all four wells are nearly independent with vanishing coupling between them [Fig.~\ref{fig:FigS-Double-Plq-Vs-Cisl}(d)]; $\Delta_{\rm SA} \approx 0 ~\text{GHz}$ and the flux dispersion near double frustration is essentially linear. 

\subsection{Double plaquette flux dispersion for different $\alpha$}

\begin{figure}
\includegraphics[width=\textwidth]{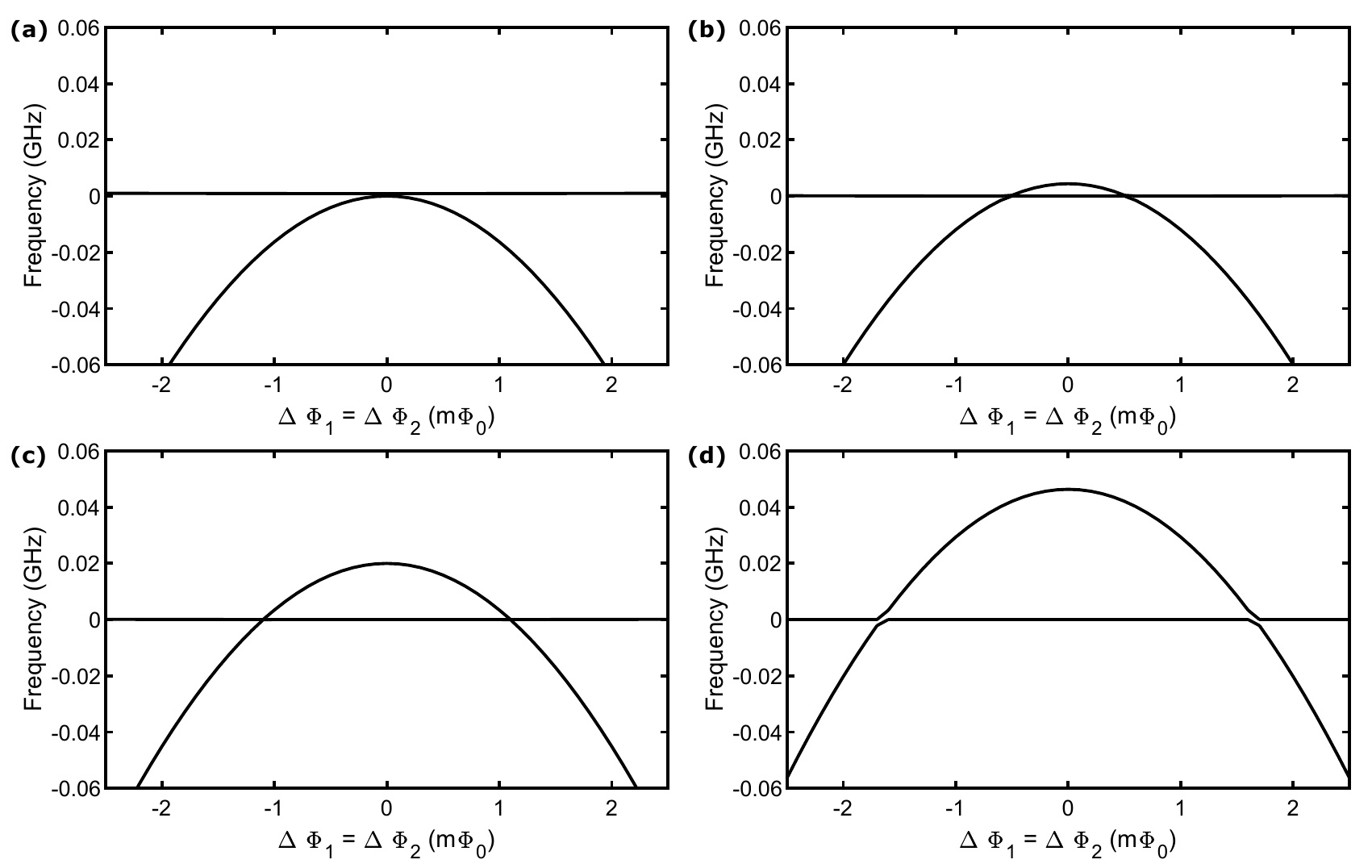}
  \caption{Flux dispersion of lowest two energy levels near double frustration with (a) $\alpha = 0$, (b) $\alpha= 0.01$, (c) $\alpha = 0.02$, (d) $\alpha = 0.03$. When $\alpha = 0$, the even- and odd-parity energy levels are nearly degenerate. As $\alpha$ becomes larger, the even-parity energy level crosses the even-parity level on either side of frustration.
\label{fig:FigS-DF-different-alpha}}
\end{figure}

As shown in Fig.~\ref{fig:FigS-DF-different-alpha}, when $\alpha = 0$, the even-parity energy level and odd-parity energy level experience the same potential, and the energy levels are nearly degenerate, with the minimum separation at exact frustration. When $\alpha \neq 0$ 
%deviates from 0, caused by asymmetries in the circuit, 
the barrier heights between the $00$, $0\pi$, $\pi0$, and $\pi\pi$ wells become different. This leads to the even-parity energy levels and odd-parity energy levels experiencing different potentials, which causes the levels to cross on either side of frustration. The crossing points move further apart for larger $\alpha$.

\subsection{Double plaquette charge dispersion at double frustration}

With effective hybridization at double frustration, the splitting between symmetric and antisymmetric levels exhibits Aharonov-Casher interference, based on the offset charge bias $Q_{\rm isl}$ of the intermediate island between the frustrated plaquettes (Fig.~\ref{fig:FigS-DF-vs-Qi-at-frustration}). When $\alpha = 0$, the symmetric/antisymmetric energy levels for both even and odd parity have $4e$ periodicity. When the symmetric and antisymmetric energy levels cross at $1e\,$mod$2e$, the gap closes [Fig.~\ref{fig:FigS-DF-vs-Qi-at-frustration}(b)], because the even- (odd-) parity wavefunctions both experience a $\cos 2\varphi$ potential. When $\alpha$ of plaquette 1 is 0.01 and $\alpha$ of plaquette 2 is 0.03 [Fig.~\ref{fig:FigS-DF-vs-Qi-at-frustration}(c)], both of the gaps for even- and odd-parity states are not fully closed at $1e\,$mod$2e$ due to incomplete destructive interference. When we bias the island charge at $0e\,$mod$2e$, the transition between the even- and odd-parity logical states is first-order insensitive to charge noise on the intermediate island.

\subsection{Structureless plaquette model}
Modeling a fully-structured three-plaquette circuit is computationally expensive, with 11 nodes [Fig.~\ref{fig:Model-circuit-schematic}(d)], and each node requires several charge states. The matrix size is 189,000$\times$189,000, and thus requires $\sim$300~GB of RAM and takes weeks to calculate the energy levels, even with processors with 40 cores. As an alternative, we can use the stuctureless plaquette model to approximate the full-structure plaquette model. 

We first connect one arm of the plaquette to form a loop, then vary the flux in this loop to obtain the potential of this arm. We next extract the Fourier components of this potential. In the structureless plaquette model, we replace the Josephson potential with the potential extracted from one arm of the plaquette. Thus, we do not need the linear inductors in the circuit, which reduces the number of effective nodes from 11 to 5, and the matrix size is now 7,000$\times$7,000. We add a renormalization factor in the junction capacitance to simulate the effect of higher internal levels. The need for this renormalization stems from the fact that the gap separating the potential energy of a single plaquette from the plaquette's higher internal states is created by a massive Dirac-like Hamiltonian, which makes the flux particle effectively chiral. This chirality impedes tunneling, but cannot be implicitly accounted for in the structureless model. We mimic that by replacing this chiral property of a flux particle by increasing its effective mass by a renormalization factor that is fitted numerically by comparing structureless and full-structure plaquettes. For high internal excited state, the phase particle is less chiral and the renormalization factor is $\sim1.5$ and low-lying internal excited states make the flux particle more chiral and yield a renormalization factor from $\sim2$ to $\sim3.5$ in the most unfavorable cases. We find the structureless plaquette model has a good agreement with the full-structure plaquette near frustration. We thus use the strucureless plaquette model for the remainder of this section. 

\subsection{Triple-frustration modeling}
% \begin{figure}
% \centering
% \includegraphics[width=6.8in]{FigS-TF-vs-Flux_Different_Qi}
%   \caption{Modeled energy level spectrum with $\alpha$ = 0 at triple frustration for simultaneous scan of flux bias to each plaquette along a line from 000 to $\pi\pi\pi$ for $C_{\rm isl}$ = (a) 1~fF, (b) 8~fF, (c) 25~fF, (d) 50~fF. 
%   {\bf Need to add $=\Delta\Phi_3$ to flux axis labels to match Fig.~4(c) of the main paper.} {\textcolor{orange}{The modeling needs to update with the correct renormalization factor.}}
%   \label{fig:triple-model}}
% \end{figure}

We model the triple-frustration flux dispersion for a simultaneous scan of the flux bias to each plaquette along a line from 000 to $\pi\pi\pi$ in Fig.~\ref{fig:Future-optimal-parameters}. In Fig.~\ref{fig:Future-optimal-parameters}(a), at tripled frustration, the computational states are a superposition of even-parity wells ($000$, $0\pi\pi$, $\pi0\pi$, $\pi\pi0$) and a superposition of odd-parity wells ($\pi\pi\pi$, $00\pi$, $0\pi0$, $\pi00$). In the ideal case of symmetric plaquettes and small flux offsets, the nearly-degenerate computational states are $(\left|000\right\rangle +\left|0\pi\pi\right\rangle +\left|\pi0\pi\right\rangle +\left|\pi\pi0\right\rangle )\pm(\left|\pi\pi\pi\right\rangle +\left|\pi00\right\rangle +\left|0\pi0\right\rangle +\left|00\pi\right\rangle )$ that correspond to even and odd charge parities. However, as we go away from the protected regime, the computational states turn into even ($\left|000\right\rangle +\left|0\pi\pi\right\rangle +\left|\pi0\pi\right\rangle +\left|\pi\pi0\right\rangle $) and odd ($\left|\pi\pi\pi\right\rangle +\left|\pi00\right\rangle +\left|0\pi0\right\rangle +\left|00\pi\right\rangle $) flux states.  A few $\rm m\Phi_0$ away from triple frustration, the lowest energy level corresponds to the wavefunction localized only in the 000 or $\pi\pi\pi$ well. The second lowest levels correspond to a superposition of $00\pi$, $0\pi0$, $\pi00$ on the left and $0\pi\pi$, $\pi0\pi$, $\pi\pi0$ on the right. The energy levels for the 0 and 1 logical states both have negative curvature with respect to flux, so the flux dispersion of the 0-1 transition is flatter compared to double frustration, thus further enhancing the protection against flux noise. %For the four different values of $C_{\rm isl}$, we observe that the gap between the symmetric and antisymmetric levels decreases and the slope of the flux dispersion increases as $C_{is}$ is increased. In Fig.~\ref{fig:triple-model}(d), the gap vanishes when $C_{\rm isl}$ is so large that there is effectively no hybridization between the plaquettes. 
%We can see the flux dependence of energy level of 000 well has 3 times slope than the energy levels of superposition of $00\pi$,$0\pi0$,$\pi00$.

\section{Fitting of energy-level spectra}

\subsection{General fitting strategy}

%Extracting correct transition frequencies is typically accomplished by fitting a Gaussian vs. frequency at each flux. However, 
For extracting the center frequency of each transition at every flux point, the complexity of the level spectrum for our device makes it not practical to implement an automated routine that fits a standard curve to each feature in the spectroscopy data. %this approach is unviable. To overcome this obstacle, 
Instead, we identify each transition feature in the data manually, then extract the maximum of the spectroscopy signal for each of these features. We then check the correspondence of the extracted transitions with our numerical model of the energy-level spectrum. We use the initial estimates for the various device parameters to calculate the energy levels, which generally match the data qualitatively. From this, we can identify most of the transitions. During the first round of fitting, where we adjust the circuit parameters about their estimated values to match with the transition data, described in detail below, we only fit the transitions that we have identified correctly in the initial step. Following this stage, the fitted energy levels typically match with the data quite well and we can identify more transitions. We then use the newly identified transitions to further refine the fit.

After extracting the plasmon, heavy fluxon, and light fluxon transitions, as well as the anticrossings from the spectroscopy flux- and charge-dependence data, we use the single (double) plaquette model for the energy-level spectra described in Sec.~\ref{Sec:model} to fit the single- (double-) frustration spectroscopy data. At single frustration, we fit $E_J$, $E_C$, $E_L$, $E_{CL}$, $C_{\rm sh}$, $\alpha$, and $L_{\rm factor}$ using the model shown in Fig.~\ref{fig:Model-circuit-schematic}(b). We fix $C_{\rm int}$ to be 1~fF, which is estimated from numerical modeling with Q3D and a theoretical estimation of the effect of the junction chain capacitance to ground. We introduce the parameter $L_{\rm factor}$ to account for variations in $L_{\rm extra}$ due to small flux offsets in the bias of the nominally unfrustrated plaquettes or SQUIDs. At double frustration, we fit the same parameter set as in the single frustration case, but with the addition of $C_{\rm isl}$. Similarly to the single frustration case, we fix $C_{\rm int}$ to be 1~fF. We assume the two plaquettes at frustration share the same set of parameters, because the actual parameters between the two plaquettes are typically only different by a few percent based on our test structures during the device fabrication. This allows us to reduce the number of fitting parameters from 15 to 8, and thus makes the fitting more practical.

The cost function of our fitting procedure is $\sum_n W_n \Delta f_n^2$, where $\Delta f_n$ is the difference between the modeled and experimental frequencies for transition $n$, and $W_n$ is the weight that we assign to transition $n$. The goal of the fitting process is to minimize the cost function and find the parameter set that has less than a 10\% difference between the modeled transitions and the experimental transitions. We use the scipy.optimize.minimize function in Python to do the fitting. We have 7 and 8 parameters for single- and double-frustration fitting, respectively. We find that the Nelder-Mead method performs better for this fitting than gradient descent methods in terms of avoiding local minima.

With this high-dimensional fit, we need to choose the initial parameters carefully. We use the initial $E_J$ and $E_L$ values calculated from the Ambegaokar-Baratoff relation using the on-chip test junction resistances. The initial $\alpha$ of the junctions is estimated from our test chips that each contain 6 identical junctions. As mentioned earlier, we define the charging energy as $E_C \equiv (2e)^2/2C$. The initial $E_C$ and $E_{CL}$ values are calculated from the the relevant junction areas measured with scanning electron microscopy with a total specific capacitance 70~fF/$\mu$m$^2$. 
%{\bf (*Are we sure this is the correct value to quote here?)}
The initial $C_{\rm sh}$ is estimated from Q3D simulation. The initial $L_{\rm factor}$ is set to 1 because our unfrustrated plaquettes are nominally biased at unfrustration. We choose the initial simplex for the minimization so that it covers the possible range for each parameter, which is typically $\pm 5 \%$ to $\pm 30\%$ of the initial values. 

\subsection{Single frustration fitting}

\begin{figure*}
\centering
\includegraphics[width=\textwidth]{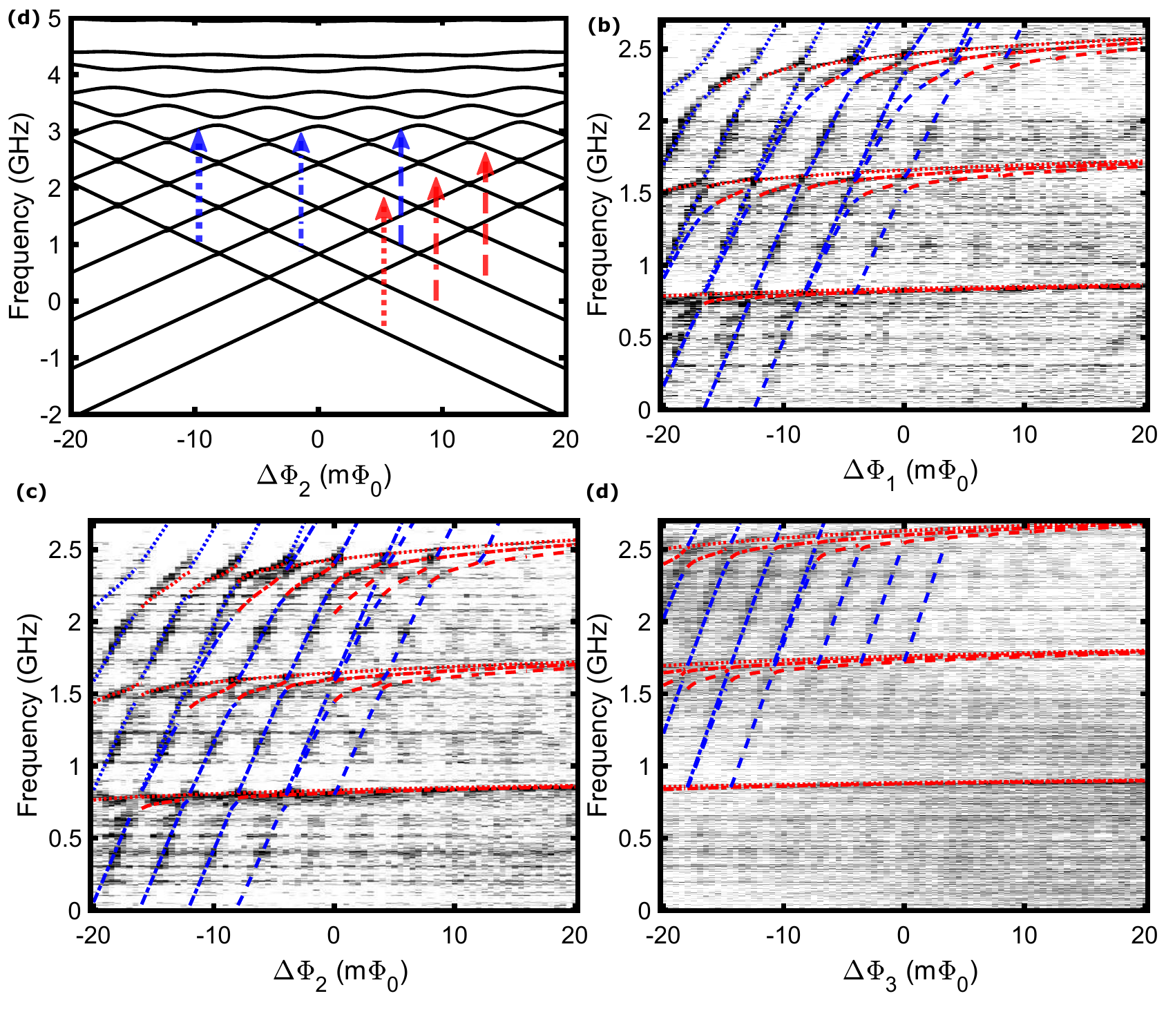}
  \caption{(a) Fitted energy levels for plaquette 2 single frustration with arrows indicating corresponding transitions probed in spectroscopy measurements. Fitting of (b) Plaquette 1 , (c) Plaquette 2  and (d) Plaquette 3 spectroscopy data. The red lines are plasmon transitions, blue lines are fluxon transitions. The dotted lines are transitions out of the $\ket{0_\pi}$ state. The dash-dotted lines are transitions out of the $\ket{1_\pi}$ state. The dashed lines are transitions out of the $\ket{2_\pi}$ state.
  \label{fig:FigS-Single-Frus}}
\end{figure*}

%Initially, 
We fit the single frustration data with our single plaquette model [Fig.~\ref{fig:Model-circuit-schematic}(b)]. We put equal weight on different transitions by setting $W_n = 1$ for the fitting. The fit runs on a computer with a 12-core processor and takes $\sim$1 day and 500 iterations to converge. In Fig.~\ref{fig:FigS-Single-Frus}(c), we show the fitting of plaquette 2 single frustration. The red lines correspond to the fitted plasmon transitions, and the blue lines correspond to the fitted heavy fluxon transitions. The dotted lines are transitions out of the $\ket{0_\pi}$ state, corresponding to the 0 level of the $\pi$ well. The dash-dotted lines are transitions out of the $\ket{1_\pi}$ state. The dashed lines are transitions out of the $\ket{2_\pi}$ state. The transitions match with the data within 10\% error. In Fig.~\ref{fig:FigS-Single-Frus}(a), we show the modeled energy levels using the fitting parameters and indicate some example plasmon and fluxon transitions out of the $\ket{0_\pi}$, $\ket{1_\pi}$ and $\ket{2_\pi}$ states.

In Fig.~\ref{fig:FigS-Single-Frus}(b-d), we show the single frustration fitting of plaquette 1, 2, 3 single frustration. Plaquette 1 behaves similarly to plaquette 2, and thus the fitted transitions and parameter values are similar. Plaquette 3 single frustration behaves somewhat differently, and the fitted energy levels and parameters differ by a larger amount compared to plaquettes 1 and 2. The fitted parameters are listed in Table~\ref{tab:fitting-SF-fit-para} and they are within 20\% of the parameters that we estimate from the design and fabrication tests, although a few of the parameters for plaquette 3 have a slightly larger discrepancy. The spectroscopy measurements at plaquette 3 single frustration are not as clean as for plaquettes 1 and 2 single frustration, thus potentially accounting for the larger variation with the estimated values.

\begin{table}
    \centering
\begin{tabular}{|c|c|c|c|c|c|c|c|}
\hline
\multicolumn{1}{|c|}{}                                                                & \multicolumn{1}{c|}{\begin{tabular}[c]{@{}c@{}}$E_J$ \\ (K)\end{tabular}}           & \multicolumn{1}{c|}{\begin{tabular}[c]{@{}c@{}}$E_C$\\  (K)\end{tabular}}         & \multicolumn{1}{c|}{\begin{tabular}[c]{@{}c@{}}$E_L$\\  (K)\end{tabular}}           & \multicolumn{1}{c|}{\begin{tabular}[c]{@{}c@{}}$E_{CL}$ \\ (K)\end{tabular}}  & \multicolumn{1}{c|}{\begin{tabular}[c]{@{}c@{}}$C_{\rm sh}$ \\ (fF)\end{tabular}}       & \multicolumn{1}{c|}{$\alpha$}                   & $L_{\rm factor}$                                                   \\ \hline
Plaquette 1          & 1.65      & 3.65      & 1.12      & 5.60          & 1160         & 0.03    & 1.1       \\ \hline
Plaquette 2          & 1.65      & 3.67      & 1.11      & 6.36         & 1190         & 0.02    & 1.1       \\ \hline
Plaquette 3          & 1.97       & 4.00      & 1.27      & 6.66         & 1440         & 0.04     & 0.91       \\ \hline
\begin{tabular}[c]{@{}c@{}}Estimated\\  parameters\end{tabular}  & 1.45      & 3.82        & 1.39      & 6.46          & 1000         & 0.02    & 1.0          \\ \hline
\end{tabular}
    \caption{Single frustration fit parameters and estimated parameters from design and fabrication tests.}
    \label{tab:fitting-SF-fit-para}
\end{table}

\subsection{Double frustration fitting}

%\begin{figure*}
%\centering
%\includegraphics[width=\textwidth]{Fitting-DF-Qi.pdf}
%  \caption{(a) Circuit image. (b) Plaquette 1/2 double frustration charge modulation data of $\ket{0_{ES}} \rightarrow \ket{3_{ES}}$ transition at 17 m$\Phi_0$. The red and blue dotted lines are the fitted transitions that correspond to different quasiparticle parity on intermediate Island 1. (c) Plaquette 2/3 double frustration charge modulation data of $\ket{0_{ES}} \rightarrow \ket{3_{ES}}$ transition at 11 m$\Phi_0$. The red and blue dotted lines are the fitted transitions that correspond to different quasiparticle parity on intermediate Island 2.
%  \label{fig:Fitting-DF-Qi}}
%\end{figure*}

\begin{figure*}
\centering
\includegraphics[width=\textwidth]{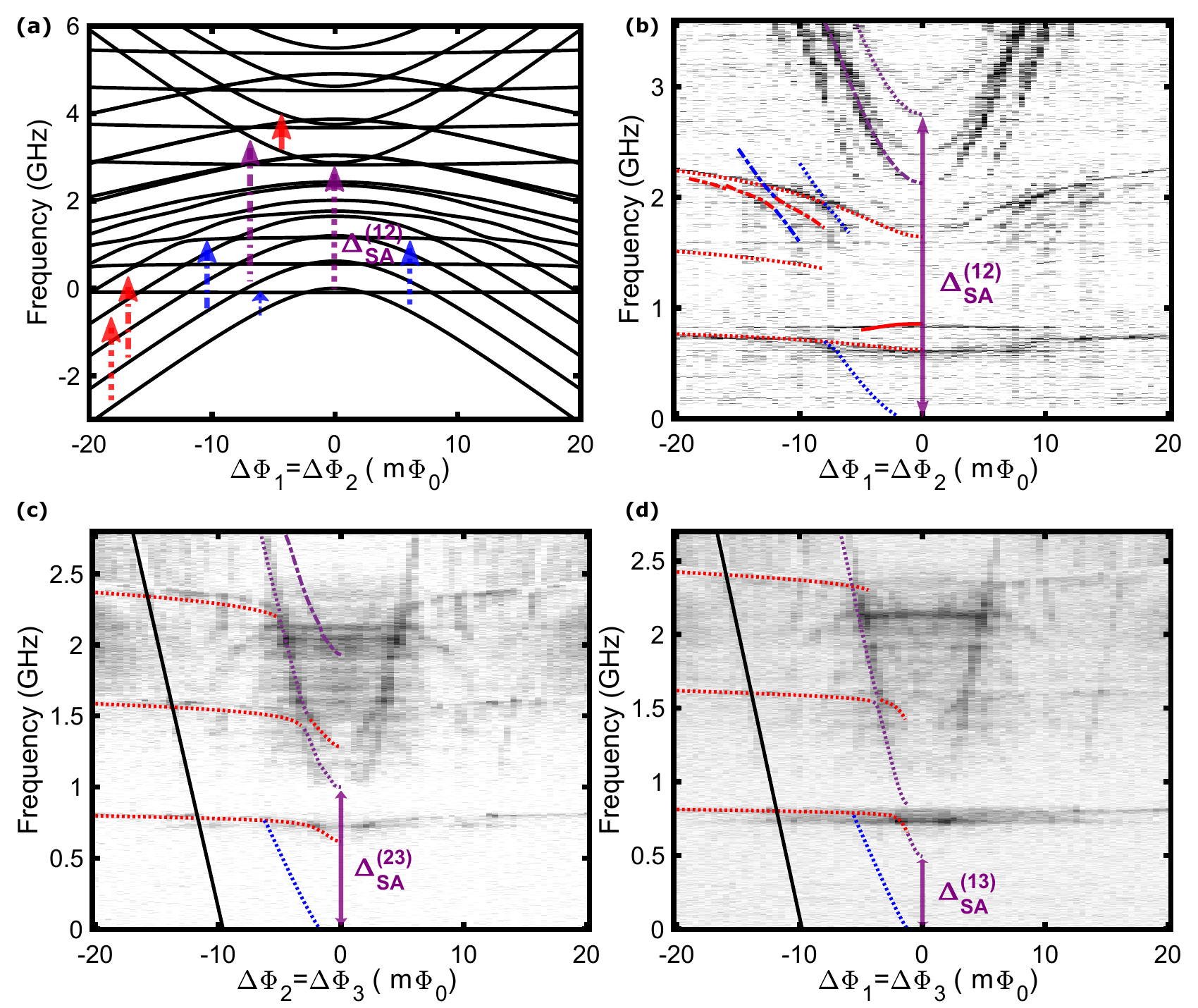}
  \caption{(a) Calculated energy-level spectrum using the fitting parameters with arrows indicating the various corresponding transitions from the spectroscopy data for plaquette (12) double frustration. Flux spectroscopy data and fitting of (b) plaquette (12), (c) plaquette (23) and (d) plaquette (13) double frustration. The red lines are the fitted plasmon transitions, the blue lines are the fitted heavy fluxon transitions, the purple lines are the fitted light fluxon transitions. The dotted lines are transitions out of the 0 state of the symmetric energy levels in the even-parity wells, which we denote as $\ket{0_{ES}}$. The dash-dotted lines are transitions out of the 1 state of the symmetric energy levels in the even-parity wells, $\ket{1_{ES}}$. The dashed lines are transitions out of the $\ket{2_{ES}}$ state. The solid black lines between -20 to -10 m$\Phi_0$ in the plaquette (23) and (13) double frustration plots correspond to transitions involving an excitation in the readout cavity: $\ket{0_{ES,n}} \rightarrow \ket{1_{EA,n-1}}$
%  $\ket{0_{ES},\,n} \rightarrow \ket{1_{EA},\,n-1}$, 
  where $n$ and $n-1$ 
%  inside the ket are 
  indicate photon number in the cavity.
  %{\bf We should add the superscripts to the $\Delta_{SA}$ labels, so, (12) for (b), (23) for (c), and (13) for (d), like in Fig. 4(b) of the main paper.} {\textcolor{orange}{Done.}}
  \label{fig:FigS-Double-Frus}}
\end{figure*}

We next use the model in Fig.~\ref{fig:Model-circuit-schematic}(c) to fit the double frustration data. Because the Hilbert space is $\sim$11 times larger at double frustration using this model, we use a 48-core processor to do the fitting. This process takes between 4-7 days and $\sim$300 iterations to converge. The anticrossings between different transitions are important features to fit because they determine the coupling between the computational states, so we put $\sim$20~times more weight for the regions in the spectroscopy data that exhibit significant anticrossings. We also simultaneously fit the corresponding charge modulation data and we only use the minimum and maximum of the charge modulation data for fitting. In order to compensate for the relatively small number of charge modulation data points, we put $\sim$50 times more weight for these features in the fitting.

In Fig.~\ref{fig:FigS-Double-Frus}(b), we show the plaquette (12) double frustration data and fitted transitions. The red lines are the fitted plasmon transitions, the blue lines are the fitted heavy fluxon transitions, and the purple lines are the fitted light fluxon transitions. The dotted lines are transitions out of the $\ket{0_{ES}}$ state, where $E$ corresponds to the even-parity hybridized well between plaquettes 1 and 2, $S$ corresponds to the symmetric hybridized energy level of plaquettes 1 and 2, and 0 corresponds to the lowest energy level with these conditions. The dash-dotted lines are transitions out of the $\ket{1_{ES}}$ state and the dashed lines are transitions out of the $\ket{2_{ES}}$ state. The red solid line is $\ket{0_{EA}} \rightarrow \ket{1_{EA}}$, which is the transition out of the 0 state of the antisymmetric energy levels in the even-parity wells to the 1 state of the antisymmetric energy levels in the even-parity wells. We see this transition because there is fast quasiparticle poisoning on the intermediate islands that is faster than our measurement timescale. When we prepare the qubit in the $\ket{0_{ES}}$ state, the fast quasiparticle poisoning closes and opens the symmetric and antisymmetric gap randomly, which allows the system to occasionally transfer population from the $\ket{0_{ES}}$ to $\ket{0_{EA}}$ states, thus leaving population in the excited antisymmetric state. This results in the transition indicated by the solid red line in Fig.~\ref{fig:FigS-Double-Frus}(b). In Fig.~\ref{fig:Fitting-DF-Qi}(b), we show the plaquette (12) double frustration charge modulation data of the $\ket{0_{ES}} \rightarrow \ket{3_{ES}}$ transition at 17~m$\Phi_0$. We can clearly see two quasiparticle bands, which we indicate by the red and blue dotted lines for the fitted transitions. We show the flux dependence of the fitted energy levels and corresponding transitions in Fig.~\ref{fig:FigS-Double-Frus}(a). In Fig.~\ref{fig:Fitting-DF-Qi}(c), we show the plaquette (23) double frustration charge modulation data of the $\ket{0_{ES}} \rightarrow \ket{3_{ES}}$ transition at 11~m$\Phi_0$.

In Fig.~\ref{fig:FigS-Double-Frus}(b-d), we show the fit results for the flux spectroscopy at plaquette (12), (23), and (13) double frustration. The $\Delta_{\rm SA}$ for (12), (23), and (13) double frustration are $\sim$ 2.7, 1.0, and 0.5~GHz, respectively, as expected for a decreasing $\Delta_{\rm SA}$ and progressively weaker hybridization for a larger intermediate island capacitance to ground. The fitted curves capture the transitions, anticrossings, and charge modulation to within 10\%. The fitted parameters are shown in Table~\ref{tab:fitting-DF-fit-para} and are in reasonable agreement with our estimated parameters.

In Fig.~\ref{fig:FigS-Double-Frus}(c,d), the solid black lines between -20 to -10 m$\Phi_0$ in the plaquette (23) and (13) double frustration plots correspond to transitions involving the readout cavity: $\ket{0_{ES,n}} \rightarrow \ket{1_{EA,n-1}}$, 
%$\ket{0_{ES},\,n} \rightarrow \ket{1_{EA},\,n-1}$
where $n$ and $n-1$ indicate photon number in the cavity. 

Near double frustration, as described in the schematic plots in Fig.~1(k) in the main paper, the parity of the initial state and the direction of the scan through the 2D flux-bias space determines which levels will disperse with respect to flux and which will be flat. For example, preparation in an even-parity state, say, $\pi\pi$, followed by a scan in the even-parity flux-bias direction, so, going from a flux bias where the $\pi\pi$ well is the global minimum to a flux where the 00 well is the global minimum, should result in the even-parity levels dispersing with flux while the odd-parity levels should be flat. This behavior should swap for the opposite parity of the initialized state and scan direction.

\begin{figure}[htb!]
\centering
\includegraphics[width=\textwidth]{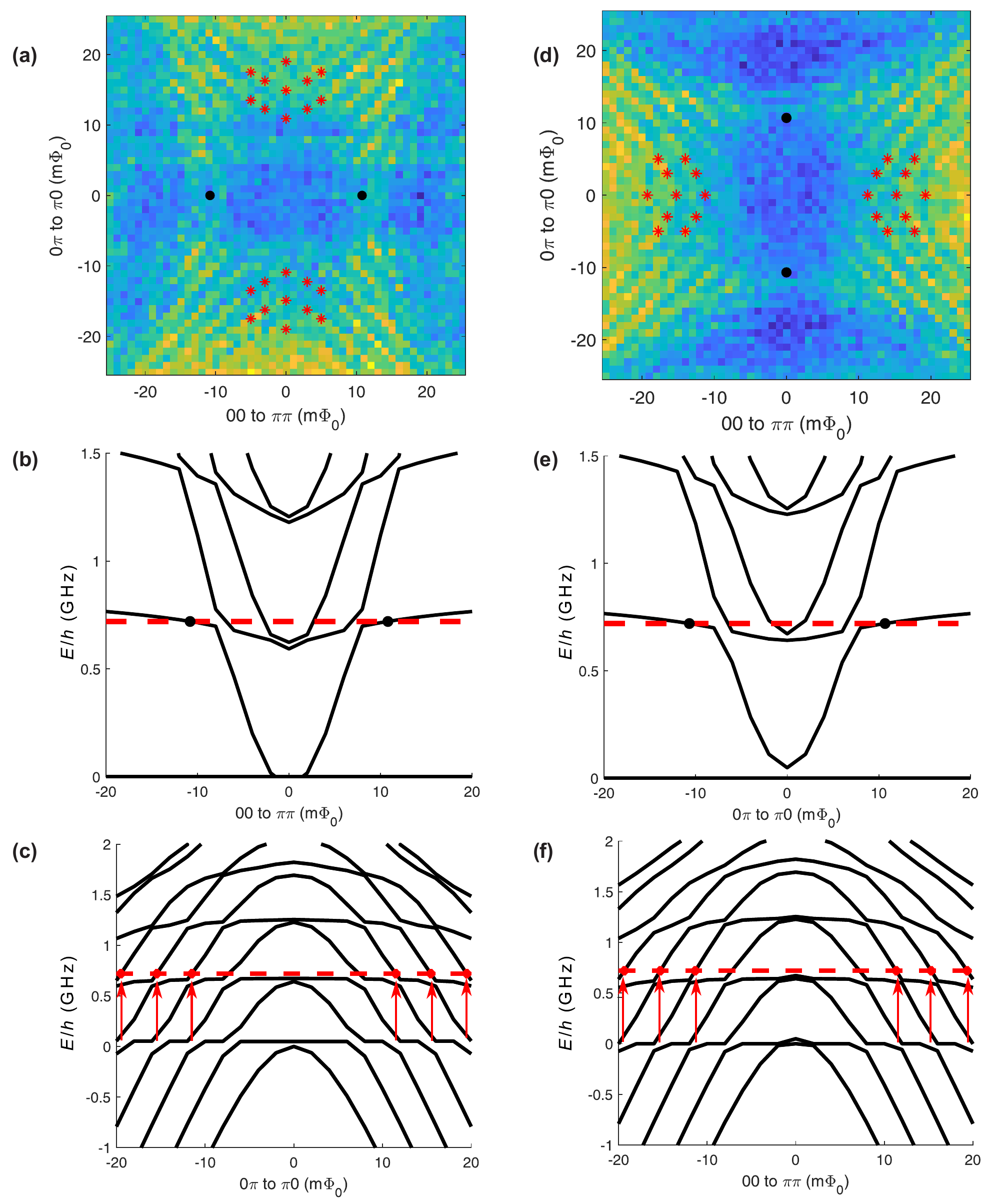}
  \caption{(a) Single-point spectroscopy measurement at 720~MHz while initializing in the $\pi\pi$ well and scanning in the even- and odd-parity flux-bias directions near plaquette (12) double frustration. (b) Simulated transitions initialized in $\pi\pi$ well and scanning in the $00$ to $\pi\pi$ even-parity flux-bias direction. (c) Modeled energy levels for initialization in even parity and scanning in the $0\pi$ to $\pi0$ odd-parity flux-bias direction. (d) Initializing in $\pi0$ well and repeating the 2D single-frequency spectroscopy scan from (a) near double frustration. (e) Simulated transitions initialized in $\pi0$ well and scanning in the $0\pi$ to $\pi0$ odd-parity flux-bias direction. (f) Modeled energy levels for initialization in odd parity and scanning in the $00$ to $\pi\pi$ even-parity direction. Details of symbols on spectroscopy plots and lines, arrows, and symbols on modeled transition plots and level diagrams are described in main text of this section.
\label{fig:Qubit_pipi_v_pi0}}
\end{figure}

To examine this difference, we measure single-frequency spectroscopy scans in the 2D flux-bias space near plaquette (12) double frustration for both even- and odd-parity initializations [Fig.~\ref{fig:Qubit_pipi_v_pi0}]. For each point in Fig.~\ref{fig:Qubit_pipi_v_pi0}(a), we initialize in the $\pi\pi$ well near double frustration, then bring the flux bias to the point indicated in the 2D flux space, apply a spectroscopy pulse at 720~MHz, and then bring the flux bias to double frustration to read out [Fig.~\ref{fig:Qubit_pipi_v_pi0}(a)]. The axes of the plot correspond to the even- and odd-parity flux directions. The black dots along the 00 to $\pi\pi$ direction indicate the points where we see transitions cross 720~MHz in our simulation of the spectrum [Fig~\ref{fig:Qubit_pipi_v_pi0}(b)]. These points correspond to the transitions that are visible in the spectroscopy scan in Fig.~4(b) of the main paper. To identify the somewhat hyperbolic features away from the even-parity flux-bias axis in the odd-parity flux-bias direction, we consider the modeled level spectrum for a preparation in the even-parity well, with a scan along the odd-parity flux-bias direction [Fig~\ref{fig:Qubit_pipi_v_pi0}(c)]. For flux-bias points of 10~m$\Phi_0$ or more away from frustration, there are many transitions out of the flat even-parity level up to the dispersing odd-parity levels. The transitions that match the 720~MHz spectroscopy frequency are labeled by red dots. If the odd-parity flux-bias scan is done for a nonzero offset along the even-parity flux-bias axis, the points that correspond to this same spectroscopy frequency shift in position, as shown again by the red dots.

If we now initialize in an odd-parity well, the $\pi0$ well in this case, then repeat the same measurement process [Fig.~\ref{fig:Qubit_pipi_v_pi0}(d)], we see that all of the features from the scan with even-parity initialization repeat, but are now rotated by 90 degrees. We again perform the same numerical modeling process for the level structure, but now with the $\pi0$ initial state and the parity of the simulated flux-bias axes swapped [Fig.~\ref{fig:Qubit_pipi_v_pi0}(e,f)]. Again, the features visible in this 2D spectroscopy scan match up with the various transitions that we identify in the modeled level spectrum.
%We see features similar to those seen when prepared in the $\pi\pi$ well except reflected on the other axis. This demonstrates the symmetry of preparing in the even-parity state versus odd-parity state.

\begin{table}[]
\centering
\begin{tabular}{|c|c|c|c|c|c|c|c|c|}
\hline
\multicolumn{1}{|c|}{}                                                                & \multicolumn{1}{c|}{\begin{tabular}[c]{@{}c@{}}$E_J$ \\ (K)\end{tabular}}           & \multicolumn{1}{c|}{\begin{tabular}[c]{@{}c@{}}$E_C$\\  (K)\end{tabular}}         & \multicolumn{1}{c|}{\begin{tabular}[c]{@{}c@{}}$E_L$\\  (K)\end{tabular}}           & \multicolumn{1}{c|}{\begin{tabular}[c]{@{}c@{}}$E_{CL}$ \\ (K)\end{tabular}}  & \multicolumn{1}{c|}{\begin{tabular}[c]{@{}c@{}}$C_{\rm sh}$ \\ (fF)\end{tabular}}       & \multicolumn{1}{c|}{$\alpha$}                                                        & \multicolumn{1}{c|}{\begin{tabular}[c]{@{}c@{}}$C_{\rm isl}$\\ (fF)\end{tabular}}  & $L_{\rm factor}$                                                   \\ \hline
\begin{tabular}[c]{@{}c@{}}Plaquette\\  (12)\end{tabular} & 1.75      & 3.54      & 1.20       & 6.34         & 1240         & 0.03    & 1.5      & 0.99       \\ \hline
\begin{tabular}[c]{@{}c@{}}Plaquette\\  (23)\end{tabular}& 1.76      & 3.53      & 0.900       & 7.40          & 1290         & 0.04    & 5.7      & 1.0      \\ \hline
\begin{tabular}[c]{@{}c@{}}Plaquette\\  (13)\end{tabular}& 1.73      & 3.48      & 0.903      & 6.62         & 1310         & 0.03    & 8.1     & 0.98       \\ \hline
\begin{tabular}[c]{@{}c@{}}Estimated\\  parameters\end{tabular}  & 1.45      & 3.82        & 1.39      & 6.46           & 1000         & 0.02    & Vary*      & 1.0          \\ \hline
\end{tabular}
    \caption{Double frustration fit parameters and estimated parameters. *The estimated $C_{\rm isl}$ for plaquette (12), (23), and (13) 
    double frustration are 0.81, 5.0, 5.8~fF. 
%    {\bf Can we really make these $C_{\rm isl}$ estimates to 3 sig figs? 2 sig figs would seem more appropriate; can we also estimate $C_{\rm isl}$ for (13) double frustration?}{\textcolor{orange}{Done}}} 
}
    \label{tab:fitting-DF-fit-para}
\end{table}

\begin{figure*}
\centering
\includegraphics[width=\textwidth]{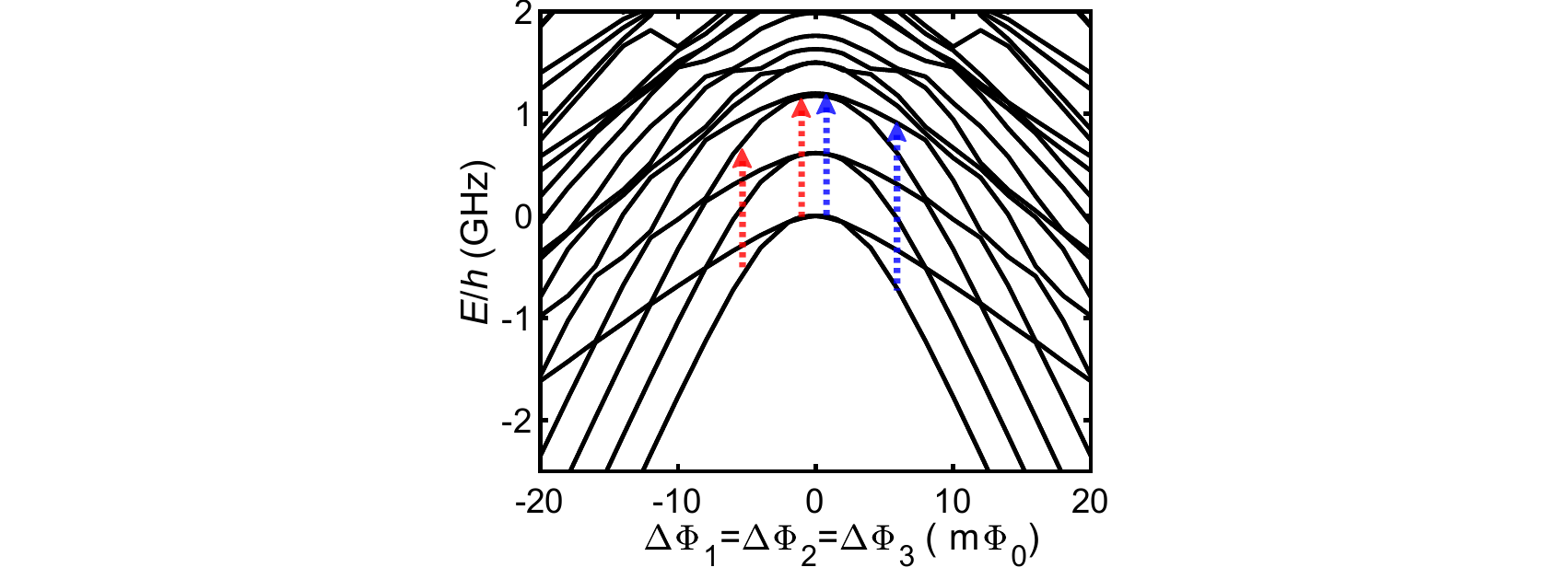}
  \caption{Modeled energy levels near triple frustration for experimental device. The dotted red arrows are example plasmon transitions and the dotted blue arrows are example heavy fluxon transitions. Within $\sim\pm2$~m$\Phi_0$ of triple frustration, the heavy fluxon transitions and plasmon transitions are difficult to distinguish. Because in both cases the initial and final levels have similar curvature with respect to flux, the transition between them has a rather flat flux dispersion.
  %, just as the even-parity and odd-parity energy levels likewise inseparable.
  \label{fig:FigS-TF-En}}
\end{figure*}

\begin{figure*}
\centering
\includegraphics[width=\textwidth]{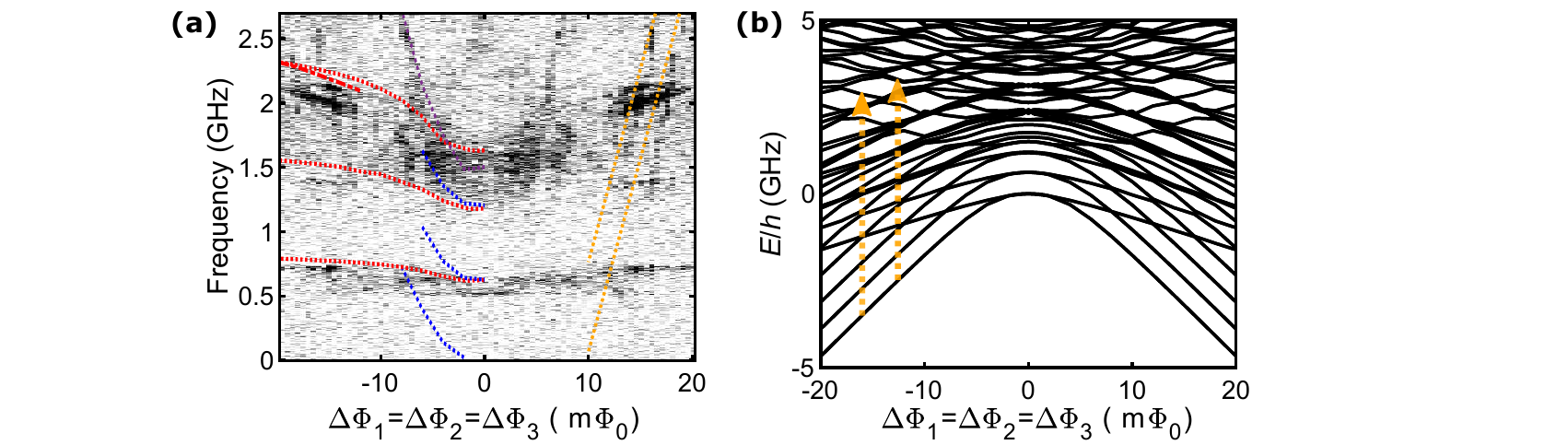}
  \caption{(a) Triple frustration spectroscopy plot similar to Fig.~4(c) in the main paper) and corresponding modeled energy levels. Lines indicate modeled transitions with: red = plasmons, blue = heavy fluxons, purple = light fluxons, dotted = transitions out of 0 level, dash-dotted = transitions out of 1 level, dashed = transitions out of 2 level, solid red line = plasmon transition between antisymmetric levels in even-parity well, orange = light fluxon plus cavity photon. (b) Triple frustration modeled energy levels. Transitions indicated by orange arrows correspond to light fluxon plus cavity photon.
  %, just as the even-parity and odd-parity energy levels likewise inseparable.
  \label{FigS-TF-traisntion involves cavity}}
\end{figure*}

\subsection{Triple frustration modeling}
% \begin{figure*}
% \centering
% \includegraphics[width=\textwidth]{Fitting-TF2.pdf}
%   \caption{(a) Device image. (b) Flux spectroscopy data and fitting of triple frustration data. The red lines are the modeled plasmon transitions, the blue line is the modeled heavy fluxon transition, the purple line is the modeled light fluxon transition. The dotted lines are transitions out of the 0 state of the symmetric energy levels in the even-parity wells. {\bf (*Do we still need to include this figure since it is the same as Fig. 4(c) in the main paper?)}
%   \label{fig:Fitting-triple-frustration}}
% \end{figure*}
We model the triple frustration data with the circuit in Fig.~\ref{fig:Model-circuit-schematic}(d). The Hilbert space of this triple frustration model is $\sim$100 times larger than for the double frustration model. Modeling one flux point takes $\sim$10 days, so it is impossible to model several flux points and fit to the triple frustration data. We model the triple frustration energy levels by simultaneously modeling different flux points on four virtual machines in parallel, which have 48-, 24-, 12- and 12-core processors, respectively. In each of the virtual machines, we use the multiprocess function in Python to model different flux points in parallel so that it uses all the computational power in that virtual machine. Because fitting is not practical here, we use the fitted parameters from the single- and double-frustration modeling, and only adjust the $L_{\rm factor}$ to account for the SQUIDs not being at exact unfrustration. Figure~4(c) in the main paper shows the triple frustration flux spectroscopy data and modeled transitions. The red lines are the modeled plasmon transitions, the blue line is the modeled heavy fluxon transition, and the purple line is the modeled light fluxon transition. The dotted lines are transitions out of the 0 state of the symmetric energy levels in the even-parity wells. Based on our modeling, we find the plasmon and light fluxon transitions are in good agreement between our modeled curves and the triple frustration data. In general, we observe that the transitions at triple frustration have an even flatter dispersion with respect to flux compared to double frustration. As shown in Fig.~\ref{fig:FigS-TF-En}, within $\sim\pm2$~m$\Phi_0$ of triple frustration, the heavy fluxon transitions and plasmon transitions are difficult to distinguish, as the even-parity and odd-parity energy levels are likewise nearly on top of each other. Because of this, the dispersion of the various heavy fluxon transitions, $\ket{0_E}\rightarrow\ket{0_O}, \ket{1_O}, \ket{2_O}, \ket{3_O}$, is comparable and rather flat.

Figure~\ref{FigS-TF-traisntion involves cavity} shows the light fluxon transition plus cavity photon, which is visible near the edges of our triple frustration spectroscopy data, and the comparison with the modeled level spectrum. The difference between the levels indicated by arrows minus the readout cavity frequency matches these particular features in our data.

\begin{figure*}
\centering
\includegraphics[width=\textwidth]{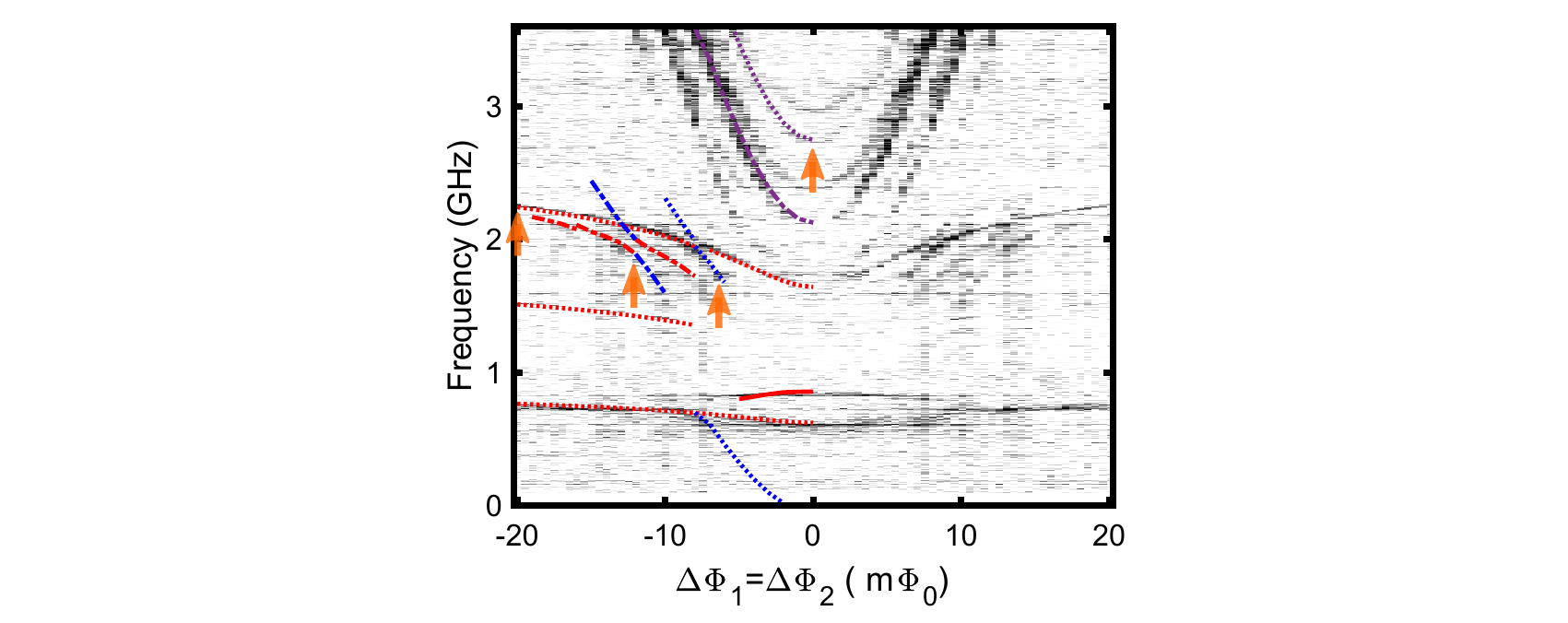}
  \caption{Flux spectroscopy data and fitting of plaquette (12) double frustration. The orange arrows point to the four transitions we use to estimate the  uncertainty in the fit parameters.
  \label{fig:Fitting-Plaq12-error-estimation}}
\end{figure*}

\begin{table}[]
\centering
\begin{tabular}{|cccccccccc|}
\hline

\multicolumn{9}{|c|}{{\bf Parameter uncertainty estimates within a 10\% range based on:}}\\ \hline
\multicolumn{1}{|c|}{\begin{tabular}[c]{@{}c@{}}$E_J$ \\ (K)\end{tabular}}           & \multicolumn{1}{c|}{\begin{tabular}[c]{@{}c@{}}$E_C$\\  (K)\end{tabular}}         & \multicolumn{1}{c|}{\begin{tabular}[c]{@{}c@{}}$E_L$\\  (K)\end{tabular}}           & \multicolumn{1}{c|}{\begin{tabular}[c]{@{}c@{}}$E_{CL}$ \\ (K)\end{tabular}}  & \multicolumn{1}{c|}{\begin{tabular}[c]{@{}c@{}}$C_{\rm sh}$ \\ (fF)\end{tabular}}       & \multicolumn{1}{c|}{$\alpha$}                                                        & \multicolumn{1}{c|}{\begin{tabular}[c]{@{}c@{}}$C_{\rm isl}$\\ (fF)\end{tabular}} & \multicolumn{1}{c|}{\begin{tabular}[c]{@{}c@{}}$C_{\rm int}$\\ (fF)\end{tabular}}       & \multicolumn{1}{c|}{$L_{\rm factor}$ }                                                      \\ \hline
\hline
\multicolumn{9}{|c|}{(a) Plasmon transition ($\ket{0_{ES}} \rightarrow \ket{3_{ES}}$ transition at 20~m$\Phi_0$)}\\ \hline

\multicolumn{1}{|c|}{\begin{tabular}[c]{@{}c@{}}1.65 \\ $\sim$ \\ 1.85\end{tabular}} & \multicolumn{1}{c|}{\begin{tabular}[c]{@{}c@{}}3.00 \\ $\sim$ \\ 4.00\end{tabular}} & \multicolumn{1}{c|}{\begin{tabular}[c]{@{}c@{}}0.770 \\ $\sim$ \\ 1.35\end{tabular}} & \multicolumn{1}{c|}{\begin{tabular}[c]{@{}c@{}}3.50 \\ $\sim$ \\ 15.0\end{tabular}} & \multicolumn{1}{c|}{\begin{tabular}[c]{@{}c@{}}1100\\  $\sim$ \\ 1400\end{tabular}} & \multicolumn{1}{c|}{\begin{tabular}[c]{@{}c@{}}0.0\\  $\sim$\\  0.1\end{tabular}} & \multicolumn{1}{c|}{\begin{tabular}[c]{@{}c@{}}0.200 \\ $\sim$ \\ 6.00\end{tabular}} & \multicolumn{1}{c|}{\begin{tabular}[c]{@{}c@{}}0.800 \\ $\sim$ \\ 2.10\end{tabular}} & \multicolumn{1}{c|}{\begin{tabular}[c]{@{}c@{}}0.83 \\ $\sim$ \\ 1.1\end{tabular}} \\ \hline

\hline
\hline
\multicolumn{9}{|c|}{(b) Heavy fluxon transition ($\ket{0_{ES}} \rightarrow \ket{2_{OS}}$ transition at 6.5~m$\Phi_0$)}                                                                        \\ \hline

% \multicolumn{1}{|c|}{}                                                                & \multicolumn{1}{c|}{\begin{tabular}[c]{@{}c@{}}$E_J$ \\ (K)\end{tabular}}           & \multicolumn{1}{c|}{\begin{tabular}[c]{@{}c@{}}$E_C$\\  (K)\end{tabular}}         & \multicolumn{1}{c|}{\begin{tabular}[c]{@{}c@{}}$E_L$\\  (K)\end{tabular}}           & \multicolumn{1}{c|}{\begin{tabular}[c]{@{}c@{}}$E_{CL}$ \\ (K)\end{tabular}}  & \multicolumn{1}{c|}{\begin{tabular}[c]{@{}c@{}}$C_{\rm sh}$ \\ (fF)\end{tabular}}       & \multicolumn{1}{c|}{$\alpha$}                                                        & \multicolumn{1}{c|}{\begin{tabular}[c]{@{}c@{}}$C_{\rm isl}$\\ (fF)\end{tabular}} & \multicolumn{1}{c|}{\begin{tabular}[c]{@{}c@{}}$C_{\rm int}$\\ (fF)\end{tabular}}       & $L_{\rm factor}$       \\ \hline

\multicolumn{1}{|c|}{\begin{tabular}[c]{@{}c@{}}1.73 \\ $\sim$ \\ 1.77\end{tabular}} & \multicolumn{1}{c|}{\begin{tabular}[c]{@{}c@{}}3.30 \\ $\sim$ \\ 3.70\end{tabular}} & \multicolumn{1}{c|}{\begin{tabular}[c]{@{}c@{}}1.10 \\ $\sim$ \\ 1.30\end{tabular}} & \multicolumn{1}{c|}{\begin{tabular}[c]{@{}c@{}}5.50 \\ $\sim$ \\ 8.00\end{tabular}} & \multicolumn{1}{c|}{\begin{tabular}[c]{@{}c@{}}1050\\  $\sim$ \\ 1300\end{tabular}} & \multicolumn{1}{c|}{\begin{tabular}[c]{@{}c@{}}0.00\\  $\sim$\\  0.04\end{tabular}} & \multicolumn{1}{c|}{\begin{tabular}[c]{@{}c@{}}1.26 \\ $\sim$ \\ 2.09\end{tabular}} & \multicolumn{1}{c|}{\begin{tabular}[c]{@{}c@{}}0.880 \\ $\sim$ \\ 1.13\end{tabular}} & \multicolumn{1}{c|}{\begin{tabular}[c]{@{}c@{}}0.83 \\ $\sim$ \\ 1.1\end{tabular}} \\ \hline

\hline
\hline
\multicolumn{9}{|c|}{(c) Light fluxon transition ($\ket{0_{ES}} \rightarrow \ket{0_{EA}}$ transition at 0~m$\Phi_0$)}                                                                        \\ \hline
% \multicolumn{1}{|c|}{}                                                                & \multicolumn{1}{c|}{\begin{tabular}[c]{@{}c@{}}$E_J$ \\ (K)\end{tabular}}           & \multicolumn{1}{c|}{\begin{tabular}[c]{@{}c@{}}$E_C$\\  (K)\end{tabular}}         & \multicolumn{1}{c|}{\begin{tabular}[c]{@{}c@{}}$E_L$\\  (K)\end{tabular}}           & \multicolumn{1}{c|}{\begin{tabular}[c]{@{}c@{}}$E_{CL}$ \\ (K)\end{tabular}}  & \multicolumn{1}{c|}{\begin{tabular}[c]{@{}c@{}}$C_{\rm sh}$ \\ (fF)\end{tabular}}       & \multicolumn{1}{c|}{$\alpha$}                                                        & \multicolumn{1}{c|}{\begin{tabular}[c]{@{}c@{}}$C_{\rm isl}$\\ (fF)\end{tabular}} & \multicolumn{1}{c|}{\begin{tabular}[c]{@{}c@{}}$C_{\rm int}$\\ (fF)\end{tabular}}       & $L_{\rm factor}$        \\ \hline

\multicolumn{1}{|c|}{\begin{tabular}[c]{@{}c@{}}1.73 \\ $\sim$ \\ 1.77\end{tabular}} & \multicolumn{1}{c|}{\begin{tabular}[c]{@{}c@{}}3.45 \\ $\sim$ \\ 3.62\end{tabular}} & \multicolumn{1}{c|}{\begin{tabular}[c]{@{}c@{}}1.15 \\ $\sim$ \\ 1.22\end{tabular}} & \multicolumn{1}{c|}{\begin{tabular}[c]{@{}c@{}}5.40 \\ $\sim$ \\ 7.80\end{tabular}} & \multicolumn{1}{c|}{\begin{tabular}[c]{@{}c@{}}1000\\  $\sim$ \\ 1500\end{tabular}} & \multicolumn{1}{c|}{\begin{tabular}[c]{@{}c@{}}0.00\\  $\sim$\\  0.05\end{tabular}} & \multicolumn{1}{c|}{\begin{tabular}[c]{@{}c@{}}1.48 \\ $\sim$ \\ 1.66\end{tabular}} & \multicolumn{1}{c|}{\begin{tabular}[c]{@{}c@{}}0.970 \\ $\sim$ \\ 1.06\end{tabular}} & \multicolumn{1}{c|}{\begin{tabular}[c]{@{}c@{}}0.71 \\ $\sim$ \\ 1.1\end{tabular}} \\ \hline

\hline
\hline
\multicolumn{9}{|c|}{\begin{tabular}[c]
{@{}c@{}} (d) Anticrossing between \\
the $\ket{1_{ES}} \rightarrow \ket{4_{ES}}$ transition and the $\ket{1_{ES}} \rightarrow \ket{2_{OS}}$ transition at 12~m$\Phi_0$ \end{tabular}}                                                                        \\ \hline

% \multicolumn{1}{|c|}{}                                                                & \multicolumn{1}{c|}{\begin{tabular}[c]{@{}c@{}}$E_J$ \\ (K)\end{tabular}}           & \multicolumn{1}{c|}{\begin{tabular}[c]{@{}c@{}}$E_C$\\  (K)\end{tabular}}         & \multicolumn{1}{c|}{\begin{tabular}[c]{@{}c@{}}$E_L$\\  (K)\end{tabular}}           & \multicolumn{1}{c|}{\begin{tabular}[c]{@{}c@{}}$E_{CL}$ \\ (K)\end{tabular}}  & \multicolumn{1}{c|}{\begin{tabular}[c]{@{}c@{}}$C_{\rm sh}$ \\ (fF)\end{tabular}}       & \multicolumn{1}{c|}{$\alpha$}                                                        & \multicolumn{1}{c|}{\begin{tabular}[c]{@{}c@{}}$C_{\rm isl}$\\ (fF)\end{tabular}} & \multicolumn{1}{c|}{\begin{tabular}[c]{@{}c@{}}$C_{\rm int}$\\ (fF)\end{tabular}}       & $L_{\rm factor}$                                                   \\ \hline

\multicolumn{1}{|c|}{\begin{tabular}[c]{@{}c@{}}1.71 \\ $\sim$ \\ 1.76\end{tabular}} & \multicolumn{1}{c|}{\begin{tabular}[c]{@{}c@{}}3.50 \\ $\sim$ \\ 3.85\end{tabular}} & \multicolumn{1}{c|}{\begin{tabular}[c]{@{}c@{}}1.14 \\ $\sim$ \\ 1.38\end{tabular}} & \multicolumn{1}{c|}{\begin{tabular}[c]{@{}c@{}}5.80 \\ $\sim$ \\ 12.5\end{tabular}} & \multicolumn{1}{c|}{\begin{tabular}[c]{@{}c@{}}1000\\  $\sim$ \\ 1500\end{tabular}} & \multicolumn{1}{c|}{\begin{tabular}[c]{@{}c@{}}0.02\\  $\sim$\\  0.06\end{tabular}} & \multicolumn{1}{c|}{\begin{tabular}[c]{@{}c@{}}0.980 \\ $\sim$ \\ 1.67\end{tabular}} & \multicolumn{1}{c|}{\begin{tabular}[c]{@{}c@{}}0.880 \\ $\sim$ \\ 1.01\end{tabular}} & \multicolumn{1}{c|}{\begin{tabular}[c]{@{}c@{}}0.91 \\ $\sim$ \\ 1.2\end{tabular}} \\ \hline

\hline
\hline
\multicolumn{9}{|c|}{(e) Intersection of fitted errors estimated from four types of transitions}                                                                        \\ \hline

% \multicolumn{1}{|c|}{\begin{tabular}[c]{@{}c@{}}$E_J$ \\ (K)\end{tabular}}           & \multicolumn{1}{c|}{\begin{tabular}[c]{@{}c@{}}$E_C$\\  (K)\end{tabular}}         & \multicolumn{1}{c|}{\begin{tabular}[c]{@{}c@{}}$E_L$\\  (K)\end{tabular}}           & \multicolumn{1}{c|}{\begin{tabular}[c]{@{}c@{}}$E_{CL}$ \\ (K)\end{tabular}}  & \multicolumn{1}{c|}{\begin{tabular}[c]{@{}c@{}}$C_{\rm sh}$ \\ (fF)\end{tabular}}       & \multicolumn{1}{c|}{$\alpha$}                                                        & \multicolumn{1}{c|}{\begin{tabular}[c]{@{}c@{}}$C_{\rm isl}$\\ (fF)\end{tabular}} & \multicolumn{1}{c|}{\begin{tabular}[c]{@{}c@{}}$C_{\rm int}$\\ (fF)\end{tabular}}       & $L_{\rm factor}$                                                   \\ \hline

\multicolumn{1}{|c|}{\begin{tabular}[c]{@{}c@{}}1.73 \\ $\sim$ \\ 1.76\end{tabular}} & \multicolumn{1}{c|}{\begin{tabular}[c]{@{}c@{}}3.50 \\ $\sim$ \\ 3.62\end{tabular}} & \multicolumn{1}{c|}{\begin{tabular}[c]{@{}c@{}}1.15 \\ $\sim$ \\ 1.22\end{tabular}} & \multicolumn{1}{c|}{\begin{tabular}[c]{@{}c@{}}5.80 \\ $\sim$ \\ 7.80\end{tabular}} & \multicolumn{1}{c|}{\begin{tabular}[c]{@{}c@{}}1100\\  $\sim$ \\ 1300\end{tabular}} & \multicolumn{1}{c|}{\begin{tabular}[c]{@{}c@{}}0.02\\  $\sim$\\  0.04\end{tabular}} & \multicolumn{1}{c|}{\begin{tabular}[c]{@{}c@{}}1.48 \\ $\sim$ \\ 1.66\end{tabular}} & \multicolumn{1}{c|}{\begin{tabular}[c]{@{}c@{}}0.970 \\ $\sim$ \\ 1.01\end{tabular}} & \multicolumn{1}{c|}{\begin{tabular}[c]{@{}c@{}}0.91 \\ $\sim$ \\ 1.1\end{tabular}} \\ \hline

\end{tabular}
\caption{Fit parameter error estimation based on (a) plasmon transition, (b) heavy fluxon transition, (c) light fluxon transition, and (d) anticrossing at flux points indicated in Fig.~\ref{fig:Fitting-Plaq12-error-estimation}. (e) Intersection of fitted errors estimated from four types of transitions in (a-d). 
%{\bf It would be better to arrange the formatting so that this table and the next four are grouped together, with just one caption underneath the set of 5 tables.}{\textcolor{orange}{Done.}}
}
\label{tab:Fitting-error-estimation}
\end{table}

\subsection{Fit parameter error estimation}
\label{sec:Fitting-error-estimation}

In general, each type of transition has a different sensitivity to the various fitting parameters for the device. In order to estimate the uncertainty in each fitted parameter, we compute the energy level spectrum while varying each parameter one at a time and keeping the other parameters at their best-fit values. We thus find the range over which each parameter can be varied while keeping the transition frequencies within 10\% of the measured values. In Fig.~\ref{fig:Fitting-Plaq12-error-estimation}, we show the four transitions we choose to estimate the fitted errors. In Table~\ref{tab:Fitting-error-estimation}(a), we list the fitted errors obtained with this method for the $\ket{0_{ES}} \rightarrow \ket{3_{ES}}$ plasmon transition at 20~m$\Phi_0$. From this table, we see the plasmon transition is sensitive to $E_J$, $E_C$, $E_L$, $C_{\rm sh}$ and $L_{\rm factor}$, so we are confident of the parameters extracted from fitting the plasmon transitions. In Table~\ref{tab:Fitting-error-estimation}(b), we list the fitted errors for the $\ket{0_{ES}} \rightarrow \ket{2_{OS}}$ heavy fluxon transition at 6.5~m$\Phi_0$. The heavy fluxon transition is especially sensitive to $E_J$, $E_C$, while moderately sensitive to the rest of the parameters, so by fitting to the heavy fluxon transition, we have high confidence in the $E_J$, $E_C$ values, with moderate confidence in the rest of the parameters. In Table~\ref{tab:Fitting-error-estimation}(c), we list the fitted errors for the $\ket{0_{ES}} \rightarrow \ket{0_{EA}}$ light fluxon transition at 0~m$\Phi_0$. The light fluxon transition is quite sensitive to $E_J$, $E_C$, $E_L$, $E_{CL}$, $C_{\rm isl}$ and $C_{\rm int}$, so these fitted parameters extracted from fitting the light fluxon transition have small errors. In Table~\ref{tab:Fitting-error-estimation}(d), we list the fitted errors for the anticrossing between the $\ket{1_{ES}} \rightarrow \ket{4_{ES}}$ transition and the $\ket{1_{ES}} \rightarrow \ket{2_{OS}}$ transition at 12~m$\Phi_0$. The anticrossing is highly sensitive to $E_J$, $E_C$, $E_L$, so the errors for these fitted parameters by fitting the anticrossing are quite small. Our fitting method fits all four types of transitions, so we are confident that the parameters of the actual chip are within the intersection of the estimated errors extracted from these four types of transitions, as shown in Table~\ref{tab:Fitting-error-estimation}(e).

\section{Prospects for implementing protected qubits}
\label{sec:optimal}

Our experiment presented in the main paper demonstrates successful concatenation of 
$\pi$-periodic plaquettes through measurements of large $\Delta_{\rm SA}^{(ij)}$ and offset-charge tuning, characteristic of strong hybridization of frustrated plaquettes. In order to implement a protected qubit based on this approach, we need a device that robustly maintains nearly degenerate computational states while pushing the remaining fluxon and plasmon energy levels significantly upwards. 
%\sout{large $\Delta_{SA}^{(ij)}$ while also having higher energy scales for plasmon excitations well above the ground-state doublet. The ideal device will have (1) large $\Delta_{SA}^{(ij)}$, (2) large plasmon transition frequencies, and (3) vanishingly small ground-state doublet gap $\Delta_{\rm EO}$ between the qubit logical states. The competing effects on these three constraints from varying the device parameters ($E_J$, $E_C$, $E_L$, $C_{\rm sh}$, $C_{isl1}$, $C_{isl2}$, and $\alpha$) are summarized in Table... {\bf (Andrey will move his discussion of contraints and effects of parameter variations, plus table, here.)}.} 
This means that for qubit design, we want to gain control over the three major energy scales: 
(1) Heavy fluxon gap $\varDelta_{\rm EO}$, that is also the computational gap that protects from $T_1$-processes and charge noise, needs to be kept small; 
(2) Light fluxon gap $\varDelta_{\rm SA}$, that determines effectiveness of concatenation and protects from flux noise and $T_2$-processes, needs to be kept large; 
(3) Plasmon gap $\omega_\text{pl}$ that determines energies of unwanted low-lying plasmon states that deteriorate initialization fidelity and facilitate thermal excitations, needs to be kept high. 
%On the other hand, we have 
For determining these energy scales, there are three primary device parameters that can be adjusted in the design: 
%three major control parameters:
Shunt capacitance $C_{\rm sh}$;
Effective intermediate island charging energy $E_C^{\rm 
isl}$, that is determined by both the geometric ground capacitance of the intermediate island and the effective screened capacitances of the Josephson junctions that also indirectly couple the intermediate island to ground;
$E_2$ of an individual plaquette, that is mainly determined by a combination of $E_L, E_J, E_{C}$.
The table below demonstrates how the three energy gaps change as the described device parameters increase. The entries of the table describe whether or not the gaps change in the desired direction, and whether they change exponentially (exp) or (sub)polynomially (poly). Each entry is also accompanied by a brief explanation. 
\begin{center}
\begin{tabular}{ | l || l | l | l |}
\hline
  & as $C_{\rm sh}$ increases
 & as $E_C^{\rm isl}$ increases
 & as $E_2$ increases\\
 
\hline
\hline
 $\varDelta_{\rm EO}$
 &
 {GOOD; exp} 
 & 
 {BAD; weak poly} 
 & 
 {GOOD, exp} \\ 
 
  lower is better
  & 
  interwell tunneling $\downarrow$ as mass $\uparrow$ 
  &  
  $E_2$ of chain slightly decreases 
  & 
  interwell tunneling $\downarrow$ as barrier $\uparrow$  \\

  &&because hybridization washes out&\\
  &&the top of $\cos2\phi$ potential&\\
 \hline
 $\varDelta_{\rm SA}$
 &
 {N/A}
 & 
 GOOD; poly
 & 
 BAD; poly  \\ 

  higher is better
  &
  no strong depenence as
  & 
  lighter fluxons hybridize better
  & 
  hybridization $\downarrow$ as barrier $\uparrow$\\

&ideally, light fluxon modes&&\\
&are decoupled from $C_{\rm sh}$&&\\
 \hline
 $\omega_\text{pl}$
 &
 BAD, poly
 & 
 BAD, weak poly
 & 
 GOOD, poly \\ 

 higher is better
 &
 $\propto\sqrt{E_2/C_{\rm sh}}$
 &
 $E_2$ of chain slightly decreases
 &
 $\propto\sqrt{E_2/C_{\rm sh}}$
 \\
 \hline
\end{tabular}
\end{center}

Note that each of these device parameters improves some 
%qubit parameters 
energy scales, while deteriorating others. Thus, the goal of optimization is not to just maximize some circuit parameters, but to compensate the negative effects by tuning other parameters. For example, increasing $E_C^{\rm isl}$ significantly improves concatenation (light fluxon modes), but negatively, albeit weakly, affects the heavy fluxon and plasmon modes. Therefore, it is always beneficial to increase $E_C^{\rm isl}$, while keeping in mind the presence of mild adverse effects. 
On the other hand, increasing $C_{\rm sh}$ and $E_2$ is exponentially beneficial for the computational gap, but (sub)polynomially adversely affects concatenation (light fluxon) and the plasmon modes. It is encouraging that adverse effects of increasing $C_{\rm sh}$ and $E_2$ are smaller than the benefits. However, these adverse effects are not negligible and need to be treated very carefully. In addition to these device parameters that can be controlled through design, there are also uncontrolled ones, such as the flux noise amplitude and junction asymmetry $\alpha$. However, improved concatenation protects the device against these types of asymmetries. This means that decreasing $\alpha$ would also increase $T_2$ polynomially -- as a power of the number of plaquettes minus one.
%\sout{Reducing $C_{isl1}$ and $C_{isl2}$ boosts $\Delta_{SA}^{(ij)}$ with no significant impact on the plasmons or $\Delta_{\rm EO}$. Minimizing $\alpha$ helps directly with constraint (3). {\bf (Should we say more here about the effects of $\alpha$?)} 
%\textcolor{orange}{(Minizing $\alpha$ does not directly helps with larger $\Delta_{SA}^{(ij)}$ and large plasmon transition frequency. But if we don't have small $\alpha$, we need to compensate the effect of large $\alpha$ by requiring larger $\Delta_{SA}^{(ij)}$ and larger $C_{shunt}$)}. 
%Increasing $E_J$ and $E_L$ raises the plasmon transitions and reduces $\Delta_{\rm EO}$, but it also reduces $\Delta_{SA}^{(ij)}$. Reducing $C_{\rm sh}$ results in higher plasmon transitions ($\propto C_{sh}^{-1/2}$), but it also increases $\Delta_{\rm EO}$ exponentially. }
%\textcolor{orange}{(Reducing $C_{\rm sh}$ increases plasmon transitions with $\sim \sqrt{C_{sh}}$, but also increases $\Delta_{\rm EO}$ exponentially.)} 

\begin{figure}[htp]
\centering
\includegraphics[width=\textwidth]{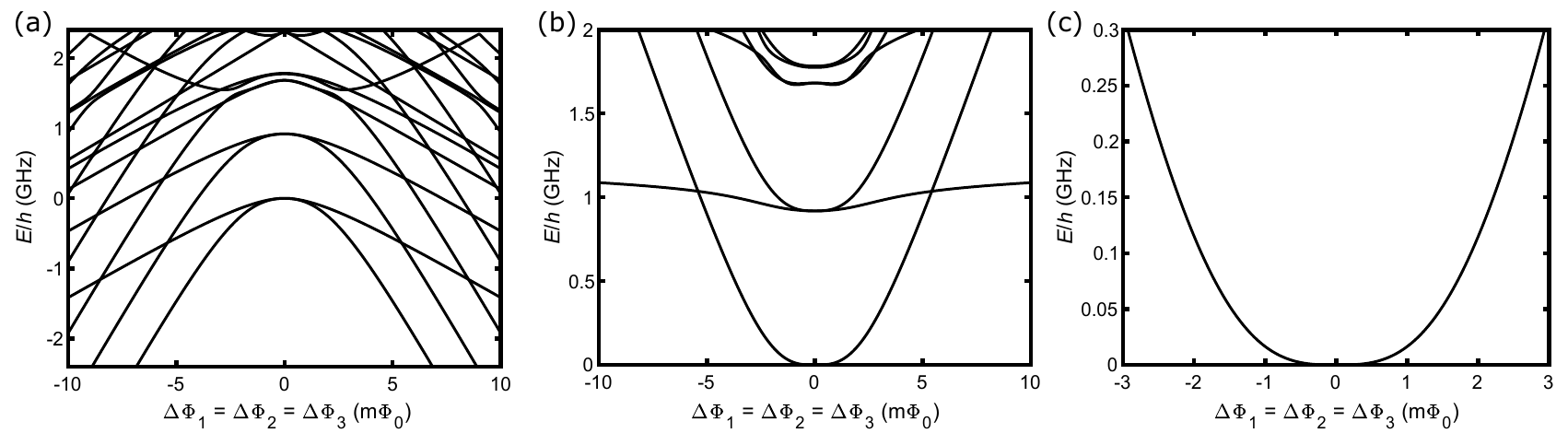}
  \caption{(a) Modeled energy levels near triple frustration with improved parameters, as described in text. (b) Modeled transitions near triple frustration. (c) Zoomed-in transition plot between 0 to 0.3~GHz.}
  \label{fig:Future-optimal-parameters}
\end{figure}

One set of parameters that provides a compromise between the competing circuit parameters needed for achieving a qubit with excellent coherence is the following: $E_J=3$~K, $E_L=2$~K, $E_C=5$~K, $C_{\rm sh}=1000$~fF, $C_{\rm isl1}=1$~fF, $C_{\rm isl2}=3$~fF, $\alpha=0.01$. Unfortunately, 
%as discussed in the main paper, 
achieving these values of $E_J$ and $E_C$ with conventional Al electrodes is not possible due to the small superconducting gap and the electronic capacitance arising when the junction plasma frequency approaches the gap \cite{SEckern1984}. Thus, protected qubits incorporating this stabilizer mechanism will need junctions fabricated from a larger gap superconductor.
%requires an enhancement of the superconducting gap of the junction electrodes to reduce the effects of the electronic capacitance. 
Reducing $\alpha$ from $\sim$0.03 in our experimental device to 0.01 will be challenging, but should be achievable with improved control over the electron-beam lithography and thin-film growth for forming the junctions. Reducing the intermediate island capacitances from our experimental device in the main paper to the values listed above requires minimizing the size of the intermediate islands as well as the length of the junction chain inductors, but the levels listed here are feasible. An alternate approach for satisfying the competing constraints involving parallel chains of concatenated plaquettes will be discussed at the end of this section. 
%{\bf So, we'll have the parallel chain discussion at the end of this section}.

In addition to optimizing the device parameters as described above, the device layout must have weaker radiative coupling to parasitic high-frequency modes to avoid spurious excitations and fast quasiparticle poisoning of the intermediate islands. Achieving this in practice is challenging because of the large $C_{\rm sh}$. On our present device, the planar capacitor layout for $C_{\rm sh}$ required an exceedingly large physical footprint, resulting in significant low-frequency parasitic modes that couple strongly to the qubit environment. For a future protected qubit implementation, switching to a parallel-plate capacitor for $C_{\rm sh}$ would allow for a much more compact structure, thus pushing parasitic antenna modes to much higher frequencies. For a conventional superconducting qubit, such as a transmon, the microwave losses for the deposited amorphous dielectric layer in such a capacitor would result in quite poor $T_1$ performance. However, for a qubit formed from concatenated $\pi$-periodic plaquettes with a large $C_{\rm sh}$, the transition matrix elements between the logical qubit states are quite small. When combined with the vanishingly small value for $\Delta_{\rm EO}$, this results is an extremely long $T_1$, even with the large loss tangent for the deposited dielectric of $C_{\rm sh}$, as will be discussed below in our treatment of the projected coherence times.

In order to estimate the coherence for a protected qubit based on this design with the parameters described earlier in this section, we first compute the level spectrum for such a device at triple frustration (Fig.~\ref{fig:Future-optimal-parameters}). With these $E_J$, $E_L$, and $C_{\rm sh}$ values, the separation between the logical qubit states is quite flat with respect to flux, with a minimum separation ($\Delta_{\rm EO}$) of $15$~kHz. The lowest plasmon transition is also pushed up to nearly 1~GHz above the ground-state doublet. At plaquette (12) and (23) double frustration, the symmetric-antisymmetric gaps are $\Delta_{\rm SA}^{(12)}=1.5$~GHz and $\Delta_{\rm SA}^{(23)}=0.9$~GHz, and so are comparable with the lowest plasmon transition. 
%{\bf (It would be nice to include these $\Delta_{SA}$ values here to keep with the logical flow in this section, but are these values something we really want to highlight? i.e., how do they compare with our experimental device?)}
%: \textcolor{red}{($E_J$, $E_C$, $E_L$, $C_{\rm sh}$, $C_{isl1}$, $C_{isl2}$ and $\alpha) = ($ 3~K, 5~K, 2~K, 1000~fF, 1~fF, 3~fF and 0.01) !!! I added brackets --otherwise it looks like $\alpha$=3K !!!}. 

Upon computing the level spectrum near triple frustration for this set of parameters, we can estimate the qubit coherence. For obtaining $T_1$, we first calculate the transition matrix element for the charge operator $\hat{N}$ between the even- and odd-parity ground states ($\psi_{\rm even},\psi_{\rm odd}$, respectively) at triple frustration for the above parameters: $\abs{\langle \psi_{\rm odd}\mid\hat{N}\mid \psi_{\rm even}\rangle} \sim 1 \times 10^{-5}$. 
%$\abs{\matrixel{O}{\hat{N}}{E}} \sim 2 \times 10^{-5}$.
$T_1$ can then be expressed as 
\begin{equation}
1/T_1 = \abs{\langle \psi_{\rm odd}\mid\hat{N}\mid\psi_{\rm even}\rangle}^2\Delta_{\rm EO}\tan\delta.    
\end{equation}
Assuming we fabricate a parallel-plate shunt capacitor using electron-beam evaporated SiO$_2$, which is compatible with our current device fabrication process, the corresponding loss tangent is $\sim$1/300 \cite{SO'Connell2008Mar}, resulting in $T_1 \sim 1 \times 10^8$~s. In general, for our architecture, the $T_1$ budget is quite high because it is determined by exponentially weak and easily suppressable interwell tunneling. The much bigger challenge is creating flat bands to enhance $T_2$.

For computing $T_2$ for the device described above, we consider the flux dispersion of the energy bands from our numerical modeling. Exactly at triple frustration, the slope with respect to flux bias vanishes and the curvature is $\sim 12$~MHz/${\rm m}\Phi_0^2$. Assuming the electronics used for supplying the flux-bias currents is capable of sufficiently fine steps, the flux resolution of our biasing will ultimately be limited by the rms flux noise level $\Phi_{\rm noise}$. We calculate the $1/f$-noise-limited $T_2$ with the formula:
\begin{equation}
    \frac{1}{T_2^\Phi} \approx \frac{A_\Phi}{h} \times  \left.\frac{\partial E}{\partial \Phi}\right|_{\Phi = \Phi_{\rm noise}} 
\end{equation}
where the flux noise is assumed to follow the power spectrum $A_{\Phi}^2/f$, we consider a Ramsey pulse performed immediately after calibration. Taking a flux-noise power spectral density of $2\,\mu\Phi_0/\sqrt{\rm{Hz}}$ at 1~Hz and integrating over 10 decades yields an rms flux noise level of 10~$\mu\Phi_0$. 
%\sout{Thus, we must also consider the flux dispersion of the energy bands for an offset from exact triple frustration by this amount.} 
%\textcolor{teal}{(actually, at this flux, the $T_2^\Phi$ is still dominate by the slope.)} 
Accounting for these various factors, we compute a flux-noise limited dephasing time $T_2^{\Phi}$ for single frustration, plaquette (12) double frustration, plaquettte (23) double frustration, and triple frustration: $1.7~\mu$s, $340~\mu$s, $190~\mu$s, $6.0 \times 10^4$~$\mu$s, respectively.
%$T_2^{\Phi}\sim 5.7 \times 10^4$~ms.
%{\bf (Yebin \& Andrey: it would be good if you could also discuss the various $T_2^{\Phi}$ values for different degrees of frustration that are used in the table for the various $\Lambda$ values in the next section}
%{\textcolor{orange}{When a device with the parameters listed above is biased at triple frustration, the flux dispersion slope is 0 and curvature at is $\sim 1 \times 10^{-3}$~GHz/m$\Phi_0^2$. Our flux bias precision limited by our equipment is $\sim$~200$\,\mu \Phi_0$, and the corresponding flux dispersion slope is $\sim 8 \times 10^{-4}$~GHz/m$\Phi_0$. Assuming we can bias the plaquettes within $100\,\mu\Phi_0$ of triple frustration and assuming a $1/f$ flux-noise amplitude of $1\,\mu\Phi_0/\sqrt{\rm{Hz}}$ at 1~Hz, the flux-noise limited $T_2$: $T_2^\Phi$, predicted by our theory collaborators is $\sim$2~ms.}}

We also consider dephasing due to charge noise at triple frustration by computing the level of fluctuations in the splitting between the computational levels. We estimate the lower limit of $T_2^Q$ with
\begin{equation}
    \frac{1}{T_2^Q} \approx A_Q \times \frac{\Delta_{\rm EO}}{e},
\end{equation}
where we assume a charge noise power spectrum $A_Q^2/f$ and we again, like in the case of flux noise, 
%$\Delta_Q$ is the charge modulation amplitude, and $A_Q$ is 
%the charge-noise spectral density and 
assume that the Ramsey pulse is performed immediately after calibration. 
For a charge-noise spectral density of $2\times 10^{-2} e/\sqrt{{\rm Hz}}$ at 1~Hz \cite{SAstafiev2006,SChristensen2019}, we estimate a charge-noise limited dephasing time, $T_2^Q>10^7$~$\mu$s. 
%{\bf (Yebin \& Andrey: it would be good if you could add a bit more detail about the calculation here.)} 
This timescale decreases for larger $\Delta_{\rm EO}$, for example, with a smaller $C_{\rm sh}$. Thus, reducing $C_{\rm sh}$ in an attempt to increase $\omega_{\rm pl}$, will eventually cause dephasing due to charge noise to become non-negligible as $\Delta_{\rm EO}$ increases. 
%{\textcolor{orange}{ At triple frustration, assuming a 1/f charge-noise amplitude of 20 m$e/\sqrt{\rm{Hz}}$, the charge-noise limited $T_2$: $T_2^Q, is \sim8~ms$. {\bf (I was using the default charge noise amplitude in Andrey's code)}. Meanwhile at triple frustration, the gap between the ground state double and the first excited state doublet, $\Delta_{plasmon}$, is $\sim 0.9$~GHz. It is larger than the current device and reduces the thermal population. While reducing the shunt capacitance can further increase the $\Delta_{plasmon}$, it also increases $\Delta_{\rm EO}$ and thus reducing $T_2^Q$.}

%(Notes:Need to increase $E_J$, $E_L$, while maintaining large $E_c$ and $E_c^{island}$ and keeping a large $C_{\rm sh}$ to suppress tunneling, to push all excited states to at least several GHz while maintaining large asymmetric/antisymmetric gap and low curvature of computational energy levels. Challenge of electronic capacitance. Also, need to reduce $C_{\rm isl}$ for both islands. Maybe we should have one set of ideal parameters and our existing $C_{\rm isl}$ values, with calculated $T_1$ and $T_2$ at TF, plus another set of ideal parameters but with reduced island capacitance, maybe $C_{\rm isl}^{(1)}\sim 1$fF and $C_{\rm isl}^{(2)}\sim 3$fF?

\begin{figure}[htp]
\centering
\includegraphics[width=3.5in]{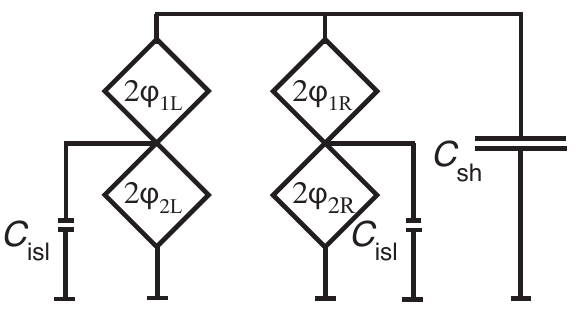}
  \caption{Schematic of qubit with two parallel chains of concatenated plaquettes.
  \label{fig:parallel-chains}}
\end{figure}

% For computing $\Lambda$ for this system, we follow two approaches: calculating the ratio between $\Delta_{SA}^{(ij)}$ with the single-plaquette flux-noise error rate, $h$, and computing the ratio of $T_2^{\Phi}$ for double or triple frustration with that at single frustration. {\bf Need to explain subtle differences between two approaches, first doesn't account for $\alpha$, second one does.} We estimate $h$ by using the rms flux-noise level described above (5$\mu\Phi_0$) and the slope of the single-plaquette flux dispersion near single frustration (XX~MHz)... {\bf (Let's check this calculation, it seems like this is more like 1.5~MHz, rather than 10~MHz, unless I made a mistake somewhere...; we should add a table of the calculated $T_2^{flux}$ for SF, DF, TF for the optimal parameters and the corresponding $\Lambda$ values. Let's use the notation $\Lambda_{\rm DS}^{(21)}$ to represent $\Lambda$ for (21) double frustration relative to single frustration.)}
In addition to the optimization of our multi-plaquette chain discussed above, there are alternative device geometries that can mitigate the adverse effects of decreasing $C_{\rm sh}$. For example, 
connecting multiple plaquette chains in parallel, as shown in Fig.~\ref{fig:parallel-chains}, results in a linear increase in the overall system $E_2$, thus raising the plasmon modes without the need for increasing $E_2$ of the individual plaquettes, which is helpful for maintaining strong concatenation. 
%
%\sout{Besides the approach for optimizing the parameters in a 3-plaquette qubit outlined above, another strategy involves attaching two multi-plaquette chains in parallel, as shown in Fig.~\ref{fig:parallel-chains}. In this case, even with relatively small $E_2$ of each plaquette to allow for strong hybridization and large $\Delta_{SA}^{(ij)}$, the connection of the plaquette chains in parallel boosts the effective $E_2$, $E_2^{eff}$ to the logical island.} 
This allows for the use of a smaller $C_{\rm sh}$ for raising the plasmon transition frequencies without sacrificing an increase in $\Delta_{\rm EO}$. Of course, such an approach with two 3-plaquette chains will require twice as many flux bias controls compared to a single chain, plus one more flux bias for the loop formed by the two chains. 
%{\bf (Andrey: does this capture everything we want to say about parallel chains here?)}

\section{$\Lambda$ calculation}
As described in the main paper, $\Lambda$ characterizes the rate at which the logical error decreases with system size. $\Lambda$ can be expressed either as the ratio of $T_2$ between the larger and smaller system sizes, or, in the specific case of a qubit based on concatenated plaquettes, can also be determined by $\Delta_{\rm SA}$ and $h_Z$, the scale of dephasing fluctuations for a single plaquette. In this section, we show that these two approaches are essentially equivalent. In addition, we compute $\Lambda$ for different degrees of frustration for a qubit with the optimal parameters described in the previous section. 
For a given level of flux noise, $T_2^{\Phi}$ will be inversely proportional to the slope of the qubit transition energy with respect to flux. Thus, computing $\Lambda$ from the ratio of the $T_2$ values will be equivalent to the inverse ratio of the energy band slopes for the two different degrees of frustration. 
%We can estimate $\Lambda$ by calculating the $T_2$. For the flux offset $\Phi_{noise}$ that is limited by the $1/f$ flux noise, the $T_2^\Phi$ is limited by the slope of flux dispersion. So we can also estimate $\Lambda$ by calculating the slope at different frustration.
% We denote $k$ as the 0-1 transition slope at single frustration. The 0-1 transition slope at double frustration is equivalent to the slope of the lower branch of an anticrossing, which is $k^2\Phi_{noise}/\sqrt{(k\Phi_{noise})^2+(\Delta_{SA}/2)^2}$. For the optimal parameter, $k$ is $\sim 100$~MHz/$m\Phi_0$, the RMS $1/f$ flux noise $\Phi_{noise}$ is 10~$\mu \Phi_0$, $\Delta_{SA}$ is $\sim$1 GHz, so $\Delta_{SA}^2 \gg (k\Phi_{noise})^2$. The transition slope at double frustration $\sim 2 k^2 \Phi_{noise}/\Delta_{SA}$. We compute the $\Lambda_{\rm DS}$ by the ratio of the slope at single and double frustration: $\Lambda_{\rm DS} = \Delta_{SA}^{(ij)}/2k\Phi_{noise}$, as shown in the last column in Table.~\ref{tab:Lambda cal}. We also compute the $\Lambda$ by comparing the $T_2^\Phi$ between different frustrations. The $\Lambda_{\rm DS}$ compute in two ways matches.}
%\subsection{$\Lambda_{\rm DS}$ estimation from $\Delta_{SA}$}

%For understanding the stabilizer interaction 
For the semi-quantitative description of the system at double frustration presented in the main paper, we can treat each plaquette as a spin-1/2 particle with an $XX$ stabilizer term. The energy of each plaquette is $H_{i} = -(k \Delta\Phi_i/{2}) Z_{i}$, where $k$ is the slope of the transition energy with respect to flux, and $\Delta \Phi_i$ is the flux offset from frustration for plaquette $i$. In this model, the double-plaquette Hamiltonian is then: 
%{\color{orange} For direct products of matrices: it's either $Z_1, Z_2, X_1 X_2$ or $Z\otimes I, I\otimes Z, X\otimes X$. I think the first format is better -- indexes imply direct product}:
\begin{gather}
\begin{aligned}
H&=-\frac{k \Delta\Phi_1}{2} Z_1-\frac{k \Delta\Phi_2}{2} Z_2-\frac{\Delta_{\rm SA}}{2} X_1 X_2. \\
% &=\left[\begin{array}{cccc}
% \frac{-k\left(\Delta\Phi_1+\Delta\Phi_2\right)}{2} & & & -\frac{\Delta_{SA}}{2} \\
% & \frac{-k\left(\Delta\Phi_1-\Delta\Phi_2\right)}{2} & -\frac{\Delta_{SA}}{2} & \\
%  & -\frac{\Delta_{SA}}{2} & \frac{k\left(\Delta\Phi_1-\Delta\Phi_2\right)}{2} & \\
% -\frac{\Delta_{SA}}{2} & & &\frac{k\left(\Delta\Phi_1+\Delta\Phi_2\right)}{2}\right).
% \end{array}\right] 
\end{aligned}
\end{gather}

\noindent Here, we assume the two plaquettes have the same $k$. The eigenvalues are 
\begin{gather}
\begin{aligned}
&E_1 =-\sqrt{\left( \frac{k \Delta\Phi_1}{2}+ \frac{k \Delta\Phi_2}{2} \right) ^2+\left(\frac{\Delta_{\rm SA}}{2}\right)^2}, \quad &E_2 =\sqrt{\left( \frac{k \Delta\Phi_1}{2}+ \frac{k \Delta\Phi_2}{2} \right) ^2+\left(\frac{\Delta_{\rm SA}}{2}\right)^2},  \\ &E_3 =-\sqrt{\left( \frac{k \Delta\Phi_1}{2}- \frac{k \Delta\Phi_2}{2} \right) ^2+\left(\frac{\Delta_{\rm SA}}{2}\right)^2}, \quad  &E_4 =\sqrt{\left( \frac{k \Delta\Phi_1}{2}- \frac{k \Delta\Phi_2}{2} \right) ^2+\left(\frac{\Delta_{\rm SA}}{2}\right)^2}.
\end{aligned}
\end{gather}

The double-plaquette circuit typically has $k \sim 300 $~MHz$/{\rm m}\Phi_0$ and $\Delta_{\rm SA} \sim 1$~GHz. Near single frustration for one of the plaquettes, the other plaquette will have $\Delta\Phi$ near 0.5 $\Phi_0$, thus $(k \Delta\Phi_1/2+ k \Delta\Phi_2/2)^2 \gg (\Delta_{\rm SA}/2)^2$. In this limit, we obtain: 

\begin{gather}
\begin{aligned}
&E_1 =-|\frac{k \Delta\Phi_1}{2}+ \frac{k \Delta\Phi_2}{2}|, \quad &E_2 =|\frac{k \Delta\Phi_1}{2}+ \frac{k \Delta\Phi_2}{2}|, \quad 
&E_3 =-|\frac{k \Delta\Phi_1}{2}- \frac{k \Delta\Phi_2}{2}|, \quad 
&E_4 =|\frac{k \Delta\Phi_1}{2}+ \frac{k \Delta\Phi_2}{2}|. \quad 
\end{aligned}
\end{gather}
Thus, the transition energy between the lowest two energy levels is $\Delta E = k\Delta\Phi$ and the corresponding slope 
%at $\Delta\Phi_{noise}$ 
is 
\begin{equation}
%    \left|\dv{E}{\Delta\Phi}\right| =k
 \left|\frac{\partial\Delta E}{\partial\Delta\Phi}\right|_{\rm S} =k,
\end{equation}
where S indicates single frustration.

Near double frustration, the slope is highest in the $\Delta\Phi_1=\Delta\Phi_2$ or $\Delta\Phi_1=-\Delta\Phi_2$ directions. Taking $\Delta\Phi_1 = \Delta\Phi_2=\Delta\Phi$, the energy levels are
\begin{gather}
\begin{aligned}
E_1 &=-\sqrt{ (k \Delta\Phi)^2+\left(\frac{\Delta_{\rm SA}}{2}\right)^2}, \quad &E_2 &=\sqrt{ (k \Delta\Phi)^2+\left(\frac{\Delta_{\rm SA}}{2}\right)^2},   \\ E_3 &=-\frac{\Delta_{\rm SA}}{2}, \quad  &E_4 &=\frac{\Delta_{\rm SA}}{2}.
\end{aligned}
\end{gather}
The transition energy between the lowest two energy levels is $\Delta_{\rm SA}/2-\sqrt{ (k \Delta\Phi)^2+(\Delta_{\rm SA}/2)^2}$, and the corresponding slope with respect to $\Delta\Phi$ is $-k^2\Delta\Phi/ \sqrt{ (k \Delta\Phi)^2+(\Delta_{\rm SA}/2)^2}$. 
For a $1/f$ flux noise level of 2~$\mu\Phi_0/\sqrt{{\rm Hz}}$ at 1~Hz, the rms flux noise amplitude is $\Phi_{\rm noise}\sim$10~$\mu\Phi_0$, and we take this to be the minimum precision with which one can set the flux bias. 
%We can bias the plaquettes with precision of $\Delta\Phi_{noise}= 10~\mu\Phi_0$, with 
For typical double-plaquette parameters, $k \sim 300 $~MHz$/{\rm m}\Phi_0$ and $\Delta_{\rm SA} \sim 1$~GHz. In this limit, $\Delta_{\rm SA}/2 \gg|(k \Delta\Phi_1/2+ k \Delta\Phi_2/2)| $, and the slope of the transition energy with respect to flux at double frustration is
\begin{equation}
%    \left|\dv{E}{\Delta\Phi}\right| = \frac{2 k^2 \Delta \Phi_{noise}}{\Delta_{SA}}
        \left|\frac{\partial\Delta E}{\partial\Delta\Phi}\right|_{\rm D} = \frac{2 k^2 \Delta \Phi_{\rm noise}}{\Delta_{\rm SA}},
\end{equation} 
where D indicates double frustration. $\Lambda_{\rm DS}$, the decrease in the logical error rate upon going from single to double frustration, is then 
\begin{equation}
    \Lambda_{\rm DS} = \left|\frac{\partial\Delta E}{\partial\Delta\Phi}\right|_{\rm S}/\left|\frac{\partial\Delta E}{\partial\Delta\Phi}\right|_{\rm D} = k \times \frac{\Delta_{\rm SA}}{2 k^2 \Delta \Phi_{\rm noise}} = \frac{\Delta_{\rm SA}}{2 k \Delta\Phi_{\rm noise}}.
    \label{eq:lambdaDS}
\end{equation}
Thus, in this approximation, $\Lambda$ defined as the ratio of $T_2$ at double and single frustration is equal to the ratio of $\Delta_{\rm SA}$ to $2h_Z$, where the factor of 2 accounts for the two plaquettes. 
%{\bf (*Need to account for factor of 2...)} \textcolor{teal}{(The factor of 2 here is because the flux modulation is $\Delta\Phi^2$ dependence, and calculating the slope we got extra factor of 2.)}

%\subsection{$\Lambda_{\rm TD}$ and $\Lambda_{\rm TS}$ estimation from $\Delta_{SA}$}
For triple frustration, the Hamiltonian for the coupled spin-1/2 model of a 3-plaquette chain is given by
\begin{equation}
    H=-\frac{k \Delta\Phi_1}{2} Z_1-\frac{k \Delta\Phi_2}{2} Z_2 -\frac{k \Delta\Phi_3}{2}  Z_3-\frac{\Delta_{\rm SA}^{(12)}}{2} X_1 X_2 -\frac{\Delta_{\rm SA}^{(23)}}{2}  X_2 X_3,
\end{equation}
assuming all three plaquettes have the same single-frustration slope $k$. 
% Again, the largest slope is along $\Delta\Phi_1 = \Delta\Phi_2= \Delta\Phi_3=\Delta\Phi$ direction. The Hamiltonian can be rewritten as
% \begin{gather}
% \begin{aligned}
% 
% \left[\begin{array}{cccccccc}
% \frac{-3 k\Delta\Phi}{2} & & & -\frac{\Delta_{SA}}{2} & & & -\frac{\Delta_{SA}}{2} & \\
% & \frac{- k\Delta\Phi}{2} & -\frac{\Delta_{SA}}{2} & & & & & -\frac{\Delta_{SA}}{2}\\
%  & -\frac{\Delta_{SA}}{2} & \frac{- k\Delta\Phi}{2}  &  & -\frac{\Delta_{SA}}{2} & & & &\\
% -\frac{\Delta_{SA}}{2} & & &\frac{k\Delta\Phi}{2}  & & -\frac{\Delta_{SA}}{2} & & \\
% & & -\frac{\Delta_{SA}}{2} & &  \frac{- k\Delta\Phi}{2} & & & -\frac{\Delta_{SA}}{2} \\
% & & & -\frac{\Delta_{SA}}{2} & & \frac{k\Delta\Phi}{2} & -\frac{\Delta_{SA}}{2} & \\
% -\frac{\Delta_{SA}}{2} & & & &  & -\frac{\Delta_{SA}}{2} & \frac{ k\Delta\Phi}{2}  &  \\
%  & -\frac{\Delta_{SA}}{2} & & & -\frac{\Delta_{SA}}{2} & & &\frac{3 k\Delta\Phi}{2}\\
% \end{array}\right] 
% \end{aligned}
% \end{gather}
Finding the eigenvalues of this Hamiltonian and expanding them to third order in the limit of $\Delta_{\rm SA}^{(12)}/2 \gg|k \Delta\Phi_1/2+ k \Delta\Phi_2/2|$ and $\Delta_{\rm SA}^{(23)}/2 \gg|k \Delta\Phi_2/2+ k \Delta\Phi_3/2|$, we find the transition energy of the lowest two levels is $\Delta E = (k\Delta\Phi)^3/\Delta_{\rm SA}^{(12)}\Delta_{\rm SA}^{(23)}$. The corresponding slope with respect to flux 
%$\Delta\Phi_{noise}= 10~\mu\Phi_0$  
is 
\begin{equation}
   \abs{\frac{\partial\Delta E}{\partial\Delta\Phi}}_{\rm T} = \frac{3 k^3\Delta\Phi_{\rm noise}^2}{\Delta_{\rm SA}^{(12)}\Delta_{\rm SA}^{(23)}},
   \label{eq:lambdaTS}
\end{equation}
where $T$ indicates triple frustration.

We can then express $\Lambda_{\rm TD}$, the reduction in logical error rate upon going from double to triple frustration by taking the ratio of the expressions in Eq.~(\ref{eq:lambdaDS}) and Eq.~(\ref{eq:lambdaTS}). Thus, for plaquette (12) double frustration, we obtain
\begin{equation}
    \Lambda_{\rm TD}^{(12)} = \frac{2 \Delta_{\rm SA}^{(23)}}{3 k \Delta\Phi_{\rm noise}}, 
\label{eq:lambdaTD12}
\end{equation}
and for plaquette (23) double frustration, we obtain
\begin{equation}
    \Lambda_{\rm TD}^{(23)} = \frac{2 \Delta_{\rm SA}^{(12)}}{3 k \Delta\Phi_{\rm noise}}, 
\label{eq:lambdaTD23}
\end{equation}
Then, $\Lambda_{\rm TS}$, the reduction in logical error rate upon going from single to triple frustration, is given by
\begin{equation}
   \Lambda_{\rm TS} = \frac{\Delta_{\rm SA}^{(12)}\Delta_{\rm SA}^{(23)}}{3 k^2\Delta\Phi_{\rm noise}^2} = \frac{3}{4}\Lambda_{\rm TD}^{(12)}\Lambda_{\rm TD}^{(23)} .
   \label{eq:lambdaTS2}
\end{equation}

\begin{table}[]
\begin{tabular}{|c|c|c|c|c|c|c|c|c|c|c|c|}
\hline
                                                          & \begin{tabular}[c]{@{}c@{}}$\Delta_{\rm EO}$ \\ at 10 $\mu\Phi_0$\\ (MHz)\end{tabular} & \begin{tabular}[c]{@{}c@{}}$\Delta_{\rm SA}$\\ (GHz)\end{tabular} & \begin{tabular}[c]{@{}c@{}}Slope \\ at 10 $\mu\Phi_0$\\ (MHz/$\rm m\Phi_0$)\end{tabular} & \begin{tabular}[c]{@{}c@{}}Curvature \\ at 10 $\mu\Phi_0$ \\ (MHz/$\rm m\Phi_0^2$)\end{tabular} & $T_2^\Phi (\mu \rm s)$ & $\Lambda_{\rm DS}^{T_2}$ & $\Lambda_{\rm DS}^{\Delta_{\rm SA}}$ & $\Lambda_{\rm TD}^{T_2}$ & $\Lambda_{\rm TD}^{\Delta_{\rm SA}}$ & $\Lambda_{\rm TS}^{T_2}$ & $\Lambda_{\rm TS}^{\Delta_{\rm SA}}$ \\ \hline
\begin{tabular}[c]{@{}c@{}}plaquette\\ 2\end{tabular}     & $3.7 $                                                                             & -                                                             & 380                                                                                  & -                                                                                           & 1.7                & -                    & -                            & -                    & -                            & -                    & -                            \\ \hline
\begin{tabular}[c]{@{}c@{}}plaquette\\ (12)\end{tabular}  & $6.7$                                                                              & 1.5                                                           & 1.8                                                                                  & $1.8 \times 10^2$                                                                           & 340                & 200                  & 200                          & 180                  & 150                          & -                    & -                            \\ \hline
\begin{tabular}[c]{@{}c@{}}plaquette\\ (23)\end{tabular}  & $5.0$                                                                              & 0.87                                                          & 3.2                                                                                  & $3.0 \times 10^2$                                                                           & 190                & 110                  & 120                          & 320                  & 270                          & -                    & -                            \\ \hline
\begin{tabular}[c]{@{}c@{}}plaquette\\ (123)\end{tabular} & $1.5 \times 10^{-2}$                                                               & -                                                             & $1.0 \times 10^{-2}$                                                                 & $1.2 \times 10^1$                                                                           & $6.0 \times 10^4$  & -                    & -                            & -                    & -                            & $3.5 \times 10^4$    & $3.1 \times 10^4$            \\ \hline
\end{tabular}
\caption{$T_2^\Phi$ calculation from slope and curvature at 10~$\mu\Phi_0$ away from frustration for a protected qubit with the optimal parameters presented in Sec.~\ref{sec:optimal}. $\Lambda^{T_2}$ is calculated from the ratio of $T_2^\Phi$ values from the modeled energy levels, as described in Sec.~\ref{sec:optimal}; $\Lambda^{\Delta_{\rm SA}}$ is computed from $\Delta_{\rm SA}$ and $h_Z$ following Eqs.~(\ref{eq:lambdaDS},\ref{eq:lambdaTD12},\ref{eq:lambdaTD23},\ref{eq:lambdaTS2}).}
\label{tab:Lambda cal}
\end{table}

%\subsection{$\Lambda$ calculation from $T_2$}
In Table~\ref{tab:Lambda cal}, we list the various energy scales and energy band parameters for the device with optimal parameters, described in Sec.~\ref{sec:optimal}, as well as the $T_2^\Phi$ values calculated from the modeled energy levels. We also list the $\Lambda$ values for the various degrees of frustration, calculated both from the ratio of $T_2^\Phi$ values from the modeled energy levels and 
from $\Delta_{\rm SA}$ and $h_Z$. 
%\textcolor{teal}{\sout{ratio {\color{purple}$\Delta_{\rm SA}/h_Z$}} ratio of slope expressed by $\Delta_{\rm SA}$ and $h_Z$ }. 
We see that both approaches for computing $\Lambda$ are in agreement, consistent with our analysis described above.
%We extract the slope and curvature at $\Delta\Phi_{noise} = 10 \mu\Phi_0$ at different frustration to estimate the $T_2$. The result is shown in Table~\ref{tab:Lambda cal}. The $\Lambda$ calculated from $T_2$ match quite well with the $\Lambda$ calculated from $\Delta_{SA}$.

\end{document}